\documentclass[review,onefignum,onetabnum,hidelinks]{siamart190516}

\usepackage{lipsum}
\usepackage{amsfonts}
\usepackage{graphicx}
\usepackage{epstopdf}
\usepackage{algorithmic,amsmath}
\ifpdf
  \DeclareGraphicsExtensions{.eps,.pdf,.png,.jpg}
\else
  \DeclareGraphicsExtensions{.eps}
\fi

\newcommand{\topagg}{\text{t}}
\newcommand{\botagg}{\text{b}}

\newsiamremark{remark}{Remark}
\newsiamremark{hypothesis}{Hypothesis}
\crefname{hypothesis}{Hypothesis}{Hypotheses}
\newsiamthm{claim}{Claim}

\headers{Platelet Aggregation in an Extravascular Injury.}{K.G. Link, et al.}

\title{A mathematical model of platelet aggregation in an extravascular injury under flow. \thanks{Submitted on March 4, 2020.
\funding{This work was funded by the National Institutes of Health under grants R01HL120728 and R61HL141794 and National Science Foundation grant CBET-1351672.}}}

\author{Kathryn G. Link\thanks{Department of Mathematics, University of Utah, Salt Lake City, UT 
  (\email{link@math.utah.edu}, \url{http://www.math.utah.edu/\string~link}).}
  \and  Matthew G. Sorrells\thanks{Department of Chemical and Biological Engineering, Colorado School of Mines, Golden, CO.}
\and Nicholas A. Danes\thanks{Department of Applied Mathematics and Statistics, Colorado School of Mines, Golden, CO.}
\and Keith B. Neeves\thanks{Departments of Bioengineering and Pediatrics, Hemophilia and Thrombosis Center, University of Colorado Denver, Anschutz Medical Campus, Aurora, CO}
\and Karin Leiderman\footnotemark[4]
\and Aaron L. Fogelson\footnotemark[2] \thanks{Department of Biomedical Engineering University of Utah, Salt Lake City, UT.}}

\usepackage{amsopn,multicol}

\ifpdf
\hypersetup{
  pdftitle={Paper},
  pdfauthor={K.G. Link, M.G. Sorrells, N.A. Danes, K.B. Neeves, K. Leiderman, A.L. Fogelson}
}
\fi

\newcommand{\ck}[1]{\color{black}#1\normalcolor}
\usepackage{textcomp}

\begin{document}

\maketitle

\begin{abstract}
We present the first mathematical model of flow-mediated primary hemostasis in an extravascular injury which can track the process from initial deposition to occlusion. The model consists of a system of ordinary differential equations (ODE) that describe platelet aggregation (adhesion and cohesion), soluble-agonist-dependent platelet activation, and the flow of blood through the injury. The formation of platelet aggregates increases resistance to flow through the injury, which is modeled using the  Stokes-Brinkman equations. Data from analogous experimental (microfluidic flow) and partial differential equation models informed parameter values used in the ODE model description of platelet adhesion, cohesion, and activation. This model predicts injury occlusion under a range of flow and platelet activation conditions.  Simulations testing the effects of shear and activation rates resulted in delayed occlusion and aggregate heterogeneity. These results validate our hypothesis that flow-mediated dilution of activating chemical ADP hinders aggregate development. This novel modeling framework can be extended to include more mechanisms of platelet activation as well as the addition of the biochemical reactions of coagulation, resulting in a computationally efficient high throughput screening tool.
\end{abstract}



\section{Introduction}
\label{sec:intro}
Hemostasis is the \ck{first line of defense upon} vascular injury (either a disruption in the subendothelial lining or a breach in the vessel wall) \ck{whereby} a \ck{blood clot (thrombus) forms} to prevent the loss of blood \cite{valentino2010blood,Versteeg2013}.  The hemostatic response consists of two intertwined processes: platelet \ck{aggregate} formation and coagulation \cite{fogelson2015fluid,rana2016blood}. These processes are triggered when reactive proteins either in the extravascular matrix or on the subendothelium \ck{are exposed to} the blood plasma. Platelets are blood cells that circulate in the human vasculature in their unactivated state 
and mediate the biophysical and biochemical aspects of thrombus formation. \ck{Platelets become activated when in contact with collagen in the subendothelium or with chemical agonists in the plasma. Activated} platelet surfaces support coagulation reactions \ck{that} produce \ck{the enzyme} thrombin. Thrombin in turn cleaves the soluble blood protein fibrinogen into fibrin monomers, which polymerize to form a gel. This gel provides structure and stability to the aggregate. The size and structure of a\ck{n} aggregate as well as the time it takes \ck{the aggregate} to form, depend not only on platelet function and coagulation, but \ck{on} the local hydrodynamic environment. Blood flowing in a vessel is subject to a pressure difference across the vascular wall and any disruption of the wall can lead to loss of blood, \ck{which} greatly affect\ck{s} the delivery and removal of platelets,  chemical agonists, and coagulation proteins.

\ck{Upon extravascular injury, two important reactive proteins are exposed: collagen and tissue factor (TF)}. Blood cells and plasma containing coagulation proteins leak from the vessel into the extravascular space. Platelets \ck{flow into the injury and} begin to deposit\ck{; they adhere} to the collagen, \ck{and bind other insoluble agonists, which} in turn triggers the activation of key integrins on platelet surfaces \cite{Andrews2007glyc,Bledzka2013, Fogelson2015, Clemetson2007platelet,Versteeg2013}. The soluble agonist ADP, \ck{which is released from platelet granules into the fluid by activated platelets themselves,} interacts with platelet receptors P2Y$_{1}$  and P2Y$_{12}$ to additionally trigger exposure of \ck{surface} integrins \cite{Baurand2001,Hechler2011,Ohlmann2010}. Th\ck{is} effect of soluble agonist ADP is an example of activation without contact with the subendothelial matrix \cite{Andersen1999,Li2010}. Activated platelets provide surfaces to which more platelets can cohere, enhancing to the growth of the platelet \ck{aggregate} \ck{to prevent further} blood leakage. \ck{Platelet} aggregate formation \ck{depends strongly on both} the hydrodynamic environment and the formation and dissociation of bonds that \ck{enable} platelet adhesion and cohesion. \ck{F}ailure of any of the processes that mediate platelet aggregation, under \ck{a variety of} flow conditions, \ck{can} result in \ck{significant} blood loss. 

In recent years there has been an increased focus on understanding the molecular basis of bleeding disorders and associated variability in levels of key coagulation factors, platelet count, and insoluble proteins. Most clinical assays used to assess bleeding risk are performed under static conditions. However, data has shown that hydrodynamic forces affect platelet \ck{aggregation} and \ck{fibrin} formation, both key components of a thrombus \cite{muthard2012blood}. This has motivated the development of mathematical models of hemostasis that integrate platelet function, coagulation, and hemodynamics to predict bleeding risk based on measurable biochemical and biophysical factors. 

Our group has developed mathematical models of flow-mediated coagulation in an intravascular injury setting to predict regulatory mechanisms of hemostasis and thrombosis \cite{elizondo2016mathematical, Fogelson2012, Fogelson2006, Kuharsky2001,Leiderman2016, Leiderman2011, leiderman2013influence, link2018local, link2019mathematical}. These models use a continuum approach and express dynamics using differential equations. Our ordinary differential equation (ODE) models of thrombosis employ a well-mixed compartment assumption that accounts for transport by flow and diffusion using simplified mass-transfer coefficients. Other models fully account for spatial variations and transport to simulate small vascular injuries under flow \cite{Leiderman2011,Leiderman2013} through the use of partial differential equations (PDEs). Despite the numerous mathematical models of flow-mediated intravascular blood clotting, there are relatively few models of hemostasis. To our knowledge, \ck{the model we present here is} the first mathematical model to incorporate key biophysical and biochemical mechanisms of primary hemostasis in the framework of an extravascular injury. \ck{Our model is} based on ODEs and is calibrated and validated using analogous experimental (\Cref{sec:appA_methods}) and PDE (\Cref{sec:appA_PDE}) models \cite{danes2019density,Schoeman2017}.

One of the many contributions of this model is its computational efficiency, which allows for systematic exploration of large sets of parameters that would be impossible to resolve using the spatial PDE model or an experimental model. Parameters that contribute to bleeding include, but are not limited to, injury size, hemodynamics, vessel wall composition, and the anatomy of the adjacent perivascular and extravascular spaces. An experimental platform for studying hemostasis with a focus on collagen-TF induced thrombus formation within a model vascular wall was previously presented in \cite{Schoeman2017}. This experimental model is an {\it{in vitro}} microfluidic flow assay (MFA) we refer to as the ``bleeding chip".  Our mathematical model design is inspired by the specific `H'-geometry that is characteristic of the bleeding chip and \ck{will be used in the future to} quantify variability of MFAs \ck{and} identify components of the clotting system that underscore the known phenotypic variability in certain bleeding disorders \cite{nogami2015phenotypic}. The goal of this work \ck{is} to create \ck{an efficient, mechanistic} model that can be used for high-throughput screening \ck{of the blood clotting process and that enables} mathematically-driven\ck{, but testable} hypotheses.

The paper is organized as follows. In \Cref{sec:model}, we present a system of ordinary differential equations that describe platelet accumulation, soluble agonist-dependent platelet activation, and flow through the injury. Flow and aggregate calculations are calibrated using occlusion times and flow rates from the bleeding chip shown in \Cref{sec:results}. Also in \Cref{sec:results} are simulations to test the effects of flow and platelet activation rates on occlusion times. The results therein suggest that inhibition of platelet activation significantly modifies primary hemostasis. Additionally, flow-mediated dilution of ADP is shown to hinder aggregate development, as discussed in \Cref{sec:discussion}.
\begin{figure}[htbp]
  \centering
   \vspace{-0.25cm}
    \hspace{-5.5cm} {\bf{(A)}}  \hspace{6.25cm} {\bf{(B)}} \\
 \includegraphics[width = \textwidth]{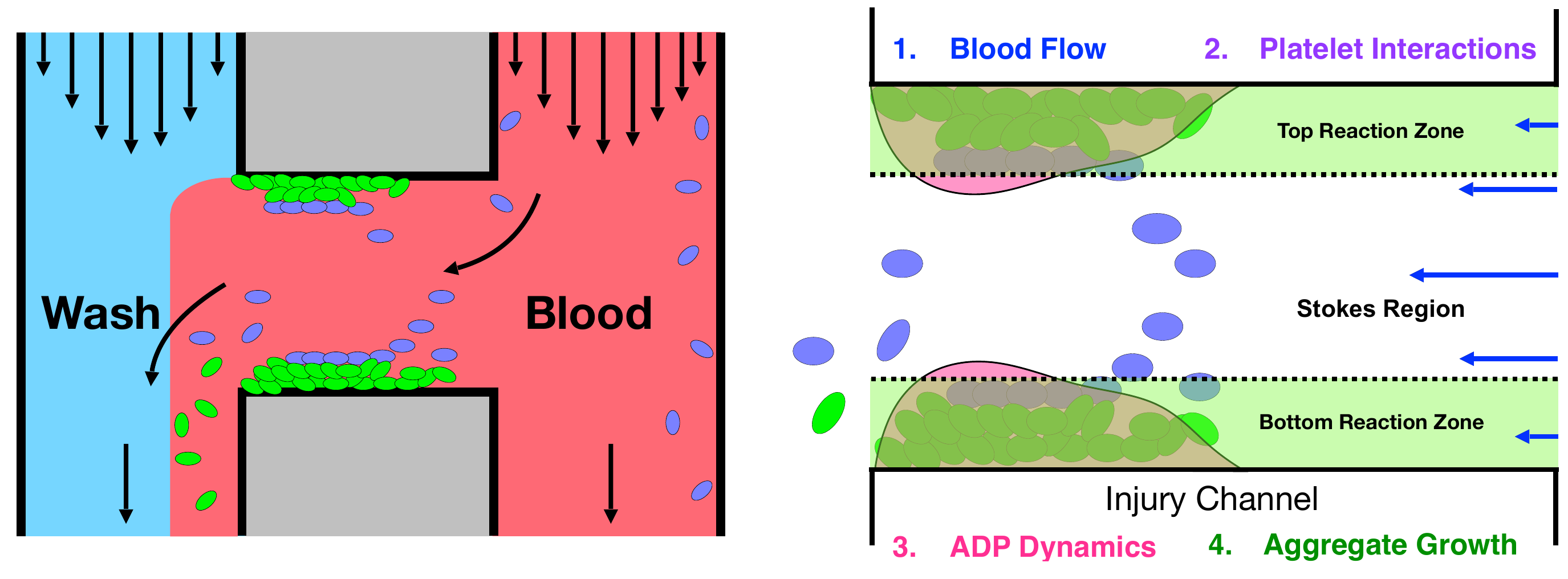}\\
 \vspace{-0.25cm}
    \caption{(A) {\bf{Bleeding chip.}} Blood and wash buffer are introduced into the two vertical channels of the bleeding chip. A horizontal injury channel connects the two vertical channels, where platelet deposition occurs \cite{Schoeman2017}. (B) {\bf{Mathematical model components.}} 1. Pressure driven flow with a Stokes-Brinkman layered velocity calculation. 2. Platelets' state (mobile/bound and unactivated/activated), mediated by activating chemical concentration [ADP], determines
adhesion to injury and cohesion to other platelets.
(3) ADP is released from bound platelets and is subject to transport by both advection and diffusion. (4) Aggregate growth is influenced by changes in thickness and porosity and feedbacks on blood flow due to increased frictional resistance.}
 \label{fig:model}
\end{figure}

\section{Mathematical Model}
\label{sec:model}
We model the situation where blood flows through a hole in the vessel
wall into the extravascular space. In this scenario, a disruption in
the endothelial layer occurs and exposes collagen to the blood.
Platelets are transported to the injury by flow and adhere to
collagen. This in turn activates them, leading to the release of
platelet agonist ADP and the recruitment of additional platelets that
cohere and form an aggregate. The model incorporates the flow of blood
through an `H'-geometry to mimic the bleeding chip, as described above
and as depicted in \cref{fig:model}A, and it explicitly models
aggregate formation under local fluid dynamic conditions. Components
of the model include descriptions of the blood flow through the
bleeding chip geometry and the injury channel as well as the platelet
interactions, soluble agonist ADP dynamics, and aggregate growth (see
\cref{fig:model}B).
\begin{figure}[!h]
  \centering
  \vspace{-0.25cm}
    \hspace{-4.0cm} {\bf{(A)}}  \hspace{3.5cm} {\bf{(B)}} \hspace{3.5cm} {\bf{(C)}}  \\
\vspace{-0.25cm}
  \begin{multicols}{3}
  \includegraphics[width = 0.315\textwidth]{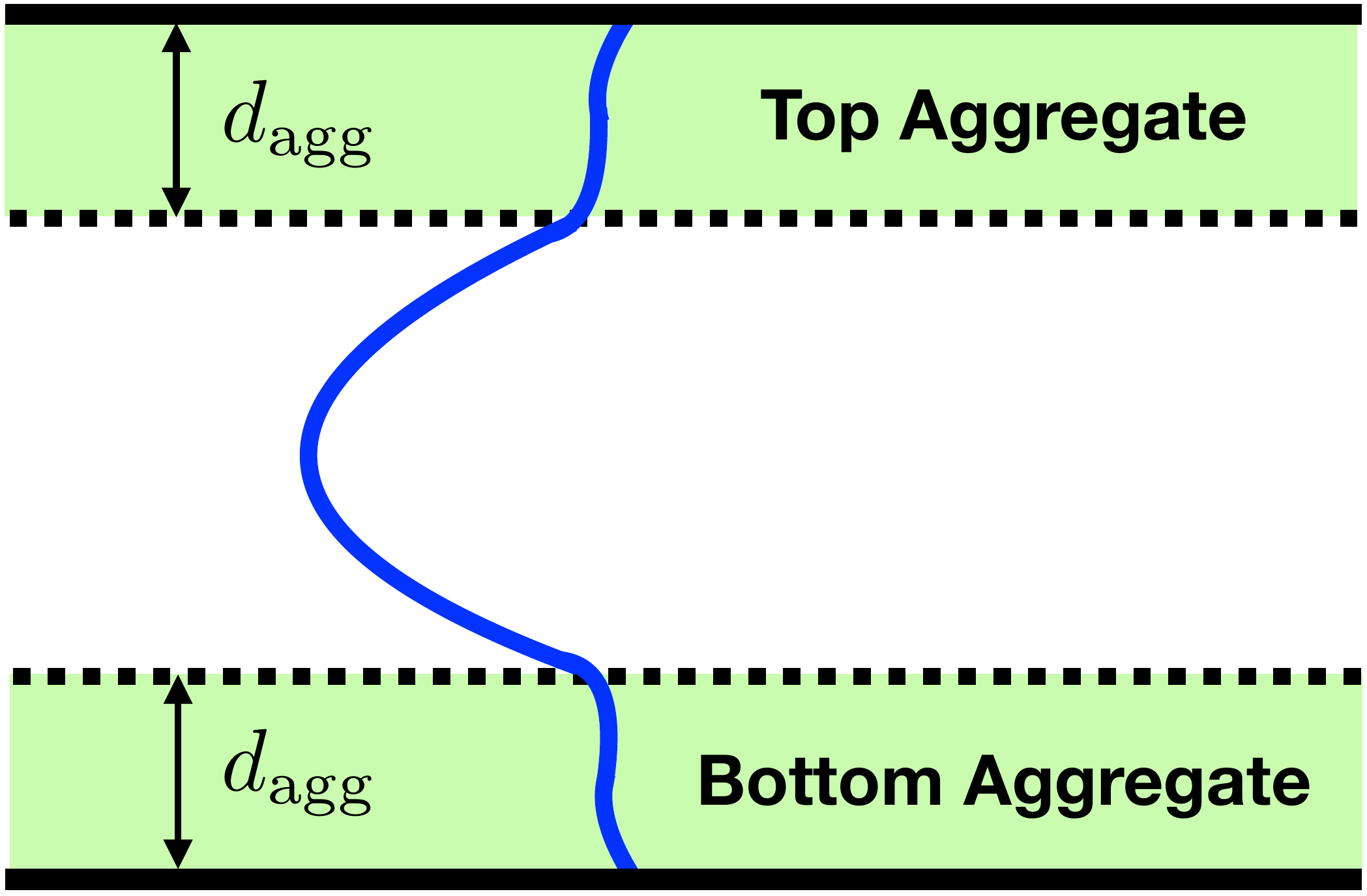}\\
   \includegraphics[width = 0.315\textwidth]{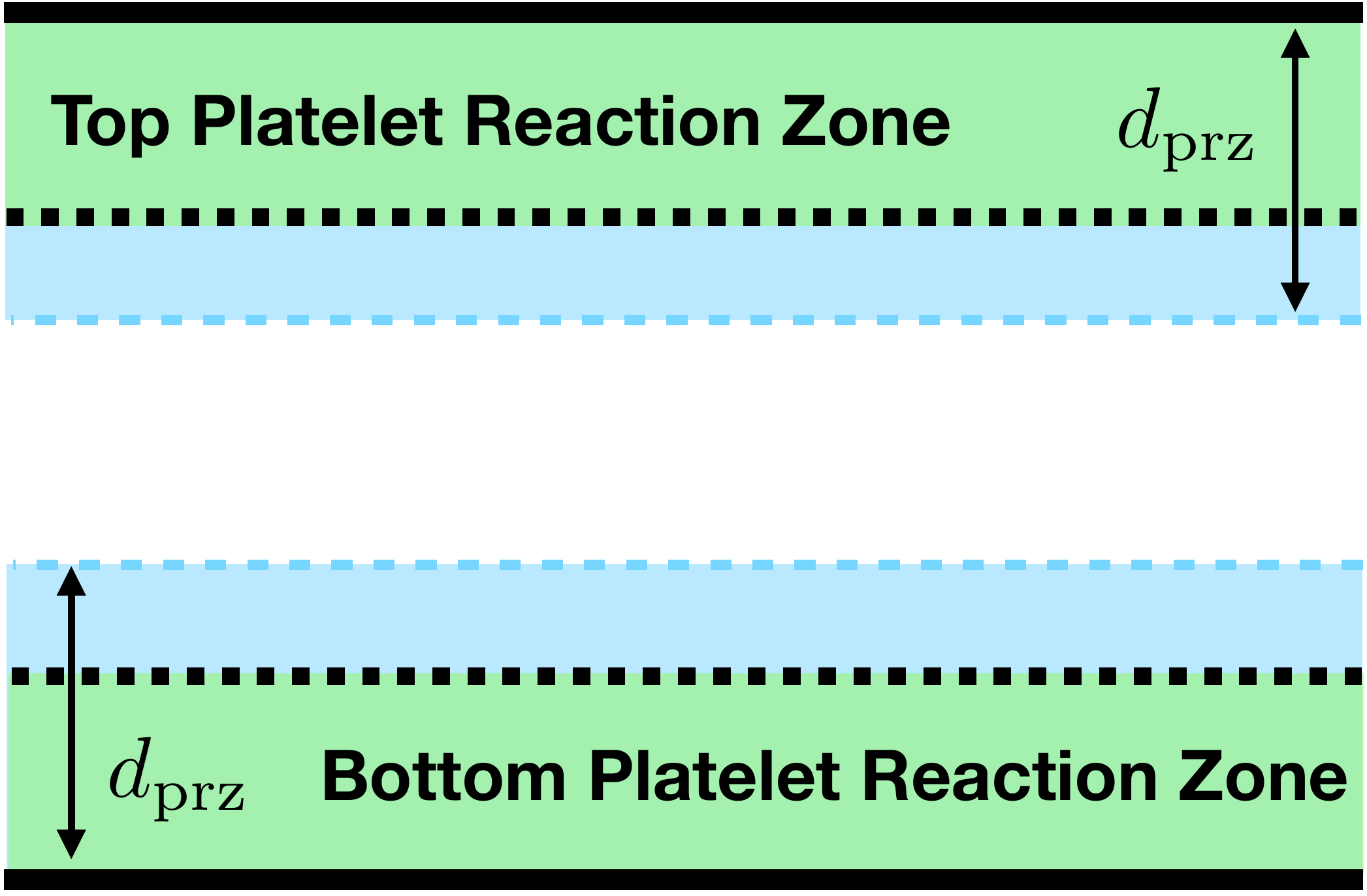}\\
   \includegraphics[width = 0.315\textwidth]{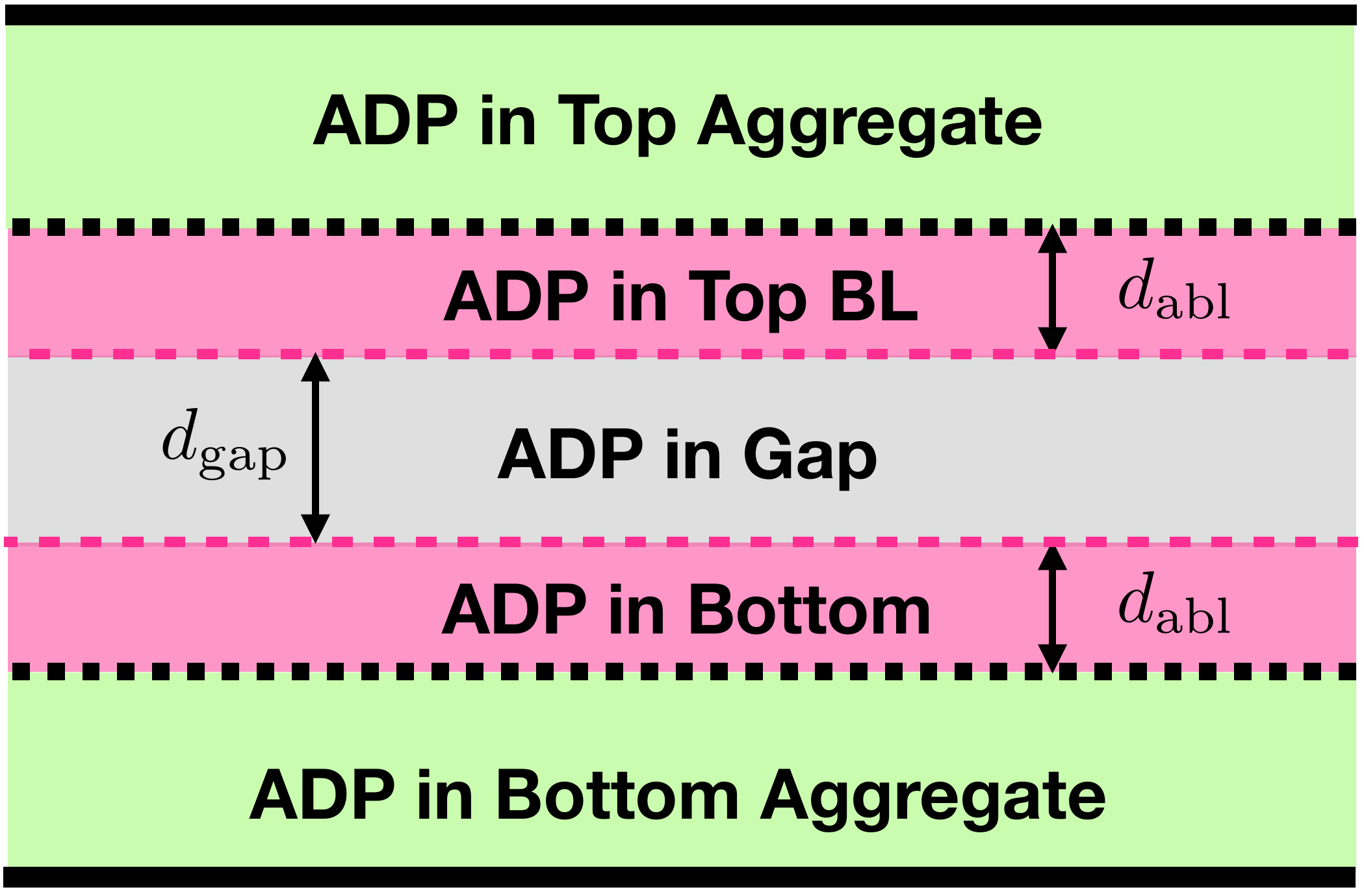}
 \end{multicols}
 \vspace{-0.5cm}
 \label{fig:model_compartments}
 \caption{{\bf{Spatial compartments of model.}} Each panel shows spatial compartments of the injury channel. (A) Two aggregate (Brinkman) compartments with thickness $d_{\text{agg}}$ with a volume of $V_{\text{agg}}$ and the Stokes region. (B) Two aggregate compartments and the platelet boundary layers (PBL), together defining platelet reaction zones (PRZ) with thicknesses of $d_{\text{prz}}$ and volumes $V_{\text{prz}}$ describe a two compartment platelet model. (C) Two aggregate compartments and the boundary layers associated with ADP (ABL) with thicknesses $d_{\text{abl}}$. The remaining compartment in the bulk of the Stokes region is denoted as the ``gap'' and has thickness $d_{\text{gap}}$. These layers define a five compartment ADP model.}
 \end{figure}

 Each component of the model involves a number of compartments which
 correspond to different spatial portions of the injury channel.  Two
 compartments are defined by the platelet aggregates on the bottom and
 top walls of the channel as shown in \cref{fig:model_compartments}A.
 Others are defined in terms of diffusive boundary layers for
 platelets (\cref{fig:model_compartments}B) and the activating
 chemical ADP (\cref{fig:model_compartments}C).  There are diffusive
 boundary layers associated with each of the aggregates.  Lastly,
 there is a compartment representing the `gap' between the boundary
 layers associated with the bottom and top aggregates shown in
 \cref{fig:model_compartments}C.  The two compartments of the platelet
 model and the five compartments describing the ADP model are defined
 precisely below.  The model unknowns are number densities for three
 populations of platelets and the concentration of ADP in the various
 compartments; the volume fraction of bound platelets in and the
 thickness of each of the aggregate compartments, and the fluid
 velocity profile across the channel.  The velocity profile is the
 only fully spatially-dependent variable in the model; it is used in
 defining the diffusive boundary layers and in defining advective
 fluxes of fluid, platelets, and ADP into and out of the various
 compartments.

\subsection{Fluid Dynamics}
\label{sec:fluid}
The mathematical representation of flow in this geometry is
nonstandard. Below, we describe how we model pressure-driven flow in
the injury channel and incorporate hinderance by the porous, growing
aggregate.  The flow of blood through the bleeding chip is described
in \Cref{sec:fluid_domain}. Output from this calculation determines
the inputs used in the Stokes-Brinkman calculation of the flow inside
of the injury channel where aggregates form. The
  resulting velocity profile is used to calculate the increased
  resistance in the injury channel and therefore alters the pressure
  drop $P_{1}-P_{2}$. Details of this calculation are found in
\Cref{sec:fluid_injury}.
\begin{figure}[htbp]
  \centering
  \begin{tabular}{ll}
{\bf{(A)}}  & {\bf{(B)}} \\
 \includegraphics[width = 0.465\textwidth]{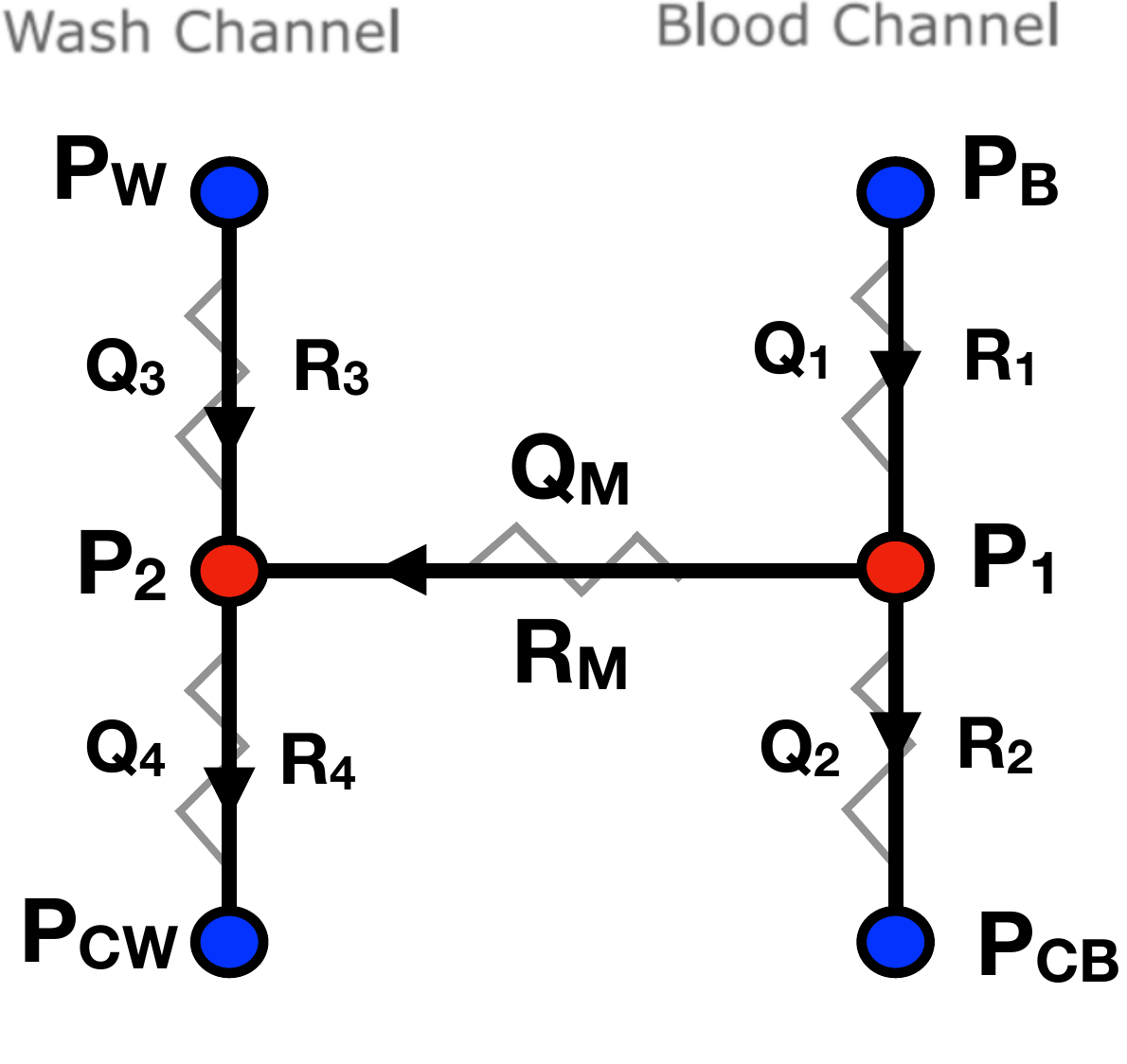}&
 \includegraphics[width = 0.465\textwidth]{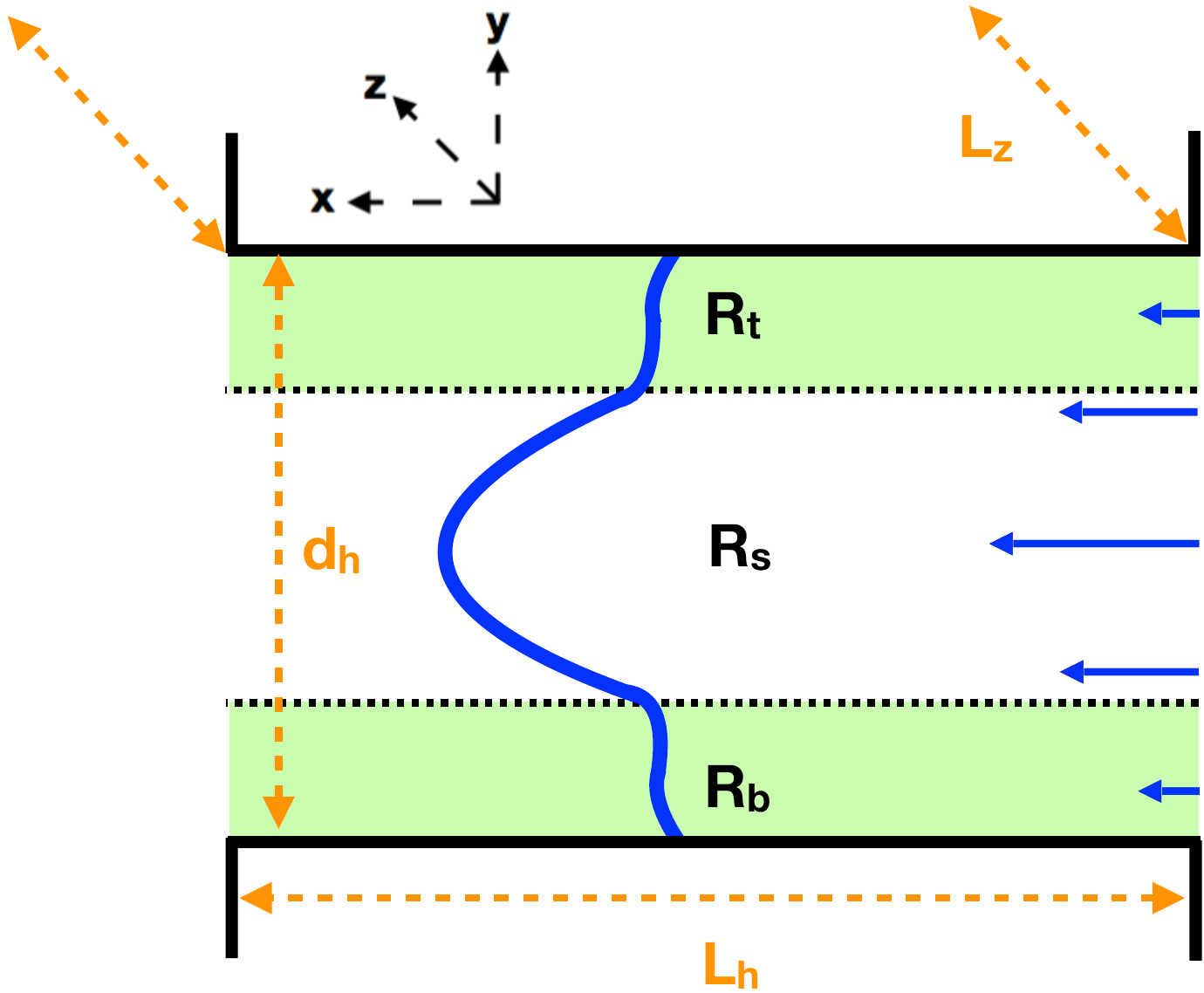}\\
 \end{tabular}
 \vspace{-0.5cm}
    \caption{{\bf{(A)}} Hydraulic circuit (HC) system describing flow through the bleeding chip. Quantities in {blue} are inputs while quantities in {red} are the desired output determined by solving system of equations in \Cref{sec:appA_circuit}. {\bf{(B)}} A schematic of the domain filled with  Brinkman regions {green} representing the bottom $R_{b}$ and top aggregates $R_{\topagg}$ and a Stokes region $R_{\text{s}}$ with a depth of $L_{z}$, width of $d_{h}$, and a length of $L_{z}$ denoted in {orange}. Flow moves from right to left and the resulting velocity profile is depicted in {blue}. }
 \label{fig:fluid}
\end{figure}
%
\subsubsection{Fluid through Extravascular Injury Domain}
\label{sec:fluid_domain} The bleeding chip contains two vertical
channels $ (L \times w \times L_{z}) = (10 \ \text{mm} \ \times \ 100 \ \mu
\text{m} \ \times 60 \ \mu \text{m})$, consisting of a `blood' channel
and a `wash' channel, connected by a horizontal `injury' channel
$(L_{h} \times d_{h} \times L_{z}) = (150 \ \mu \text{m}\ \times \ 20
\ \mu \text{m} \ \times 60 \ \mu \text{m})$, which forms an `H'-shaped
geometry (\cref{fig:model}A).  The flow rates and/or pressures
at the inlets and outlets are prescribed and blood flows into the device through the blood channel. A fraction of the blood exits through the outlet of the blood channel. The remaining blood makes its way through the injury channel (from right to left) and exits through the outlet of the injury channel into the wash channel. Two hydraulic resistors in series describe the flow in each of the blood and wash channels and a connecting resistor in parallel
describes the flow in the injury channel (\cref{fig:fluid}A). The resistances $R_{1}$, $R_{2}$, $R_{3}$, $R_{4}$ are input parameters; the resistance $R_{M}$ is obtained from our calculation of the flow in the injury channel as explained in the next section. By solving the linear system of equations for this hydraulic circuit (HC) as shown in \cref{sec:appA_circuit}, we determine the 
pressure gradient $G_{h} = (P_{1} - P_{2})/L_{h}$ across the injury
channel. The resulting pressure gradient $G_{h}$ is used to determine the
flow velocity through the injury channel. 
\subsubsection{Fluid through Injury}
\label{sec:fluid_injury}
The calculation of the fluid velocity profile at each time proceeds as
follows. We assume that we know the current thicknesses $d_{\botagg}$ and
$d_{\topagg}$ of the aggregate compartments (see
\cref{fig:model_compartments}A), and the volume fractions of bound
platelets $\theta_{\botagg}^{B}$ and $\theta_{\topagg}^{B}$ of the
aggregates. Define the intervals $R_{\botagg} = \{0 \le y \le
d_{\botagg}\}$, $R_{\text{s}} = \{d_{\botagg} \le y \le d_{\text{h}} -
d_{\topagg}\}$, and $R_{\topagg} = \{d_{\text{h}} - d_{\topagg} \le y
\le d_{\text{h}}\}$. Given that the Reynolds number in the injury
channel of the bleeding chip is $R_{e} \approx 0.0283 < < 1$
as calculated in \Cref{hca:flow_params}, the
unidirectional flow in the $x$-direction for $y \in R_{\text{s}}$ is modeled
by the Stokes equations, and the flows for $y \in R_{\botagg}$ and $y \in
R_{\topagg}$ are modeled by the Brinkman equations, as we have done in
previous studies \cite{Leiderman2016,Leiderman2011}. These are the Stokes equations
modified by the inclusion of a Brinkman drag term among the forces
acting on the fluid. The fluid dynamics equations are
\begin{align}
& 0 = -\frac{\partial p_{\botagg}}{\partial x} + \mu \frac{\partial^{2}u_{\botagg}}{\partial y^{2}} - \mu\alpha_{\botagg}(\theta_{\botagg}^{B}) u_{\botagg}, \ \ \ \  \text{ for } y \in R_{\botagg}, \label{pde1}\\
& 0 = -\frac{\partial p_{\text{s}}}{\partial x} +  \mu \frac{\partial^{2}u_{\text{s}}}{\partial y^{2}}, \ \ \ \ \ \ \ \ \ \ \ \ \ \ \ \ \ \ \ \ \ \ \  \text{ for } y \in R_{\text{s}}, \label{pde2}\\
& 0 = -\frac{\partial p_{\topagg}}{\partial x} + \mu \frac{\partial^{2}u_{\topagg}}{\partial y^{2}} - \mu\alpha_{\topagg}(\theta_{\topagg}^{B}) u_{\topagg}, \ \ \ \ \ \ \text{ for }y \in R_{\topagg}. \label{pde3}
\end{align}
\noindent where we assume the density of the fluid is $\rho = 1 \ g/cm^{3}$ and the Brinkman coefficients $\alpha_{\botagg}$ and $\alpha_{\topagg}$ are functions of $\theta_{\botagg}^{B}$ and $\theta_{\topagg}^{B}$, respectively. These differential equations are supplemented with no-slip boundary conditions $u_{\botagg}(0) = 0$ and $u_{\topagg}(d_{h}) = 0$ as well as matching conditions at the edge of each aggregate compartment. The matching conditions are that the velocity $u$ and the shear stress $\mu \frac{du}{dy}$ are continuous for all $y$. The pressure gradient $\tfrac{\partial p}{\partial x}$ in the injury channel is assumed to be independent of $y$, so it is a constant $G_{h}$ determined from the circuit calculation found in \cref{sec:appA_circuit}.

To relate the volume fraction of bound platelets $\theta^{B}_{\botagg}, \theta^{B}_{\topagg}$ to the permeability ($1 / \alpha_{\botagg}$, $1 / \alpha_{\topagg}$) of the Brinkman layers, the functions $\alpha_{\botagg}$ and $\alpha_{\topagg}$ are defined with the widely-used Kozeny-Carman relation \cite{mccabe2005unit}. 
\begin{align*}
\alpha(\theta) = C_{K}\frac{\theta^{2}}{(1-\theta)^{3}},
\end{align*} 
and $C_{K}$ is a fitted parameter whose estimation is discussed
in \Cref{sec:appC}. Because of the linearity of the differential
equations above \eqref{pde1}-\eqref{pde3}, the velocity in each
compartment is a linear combination of exponential functions of $y$,
with a total of six unknown coefficients. We determine these
coefficients by solving the linear system of equations corresponding
to the two boundary conditions and the four matching conditions (\cref{sec:appB}). For later convenience, we define the velocity as $u(y) =
(u_{\botagg}(y)\text{ for } y\in R_{\botagg}\text{, }u_{\text{s}}(y)$ $\text{ for }y\in
R_{\text{s}}\text{, }u_{\topagg}(y)\text{ for }y\in R_{\topagg})$. 

In practice, we first determine a preliminary velocity field by taking $G_{h} = 1$. This allows us to calculate $R_{M} = \tfrac{L_{h} \cdot 1}{\int_{0}^{d_{h}}u(y)dy}$ for use in the circuit analysis described in \Cref{sec:appA_circuit}. Once the actual $G_{h}$ is known from the circuit analysis, the velocity is obtained by multiplying the preliminary velocity by $G_{h}$. The velocity profile and the upstream concentration of platelets dictate the transport of mobile platelets and ADP to and from the site of injury.  Platelets and ADP both move by advection and diffusion. For ADP, diffusion is Brownian. For platelets, diffusion is used as a model for the effects on platelet motion of local flow disturbances generated by the complex motions of the deformable red blood cells that make up approximately 40\% of the
bloods volume \cite{skorczewski2013platelet}. The diffusion of
platelets and ADP to and from the walls/growing aggregates define the
regions in which platelet activation, adhesion and cohesion can
occur. For clarity and brevity, all descriptions of the dynamics of
platelets, soluble agonist ADP, and aggregate growth will be
associated with the bottom wall of the injury, although there are
similar equations for dynamics near the top wall. We therefore will
temporarily drop the subscript $b$ that denotes `bottom'.

\subsection{Platelets}
\label{sec:plts}
We consider a three species model of platelet aggregation involving
the number densities (plt/$\mu$L) of unactivated and activated mobile
platelets $P^{m,u}$, $P^{m,a}$ in the platelet reaction zone (PRZ) and
bound platelets $P^{b,a}$ in the growing aggregate region. The PRZ is
the union of the aggregate region and the adjacent platelet boundary
layer (PBL) (\Cref{fig:model_compartments}B). Note the superscripts
$m$ and $b$ are associated with a platelet's state of mobility (mobile
or bound) and the superscripts $u$ and $a$ indicate its state of
activation (unactivated or activated). The first two platelet
populations are mobile and are subject to the effects of flow within
and (sufficiently) near the growing aggregate. $P^{m,u}$, enter
upstream and leave downstream whereas $P^{m,a}$ only leave downstream
because we have assumed that no mobile, activated platelets come into
the injury channel from upstream.  $P^{m,u}$ can adhere to the wall
and therefore become bound and activated. Activated, mobile platelets
can either adhere to the wall or cohere to $P^{b,a}$. In addition to
adhesion, $P^{m,u}$ are activated by exposure to the soluble agonist
ADP. Platelet adhesion and cohesion increase the volume of bound
platelets in the aggregate and consequently increase the thickness and
volume fraction of bound platelets of each aggregate.  They thus
indirectly affect the delivery/removal of mobile platelets by flow.

\subsubsection{Coupling Flow to Platelet Transport} 
\label{sec:BL_calc} 
\begin{figure}[!h]
  \centering
  \vspace{-0.3cm}
    \hspace{-5.5cm} {\bf{(A)}}  \hspace{6.0cm} {\bf{(B)}} \\
\vspace{-0.3cm}
  \begin{multicols}{2}
  \includegraphics[width = 0.48\textwidth]{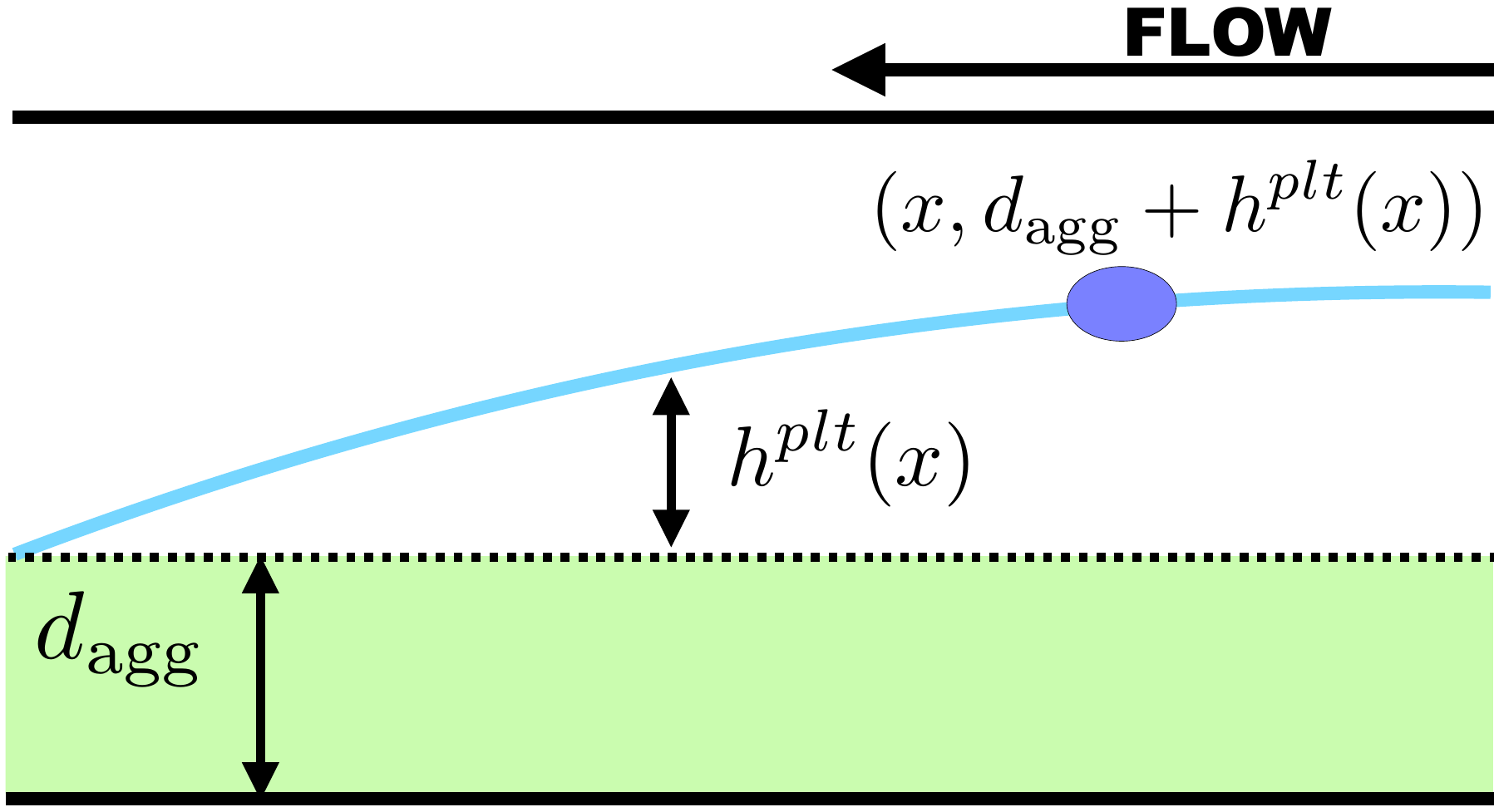}\\
   \includegraphics[width = 0.48\textwidth]{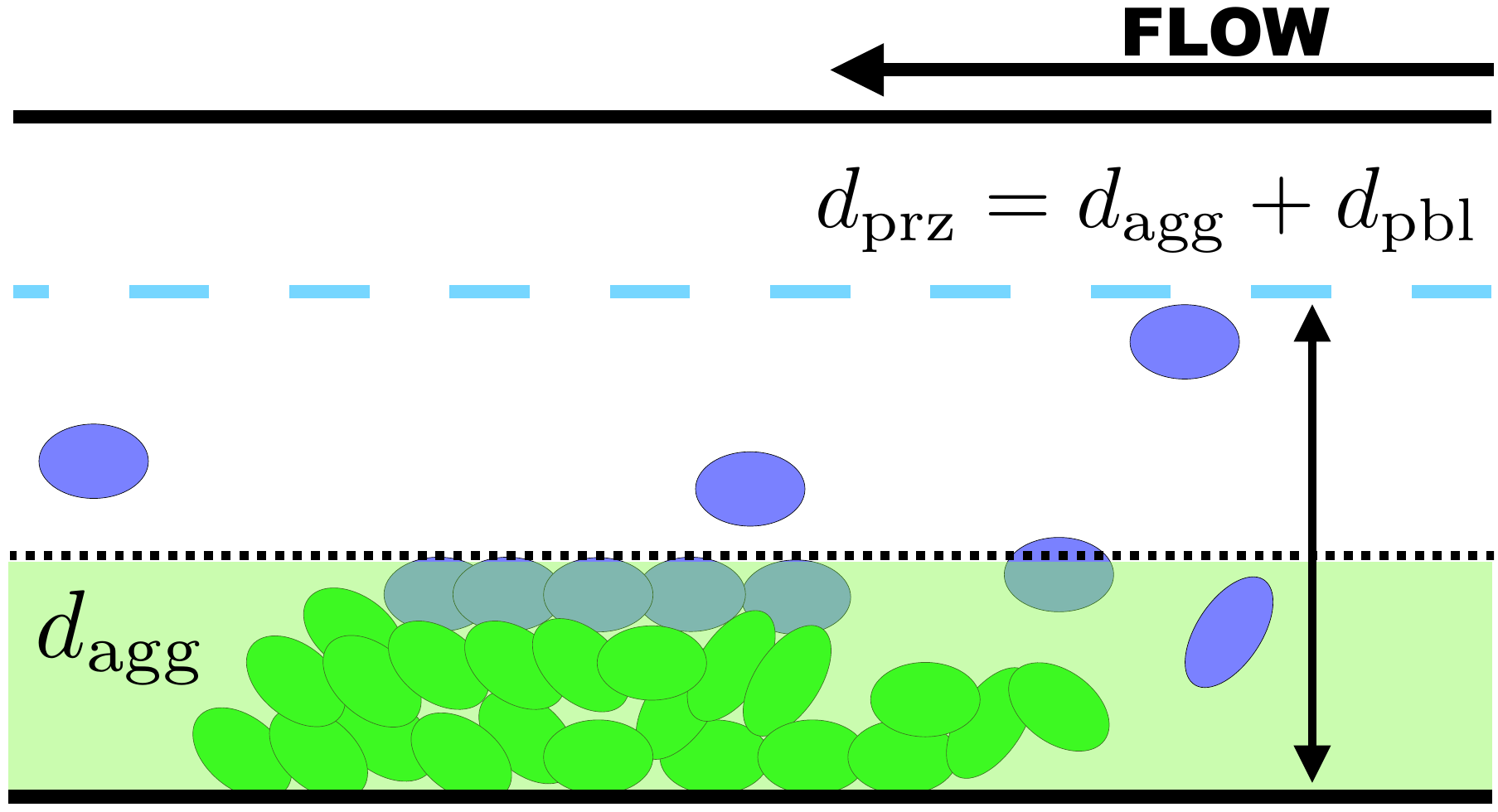}
 \end{multicols}
 \vspace{-0.5cm}
    \caption{{\bf{Boundary layers associated with the transport of platelets.}} Blood flows from right to left, entering the injury channel at $x = 0$ and exiting downstream at $x = L_{h}$. {\bf{(A)}} Consider a platelet at the position $(x,d_{\text{agg}} + h^{plt}(x))$. The distance it must travel to reach the edge of the growing aggregate is $h^{plt}(x)$ (cyan) and the thickness of the aggregate is denoted as $d_{\text{agg}}$. {\bf{(B)}} The PBL thickness at the entrance of the injury channel $h^{plt}(0)$ (cyan, dashed) defines the reaction zone for platelet interactions in that $d_{prz} = d_{\text{agg}} + d_{\text{pbl}}$, where $d_{\text{pbl}} = \min(h^{plt}(0),d_{se})$ and $d_{se} = d_{\text{plt}} = 2 \ \mu$m.}
\label{fig:BL_plt}
\end{figure}

Similarly to what we did in \cite{Fogelson2006,Kuharsky2001}, we
approximate the effects of diffusive motion in conjunction with
advective transport by defining appropriate boundary layers, the
regions from inside of which platelets can reach the growing aggregate
before being carried away downstream.  We estimate the thickness of
the boundary layer as follows (our estimates agree very well with
those in \cite{Cussler2009}).  We denote the thickness of the bottom
aggregate as $d_{\text{agg}}$ and we define a platelet boundary layer
as follows. For each $x \in (0, L_{h})$, let $h^{plt}(x)$ be defined
by equating the typical time it would take a platelet at location $(x,
d_{\text{agg}}+h^{plt}(x))$ to diffuse to the edge of the aggregate
with the time it takes that platelet to be carried to the outlet of
the injury by the flow:
$$
\frac{h^{plt}(x)^{2}}{2D_{p}} = \frac{L_{h} - x}{u(d_{\text{agg}} + h^{plt}(x))}.
$$
Solving for $h^{plt}(x)$ defines a platelet boundary layer for
platelets as shown in \cref{fig:BL_plt}A). Because the platelet has a
non-vanishing size, we modify our definition of the top edge of the
platelet boundary layer to $d_{\text{pbl}}(x) =
\min(h^{plt}(x),d_{\text{plt}})$, where $d_{\text{plt}}$ is the
diameter ($\approx 2 \mu$m) of a platelet.  We are interested in the
advective flux of platelets into and out of this compartment, shown in
\cref{fig:BL_plt}B. Transport into this compartment at the inlet of
the injury channel depends on the inlet platelet concentration and the
fluid velocity $u(y)$ for $d_{\text{agg}} < y < d_{\text{agg}} +
d_{\text{pbl}}(0)$. While the boundary layer narrows between the inlet
and the outlet of the injury channel, platelets can leave the boundary
layer by advection for this same range of $y$ values. Hence, letting
$d_{\text{pbl}} = d_{\text{pbl}}(0)$, we define the PBL compartment
associated with the bottom aggregate to be those points in the injury
channel with $d_{\text{agg}} < y < d_{\text{agg}} +
d_{\text{pbl}}$. For the bottom aggregate, the PRZ is defined by $0
\le y \le d_{\text{prz}} = d_{\text{agg}} + d_{\text{pbl}}$. A similar
construction is used to define the platelet boundary layer compartment
adjacent to the aggregate on the top injury channel wall.

The advective transport of platelets into and out of each of the
platelet compartments is hindered to a degree that depends on the
fraction of each compartment's volume that is occupied by
platelets. Let $\theta^{M} = v_{\text{plt}}(P^{m,u} + P^{m,a})$ and
$\theta^{B} = v_{\text{plt}}P^{b,a}$, where $v_{\text{plt}}$ is the
volume of an individual platelet. The quantities $\theta^{M}$ and
$\theta^{B}$ are the volume fractions occupied by mobile and bound
platelets, respectively. Note that bound platelets are found only in
the aggregate compartments while mobile platelets can be found
throughout the injury channel, but we track separately the number
densities of mobile platelets in the two PRZs. To describe the
transport of mobile, unactivated platelets into the injury channel, we
define the quantity
\begin{align*}
J_{\text{prz}}^{in} = & \Big(\int_{d_{\text{agg}}}^{d_{\text{prz}}} u(y) P^{up}(y) dy\Big)W(\theta^{M}) \ +  \ \Big(\int_0^{d_{\text{agg}}} u_{\text{}}(y) P^{up}(y) dy \Big) W(\theta^{M} + \theta^{B}),\\
\end{align*}
where $P^{up}(y)$ is the number density of mobile, unactivated
platelets at the inlet to the injury channel. In this work, we assume
a uniform and constant upstream distribution $P^{up}(y) = P^{up,*}$,
but using non-uniform distributions is straightforward (\Cref{sec:appD}). Note that the use of $P^{up,*}$ in the model
yields symmetric top and bottom aggregates. The model framework allows
for asymmetric aggregate formation if $P^{up}(y)$ is not symmetric and
simulations found in \Cref{sec:appD} explore the resulting
aggregates. The function $W$ is intended to limit platelet entry and
movement through the region in which the volume fraction of platelets
is already high. We used the specific function
$$W(\theta) = \tanh(\pi(1-(\theta / \theta^{\max}))),$$
which monotonically decreases from $W(0) = 1$ to $W(1)$ = 0
\cite{Leiderman2011}. Therefore, there is no hinderance when the
volume fraction is $0$ and total hinderance as the volume fraction
approaches $1$. In choosing this function, we assume that the ability
for mobile platelets to move into a region is only gradually impaired
until $\theta$ nears $\theta^{\max} = 0.6$ and then drops quickly. The
rate (number/time), at which mobile, unactivated platelets enter the
PRZ is $L_{z}J^{in}_{\text{prz}}$.

We define the following quantities to describe the transport of
unactivated and activated mobile platelets out of the injury
channel. The rates at which these platelets leave the PRZ are
$L_{z}J_{\text{prz}}^{out,u}$ and $L_{z}J_{\text{prz}}^{out,a}$,
respectively, where
\begin{align*}
J_{\text{prz}}^{out,u} =& \  \Bigg[\Big(\int_{d_{\text{agg}}}^{d_{\text{prz}}}u(y)  dy\Big)W(\theta^{M}) \ +  \ \Big(\int_0^{d_{\text{agg}}} u(y) dy\Big) W(\theta^{M} + \theta^{B})\Bigg] P^{m,u},\\ \label{eq:Iu1}
J_{\text{prz}}^{out,a} =& \  \Bigg[\Big(\int_{d_{\text{agg}}}^{d_{\text{prz}}}u(y)  dy\Big)W(\theta^{M}) \ +  \ \Big(\int_0^{d_{\text{agg}}} u(y) dy\Big) W(\theta^{M} + \theta^{B})\Bigg] P^{m,a}.
\end{align*} 
The flow of blood in the injury channel is coupled to the transport
of platelets into and out of the injury channel as we describe below.

\subsubsection{Cohesion, Adhesion, and Dilution}
\label{sec:coh_adh_dil}
Platelets contribute to aggregate formation through two binding
processes: adhesion and cohesion. Both unactivated and activated
mobile platelets can adhere directly to the bottom or top wall of the
injury channel. Mobile, activated platelets can cohere to already
bound, activated platelets. Platelets are activated through direct
contact with the wall or by exposure to the soluble agonist ADP. The
activation state of the platelet as well as the available space both
on the wall and in the developing aggregate determine which binding
processes can occur.

Since adhesion requires that a platelet directly contacts the wall, we
limit adhesion to those platelets within a specified distance $d_{se}$ from the walls. In the model, a platelet's location
relative to the walls is only known with respect to whether the
platelet is inside one of the PRZs, and so we enforce this limit
approximately in the adhesion rate function
\begin{equation*}
k_{\text{adh}}(d_{\text{agg}},\theta^{B}) = k_0^{adh}\min\bigg(1-\frac{d_{\text{agg}}}{d_{se}},0\bigg)\bigg(\theta^{\max} - \theta^{B}\bigg), \label{eq:kadh1}
\end{equation*}
where $ k_0^{adh}$ is a first order rate constant, the second term
represents the accessibility of the wall, and the third term
represents the porosity of the aggregate. With this function, adhesion
is further limited as the volume fraction of platelets in the
aggregate approaches its maximum value $\theta^{\max}$, because
platelets already bound to the wall occupy a portion of the wall space
available for adhesion. Both unactivated and activated mobile
platelets can adhere to the wall and hence the rate of adhesion is
proportional to the adhesion rate function and the total number
density of mobile platelets.

For mobile platelets to cohere to the aggregate, they must be
activated and sufficiently close to activated bound platelets. As
previously mentioned, activated platelets have activated integrin
receptors on their surfaces that allow for
$\alpha_{IIb}\beta_{3}$-fibrinogen-$\alpha_{IIb}\beta_{3}$ and
$\alpha_{IIb}\beta_{3}$-vWF-$\alpha_{IIb}\beta_{3}$ bond
formation. These bonds are strong, and long-lived with very small off
rates. Therefore, we assume that bond breakage is negligible and do
not allow for platelet detachment in this model. The rate of
platelet-platelet cohesion, $k_{\text{coh}}P^{b,a}P^{m,a}$, depends on the
second order rate constant $k_{\text{coh}}$ and the number densities of bound
activated platelets, $P^{b,a}$, and mobile, activated platelets
$P^{m,a}$, respectively.

In addition to changing the number of bound platelets in the aggregate
and the flow through the injury channel, the processes of adhesion and
cohesion change the total volume of the aggregate and of the adjacent
boundary layers (and therefore of the PRZs) by increasing their
thicknesses. For a given number of platelets in the PRZ, an increase
in that compartment's volume results in a decrease in the number
density of platelets in it -- an effect we refer to as dilution. For
example, the term describing this effect on mobile unactivated
platelets has the form
$$
 \Bigg(\frac{\frac{d}{dt} (d_{\text{prz}})}{d_{\text{prz}}}\Bigg) P^{m,u}_{\text{}},
$$
and appears naturally in calculating the rate at which the number of
these platelets in the platelet reaction zone
changes. 
%
\subsubsection{Evolution Equations of Platelets}
\label{sec:plt_eqs}
Here we account for the processes described in \Cref{sec:BL_calc} and
\Cref{sec:coh_adh_dil} to derive evolution equations for the platelet
number densities. Let $V_{\text{prz}} = L_{h} L_{z} d_{\text{prz}}$
denote the volume of the platelet reaction zone. Then the rate of
change of the number of mobile, activated platelets in that reaction
zone is
\begin{align*}
\underbrace{\frac{d (V_{\text{prz}}P^{m,u})}{dt}}_{\text{rate of change of \# plts}}  =& \  \underbrace{L_{z} (J_{\text{prz}}^{in} -
J_{\text{prz}}^{out,u})}_{\text{flow in/out of reaction zone}} -  \underbrace{V_{\text{prz}}k_{\text{act}} P^{m,u}_{\text{}}}_{\text{activation}} - \ \underbrace{V_{\text{prz}}k_{\text{adh}} P^{m,u}_{\text{}}}_{\text{adhesion}}
\end{align*}
Applying the product rule on the left-hand side and rearranging terms, we determine that 
\begin{align}
\frac{d  P^{m,u}}{dt}  =& \  \underbrace{\frac{(J_{\text{prz}}^{in} -
J_{\text{prz}}^{out,u})}{L_h d_{\text{prz}} }}_{\text{flow in/out of reaction zone}} -  \underbrace{k_{\text{act}}P^{m,u}_{\text{}}}_{\text{activation}}
- \ \underbrace{k_{\text{adh}} P^{m,u}_{\text{}}}_{\text{adhesion}}
 - \underbrace{\Bigg(\frac{\frac{d}{dt} (d_{\text{prz}})}{d_{\text{prz}}}\Bigg) P^{m,u}_{\text{}}}_{\text{dilution}}. \label{eq:Pu1} 
 \end{align}
 The first and third terms describe transport and adhesion. The
 second term describes activation of mobile unactivated platelets by
 exposure to ADP. Based on similar considerations, we find evolution
 equations for the number densities of the other platelet species
 $P^{m,a}$ and $P^{b,a}$.
\begin{align}
\frac{d P^{m,a}_{\text{}}}{dt}    = &  \ \underbrace{- \frac{ J_{\text{prz}}^{out,a}}{L_h d_{\text{prz}} }}_{\text{flow out of reaction zone}} \ + \ \underbrace{k_{\text{act}}P^{m,u}_{\text{}}}_{\text{activation}} - \underbrace{k_{\text{adh}} P^{m,a}_{\text{}}}_{\text{adhesion}}\\
  -& \ \underbrace{k_{\text{coh}} P^{b,a}_{\text{}}  P^{m,a}_{\text{}}}_{\text{cohesion}}
\ - \ \underbrace{\Bigg( \frac{\frac{d}{dt} (d_{\text{prz}})}{d_{\text{prz}}} \Bigg) P^{m,a}}_{\text{dilution}}, \nonumber  \label{eq:Pa1} 
\end{align}
\vspace{-0.5cm} 
\begin{align}
\frac{d P^{b,a}_{\text{}}}{dt}  = & \   \underbrace{\frac{d_{\text{prz}}}{d_{\text{agg}}} \Big(k_{\text{adh}}( P^{m,u}_{\text{}} + P^{m,a}_{\text{}})  +  k_{\text{coh}} P^{b,a}_{\text{}} P^{m,a}_{\text{}} \Big)}_{\text{adhesion and cohesion}} - \ \underbrace{\Bigg(\frac{\frac{d}{dt} (d_{\text{agg}})}{d_{\text{agg}}} \Bigg) P^{b,a}_{\text{}}}_{\text{dilution}}. 
\end{align}
%

\subsection{Soluble Agonist ADP}
\label{sec:ADP}
An unactivated platelet stores quantities of ADP in its dense granules
and releases this ADP into the surrounding fluid when the platelet is
activated.  Similar to the treatment in \cite{Leiderman2011}, we
assume this release occurs during the time interval 1-5s after
activation. Once released, that ADP moves by advection and diffusion
within and out of the aggregate where it can activate mobile
platelets. Describing these processes requires defining two boundary
layers: one within the aggregate $h^{adp}_{\text{agg}}(x)$ and one in the
Stokes region $h^{adp}_{\text{BL}}(x)$. These boundary layers are used to
define the ADP compartments shown in \cref{fig:model_compartments}C
and the ADP concentrations tracked are those inside the aggregate
$[ADP]_{\text{agg}}$, in the Stokes region boundary layer
$[ADP]_{\text{BL}}$, and in the remaining Stokes region which we
denote as $[ADP]_{\text{gap}}$.
%
\begin{figure}[!h]
  \centering
    \vspace{-0.3cm}
   \hspace{-6.0cm} {\bf{(A)}}  \hspace{6.0cm} {\bf{(B)}} \\
  \vspace{-0.3cm}
  \begin{multicols}{2}
  \includegraphics[width = 0.48\textwidth]{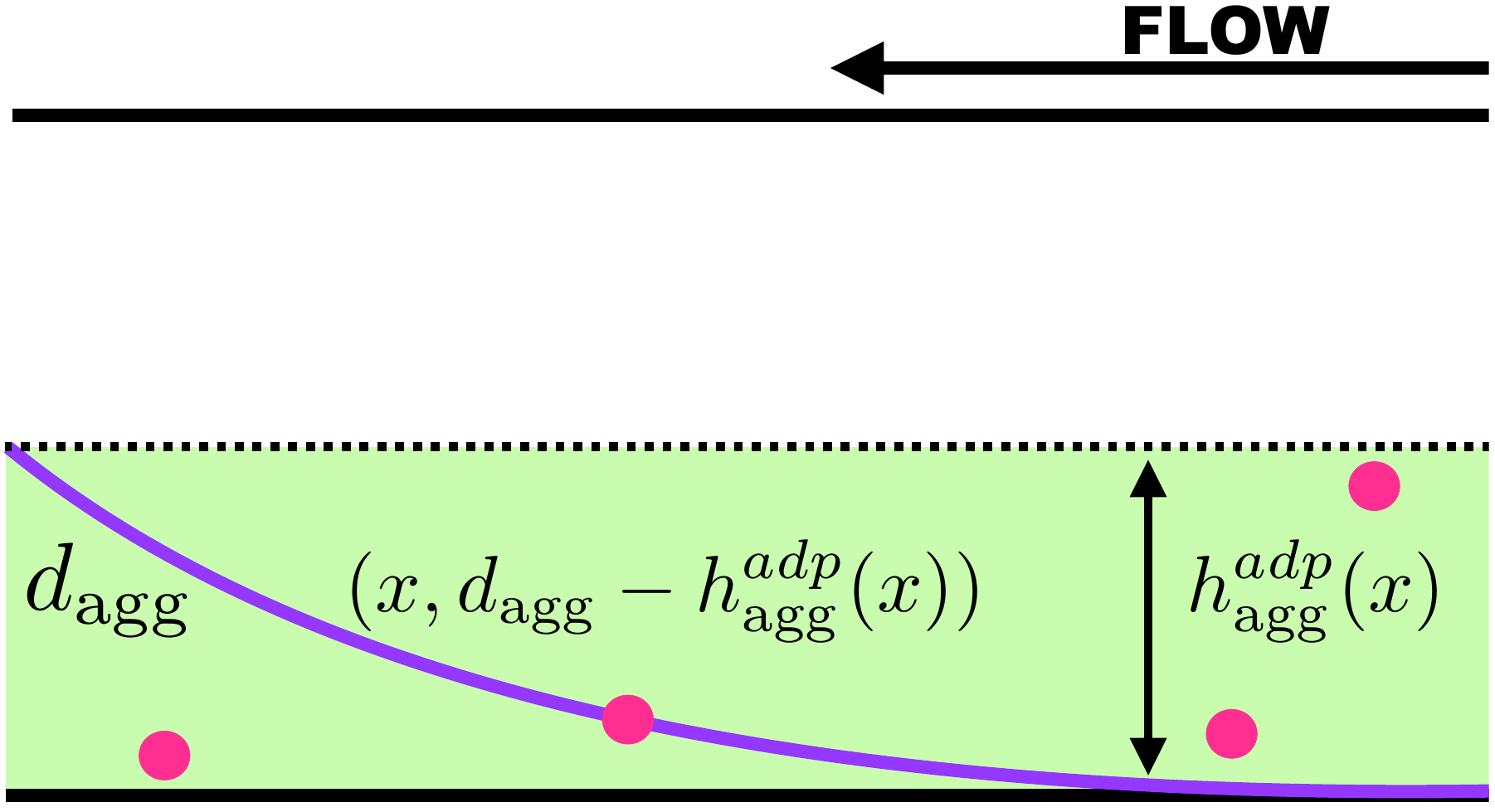}\\
   \includegraphics[width = 0.48\textwidth]{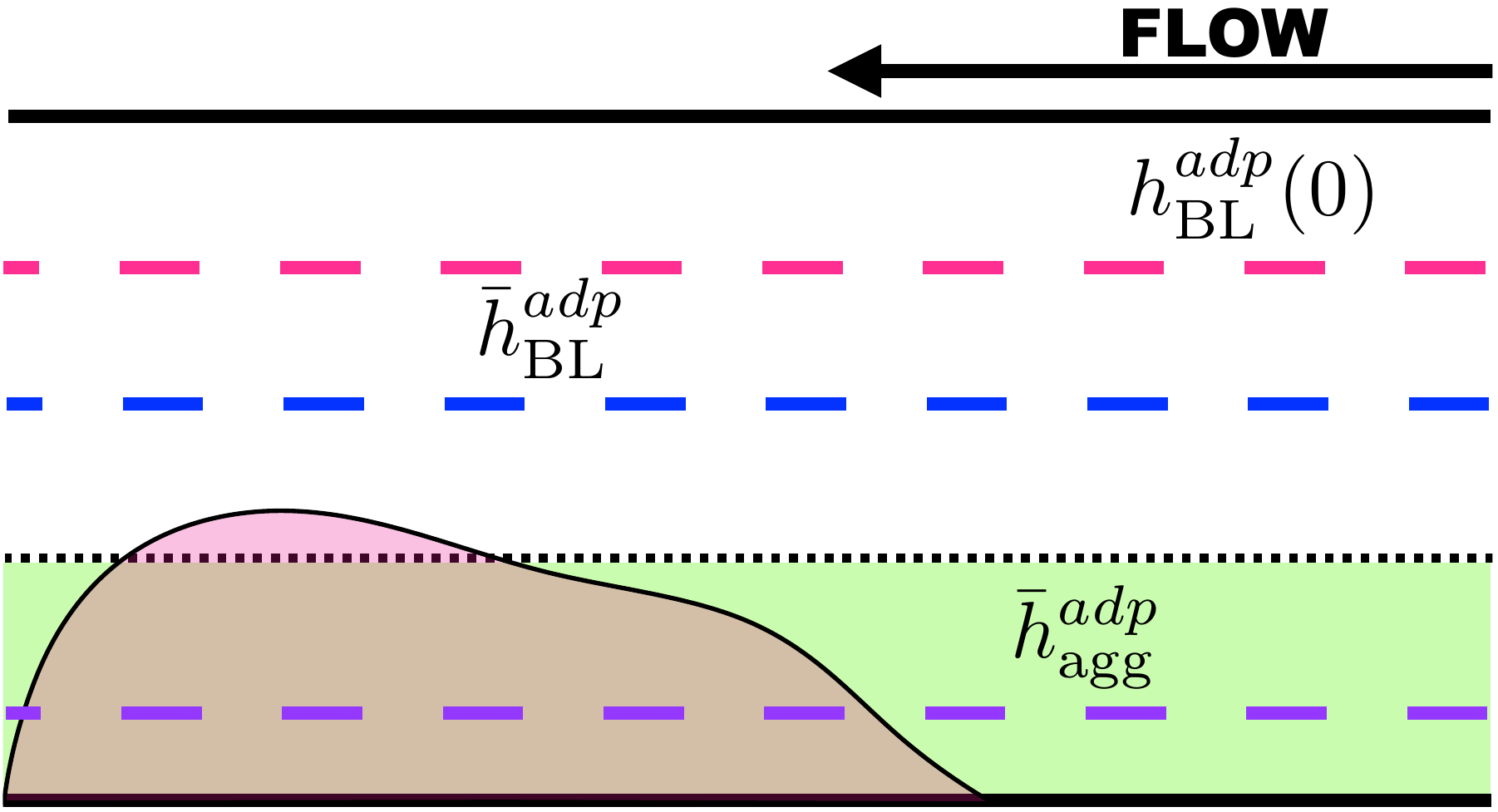}
 \end{multicols}
 \vspace{-0.7cm}
 \caption{{\bf{Boundary layers associated with the transport of
       soluble agonist ADP.}} (A) Consider an ADP molecule at the
   position $(x,d_{\text{agg}} - h^{adp}_{\text{agg}}(x))$. The
   distance it must travel to reach the edge of the growing aggregate is
   $h^{adp}_{\text{agg}}(x)$. (B) The boundary layer thickness at the
   entrance of the injury channel in the Stokes region is
   $h^{adp}_{BL}(0)$ (magenta, dashed) and defines the reaction zone for
   ADP interactions in that region (ARZ). Both the average thickness in the Stokes region
   $\bar{h}^{adp}_{BL}$ (blue, dashed) and in the growing aggregate $\bar{h}^{adp}_{\text{agg}}$ (purple, dashed) are used to define diffusive transport from the aggregate into the Stokes region.}
  \label{fig:BL_adp}
\end{figure}
\vspace{-0.25cm}
\subsubsection{Boundary Layer Calculations}
\label{sec:ADP_BL}
The first boundary layer is related to ADP movement in the Stokes
region after it has left the aggregate where it is released. The
calculation is like that for platelets in the Stokes region as
described above in \Cref{sec:BL_calc}. This boundary layer is thinnest
at the upstream end and widest at the downstream end.  Its thickness
is greater than that, $h^{plt}(x)$, of the platelet boundary layer.
This is because (i) ADP has a larger diffusivity than platelets and
(ii) the choice of velocity used in the boundary layer
calculation. For the platelets, it is a velocity in the Stokes region
$u(d_{\text{agg}}+h^{plt}(x))$ and for the ADP it is the (lower)
velocity on the edge of the Stokes region
$u(d_{\text{agg}})$. Therefore, $h^{adp}_{\text{BL}}(x) > h^{plt}(x),
\forall x \in [0, L_{h}]$. As shown in \cref{fig:BL_adp}B, the maximum
thickness of this ADP boundary layer in the Stokes region is denoted as
$h^{adp}_{\text{BL}}(0)$ 
and the average boundary layer thickness as
$\bar{h}^{adp}_{\text{BL}}$.

The second boundary layer describes ADP movement \emph{within} the aggregate
towards the Stokes region and requires the new calculation given in this
section. We do not have a separate ADP concentration in this boundary
layer; its thickness is used only in calculating the ADP concentration
gradient. Consider the starting point $(x, d_{\text{agg}} -
h_{\text{agg}}^{adp}(x))$ inside the aggregate and equate the time it
takes an ADP molecule there to diffuse into the Stokes region with the
time it would take it to be washed past the injured region by the
flow:
$$\frac{(d_{\text{agg}} - h_{\text{agg}}^{adp}(x))^{2}}{2D_{a}} = \frac{L_{h} - x}{u(d_{\text{agg}})}.$$  
We solve for the thickness $h^{adp}_{\text{agg}}(x)$ and define
$\bar{h}^{adp}_{\text{agg}}$ to be the average aggregate layer
thickness shown in \cref{fig:BL_adp}B. The ADP boundary layer
thickness $d_{\text{abl}} = h^{adp}_{\text{BL}}(0)$ and the thickness
of the growing aggregate contribute to the total thickness of the
regions in which we track the ADP concentrations:
$[ADP]_{\text{agg}}$, $[ADP]_{\text{BL}}$, and $[ADP]_{\text{gap}}$.


\subsubsection{ADP Release and Advection}
The source of ADP in this model corresponds to the release of ADP from activated, bound platelets. The rate of release is
\begin{equation*}
\sigma_{\text{release}}(t) = \int_{0}^{\infty}\hat{A}R(\tau)\frac{d}{dt}(P^{b,a})(t - \tau)d\tau, \label{eq:sigmaADP}
\end{equation*}
where $\hat{A}$ (\Cref{sec:appC}) is the total quantity of ADP
released by an activated platelet, $\hat{A}R(\tau)$ is the rate of
release at an elapsed time since activation $\tau$ and
$\int_{0}^{\infty}R(\tau)d\tau = 1$. The function $R(\tau)$ is defined
as having value zero up to one second after activation, a positive
bell-shaped function for the time interval $1 < \tau < 5$ seconds, and
a peak at three seconds. The term $\frac{d}{dt}(P^{b,a})(t-\tau)d\tau$
is the number of platelets newly activated and bound in the time
interval $[t-\tau, t-\tau + d\tau]$.  We account only for ADP release
from bound activated platelets because activated platelet that do not
become bound are carried downstream before they begin to release their
ADP.

To facilitate our describing the advective transport of ADP out of the
aggregate, the boundary layer, and the gap, we define the expressions
\begin{align*}
J_{\text{agg}}^{adp} &= \Big(\int_0^{d_{\text{agg}}} u_{\text{}}(y) dy\Big)[ADP]_{\text{agg}}, \\
J_{\text{BL}}^{adp} &=  \Big(\int_{d_{\text{agg}}}^{d_{\text{agg}} + d_{\text{abl}}} u(y) dy \Big)[ADP]_{\text{BL}},\\
J_{\text{gap}}^{adp} &=  \Big(\int_{d_{\text{agg}} + d_{\text{abl}}}^{d_{\text{agg}} + d_{\text{abl}} + d_{\text{gap}}} u(y) dy \Big)[ADP]_{\text{gap}},  \label{eq:IADP1} 
\end{align*}
respectively. Then, the rates of advective transport of ADP out of the
ADP compartments in the injury channel are
$L_{z}J^{adp}_{\text{agg}}$, $L_{z}J^{adp}_{\text{BL}}$, and
$L_{z}J^{adp}_{\text{gap}}$.  Early on, the magnitude of the velocity
and the thicknesses of the aggregate and boundary layers lead to the
ordering $J_{\text{gap}}^{adp} > J_{\text{BL}}^{adp} >
J_{\text{agg}}^{adp}$. As the aggregate grows, the boundary layers
become larger, decreasing the thickness of the gap $d_{\text{gap}}$
and resulting in a new ordering $J_{\text{BL}}^{adp} >
J_{\text{gap}}^{adp}$.  If the flow is slowed sufficiently,
$d_{\text{gap}} = 0$ and $J_{\text{gap}}^{adp} = 0$.
\begin{figure}[!h]
  \centering
  \vspace{-0.3cm}
 \hspace{-5.0cm} {\bf{(A)}}  \hspace{6.25cm} {\bf{(B)}} \\
\vspace{-0.3cm}
  \begin{multicols}{2}
  \includegraphics[width = 0.45\textwidth]{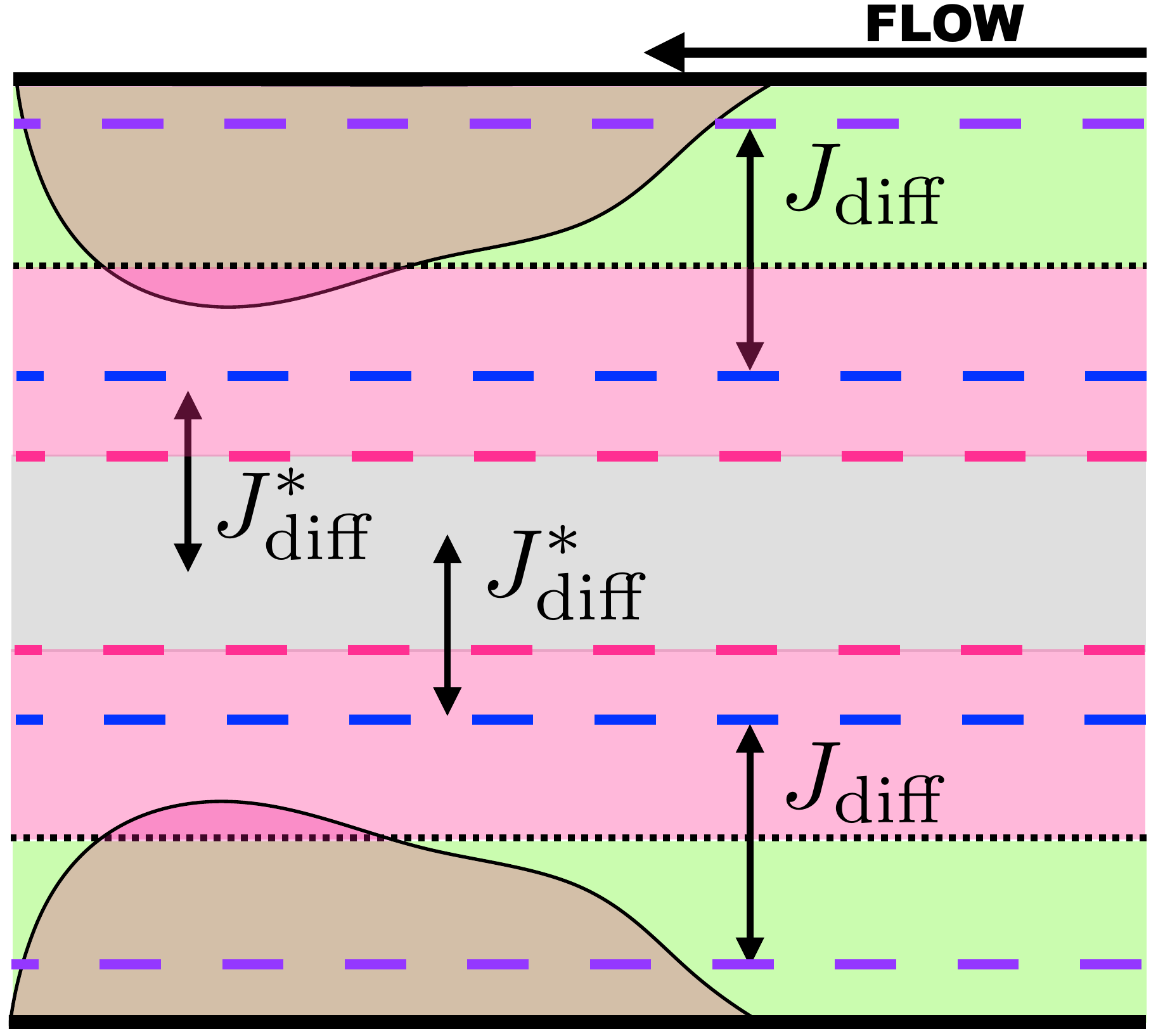}\\
  \includegraphics[width = 0.45\textwidth]{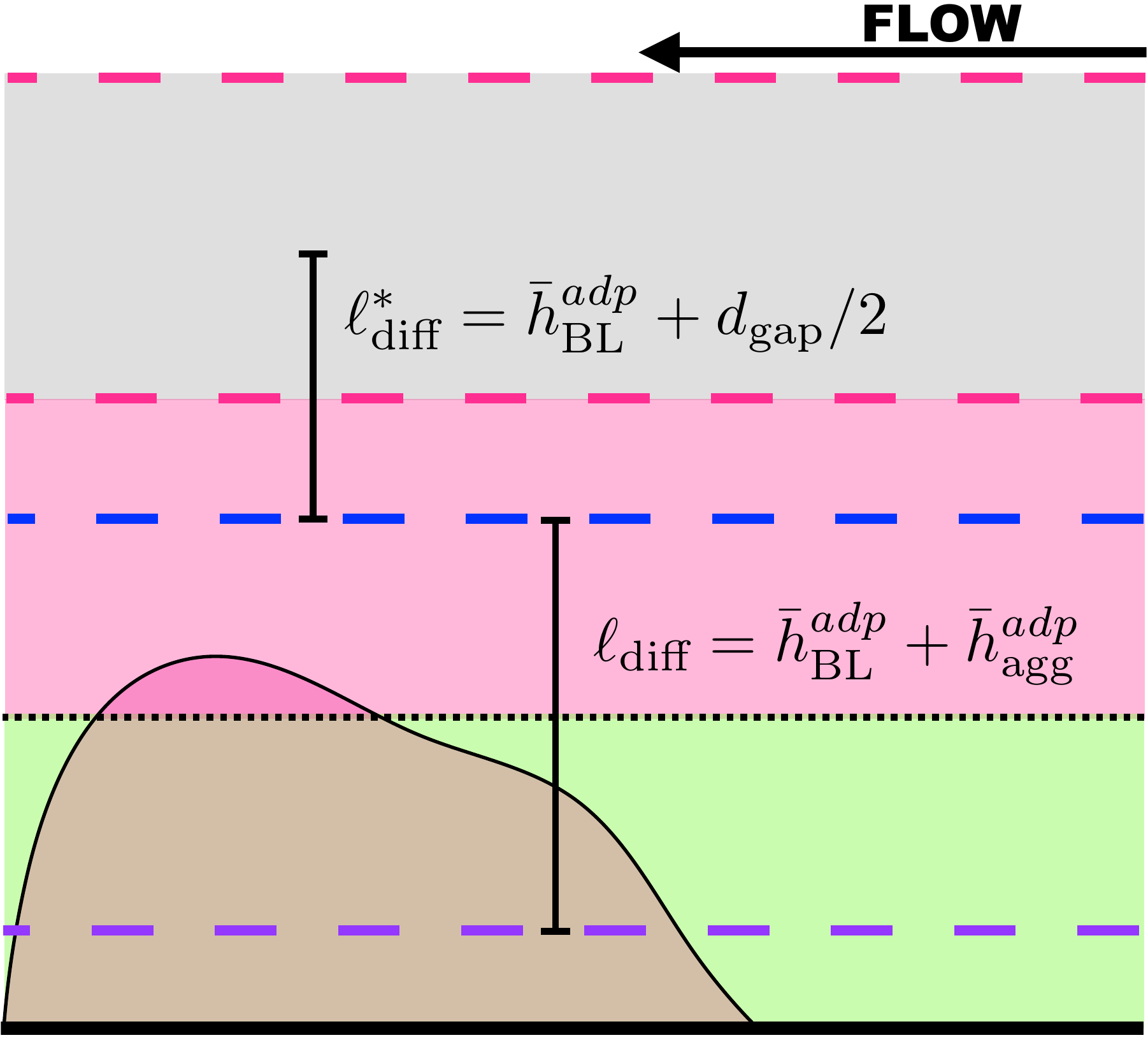}
  \end{multicols}
 \vspace{-0.5cm}
 \caption{{\bf{Compartment model describing diffusive transport of
       ADP.}}  (A) Diffusive fluxes of ADP from the aggregates (green)
   to the associated boundary layers (blue) are denoted as
   $J_{\text{diff}}$. ADP also diffuses from the boundary layers
   associated with the bottom and top aggregates to the bulk of the
   Stokes region (pink) with flux $J_{\text{diff}}^{*}$. The fluxes are subject
   to changing layer thickness that are determined by the flow velocity. (B) The dashed lines correspond to boundary layers
   associated with the bottom aggregate. The specified lengths
   $\ell_{\text{diff}}$ and $\ell_{\text{diff}}^{*}$ are used to
 define the gradients of ADP concentration between the aggregates
   and boundary layers as well as between boundary layers and bulk Stokes region, respectively.}
  \label{fig:ADP_diff}
\end{figure}
\subsubsection{Diffusive Transport of ADP}
\label{sec:adp_diff}
Diffusive transport of ADP from the aggregate into the boundary layer
and into the bulk Stokes region also changes the ADP concentrations.
The diffusive transport of ADP (\cref{fig:ADP_diff}A) between the
aggregate (green) and the boundary layer region (pink) and that
between boundary layer and the bulk Stokes region (grey) are
determined using Fick's law \cite{fick1855v}. Let $V_{\text{abl}}
= L_{h}L_{z}d_{\text{abl}}$ be the volume of the ADP boundary layer
and $\ell_{\text{diff}}$ be the length that determines the gradient of
ADP from the aggregate to the boundary layer. As shown in
\cref{fig:ADP_diff}B, this length is the distance between the edge of
the boundary layer in the aggregate (purple, dashed) and the edge of
the average boundary layer in the Stokes region (dashed, blue). The
diffusive flux of ADP from the aggregate into the boundary layer is
\begin{align*}
J_{\text{diff}}&=D_{a}\frac{([ADP]_{\text{agg}} - [ADP]_{\text{BL}})}{c_{1}\ell_{\text{diff}}}, \nonumber
\end{align*} 
and the rate of diffusion is $L_{h}L_{z}J_{\text{diff}}$, where
$L_{z}L_{h}$ is the area of the interface between the zones, $D_{a}$
is the diffusion coefficient, $([ADP]_{\text{agg}} -
[ADP]_{\text{BL}})$ is the difference in ADP concentration between the
aggregate and the boundary layer, and $\ell_{\text{diff}}$ is the
length used to define the gradient.  Similarly, the diffusive flux of
ADP from the boundary layer into the `gap' is 
\begin{align*}
J_{\text{diff}}^{*}&=D_{a}\frac{([ADP]_{\text{BL}} - [ADP]_{\text{gap}})}{c_{2}\ell_{\text{diff}}^{*}}, \nonumber
\end{align*} 
where the rate of transport flux is $L_{h}L_{z}J_{\text{diff}}^{*}$,
$([ADP]_{\text{BL}} - [ADP]_{\text{gap}})$ is the difference in ADP
concentration and $\ell_{\text{diff}}^{*}$ is half the distance
between the edge of the boundary layer and the center of the gap.  The
fitted parameters $c_{1}$ and $c_{2}$ are discussed in
\Cref{sec:appC}.

\subsubsection{Evolution Equations for ADP}
Accounting for release, transport due to advection, and diffusive
transport between the aggregates, boundary layer regions, and the gap
in the Stokes region, the ADP concentrations in the aggregate and
boundary layer region evolve according to the equations
\begin{align}
\frac{d[ADP]_{\text{agg}}}{dt} = & \ \underbrace{\sigma_{\text{rel}}(t)}_{\text{release}} \ - \ \underbrace{\frac{J_{\text{diff}}}{d_{\text{agg}}}}_{\parbox{3em}{\scriptsize diffusion into BL}} 
  \ - \ \underbrace{\frac{J_{\text{agg}}^{adp}}{L_{h} d_{\text{agg}}}}_{\text{flow out of agg}} - \underbrace{\bigg(\frac{\frac{d(d_{\text{agg}})}{dt}}{d_{\text{agg}}}\bigg)[ADP]_{\text{agg}}}_{\text{dilution}},
 \end{align}
and
 \begin{align}
 \frac{d[ADP]_{\text{BL}}}{dt} = & \underbrace{\frac{J_{\text{diff}}}{d_{\text{abl}}}}_{\parbox{3em}{\scriptsize diffusion into BL}} - \underbrace{\frac{J^{*}_{\text{diff}}}{d_{\text{abl}}}}_{\parbox{3em}{\scriptsize diffusion into gap}}
 \ - \ \underbrace{\frac{J_{\text{BL}}^{adp}}{L_{h} d_{\text{abl}}}}_{\text{flow out of BL}} - \underbrace{\bigg(\frac{\frac{d(d_{\text{abl}})}{dt}}{d_{\text{abl}}}\bigg)[ADP]_{\text{BL}}}_{\text{dilution}},
\end{align}
%
The concentration of ADP in the gap changes because of diffusive
transport from \emph{both} the bottom and top boundary layers:
\begin{align}
   \frac{d[ADP]_{\text{gap}}}{dt} = &  \underbrace{\ \ \ \frac{J^{*}_{\text{diff,b}}}{d_{\text{gap}}} \ \ \ }_{\text{\parbox{6em}{\scriptsize diffusion into gap from bottom}}} \ + \ \underbrace{\ \ \ \frac{J^{*}_{\text{diff,t}}}{d_{\text{gap}}} \ \ \ }_{\parbox{6em}{\scriptsize diffusion into gap from top}}  \ - \ \underbrace{\frac{J_{\text{gap}}^{adp}}{L_{h}d_{\text{gap}}}}_{\text{flow out of gap}}- \underbrace{\bigg(\frac{\frac{d(d_{\text{gap}})}{dt}}{d_{\text{gap}}}\bigg)[ADP]_{\text{gap}}}_{\text{dilution}},
\end{align}
where $J^{*}_{\text{diff,b}}$ and $J^{*}_{\text{diff,t}}$ are the
respective diffusive fluxes of ADP into the gap from the bottom and
top boundary layers.  
 
\subsection{Aggregate Formation}
The formation of the aggregate layer is described by its evolving
thickness, $d_{\text{agg}}$, and volume fraction of bound platelets,
$\theta^{B}$. To determine evolution equations for these quantities, one must consider the adhesion
and cohesion events that contribute to the change in volume occupied
by bound platelets. Let $V_{\text{agg}} = L_{h}L_{z}d_{\text{agg}}$.
The rate of change of the volume occupied by bound platelets is
\begin{align*}
\frac{d}{dt}(V_{\text{agg}}\theta^{B}) = \underbrace{V_{\text{prz}}k_{\text{adh}}
                              (P^{m,u}+ P^{m,a})v_{\text{plt}}}_{\text{adhesion of mobile plts.}} \ + \   \underbrace{V_{\text{prz}}
                              (k_{\text{coh}} P^{b,a} P^{m,a})v_{\text{plt}}}_{\text{cohesion of mobile, act. plts}}.
                              \label{conservation_1}
\end{align*}
Dividing by $d_{\text{agg}} \theta^{B}$ and using the product rule,
and partitioning the adhesive term and cohesion term we find the
following equations.
\begin{align}
\frac{d \theta^{B}_{\text{}}}{dt}  = & \ \gamma_{\text{}} \bigg(\frac{d_{\text{prz}}}{d_{\text{agg}}}\bigg)k^{}_{\text{adh}}
                              \Big(P^{m,u}_{\text{}} + P^{m,a}_{\text{}}\Big)v_{\text{plt}} \ +   \ \beta_{\text{}} \bigg(\frac{d_{\text{prz}}}{d_{\text{agg}}}\bigg)
                              k_{\text{coh}}\Big( P^{b,a}_{\text{}} P^{m,a}_{}\Big)v_{\text{plt}}\label{eq:thetab1}\\
&\vspace{1.0pt} \nonumber\\
\frac{d}{dt}(d_{\text{agg}})  = & \ (1-\gamma_{\text{}})(d_{\text{prz}})k_{\text{adh}} \bigg(\frac{P^{m,u}_{\text{}} + P^{m,a}_{\text{}}}{P^{b,a}_{\text{}}}\bigg) 
				\ + \  (1-\beta_{\text{}})(d_{\text{prz}})k_{\text{coh}} \bigg(\frac{P^{b,a}_{\text{}} P^{m,a}_{\text{}}}{P_{\text{}}^{b,a}}\bigg) \label{eq:d1},
\end{align}
where $\beta_{\text{}}$, $\gamma_{\text{}}$ are assumed to have the functional forms  
\begin{align*}
\gamma_{} = \begin{cases}
0, \text{ if } \theta^{B} = \theta^{\max}\\
1, \text{ if }d_{\text{agg}} \ge d_{\text{plt}}
\end{cases}   
\beta_{} = \begin{cases}
\frac{1}{2}(1 + \tanh(k_{b}(s_{b} - \theta^{B}))), \text{ if } d_{\text{gap}} > 0, \theta^{B} < \theta^{\max}\\
0, \text{ if } \theta^{B} = \theta^{\max}\\
1, \text{ if }d_{\text{gap}} = 0
\end{cases}. 
\end{align*}
The partition functions $\beta$ and $\gamma$ are phenomenological in
nature and will be partially validated by PDE
model (\Cref{sec:appA_PDE}). Details are found in \Cref{sec:appC}. Equations
\eqref{eq:Pu1}-\eqref{eq:d1} describe platelet dynamics, ADP dynamics,
and aggregate formation associated with the bottom wall of the injury
zone. There is a similar set of equations governing the concentrations
of platelets species, ADP, aggregate thickness, and aggregate volume
fraction associated with the top aggregate.

\subsection{Platelet Activation}
\label{sec:platelet_act}
As noted above, once an unactivated platelet enters the PRZ it can adhere to the wall and therefore become bound and
activated, it can be activated by ADP and possibly cohere
  to already bound platelets, or it can be carried out of the injury
channel by the flow.  While platelet activation is a complex process
triggered by a range of stimuli and made up of an ensemble of
responses \cite{Li2010}, we focus on activation by direct adhesion to
the wall and by exposure to ADP and on the resulting activation of
integrin $\alpha_{IIb}\beta_{3}$ receptors and release of ADP.  We
discussed the activation by direct adhesion above.  Activation
triggered when ADP in the plasma binds to a platelet's P2Y$_{1}$ and
P2Y$_{12}$ receptors, provides a means of activation that does not
require contact with the SE matrix.  Because ADP stimulates this
activation and additional ADP is released as a consequence, there is a
positive feedback in the system involving the concentration of ADP and
the transition of mobile, unactivated platelets to mobile, activated
platelets. To model ADP-dependent activation, we assume that the
activation trigger is a weighted average of the ADP concentrations in
the aggregate and in the boundary layer. We define the rate of ADP-induced activation
as
$$
k_{\text{act}}= k^{act}_{0}A([ADP]_{\text{agg}},[ADP]_{\text{BL}}),
$$
where 
$$
A([\text{ADP}]_{\text{agg}}, [\text{ADP}]_{\text{BL}}) =
\bigg(\frac{d_{\text{agg}}}{d_{\text{arz}}}\bigg)\frac{[\text{ADP}]_{\text{agg}}}{[\text{ADP}]^{*} + [\text{ADP}]_{\text{agg}}} + \bigg(\frac{d_{\text{pbl}}}{d_{\text{arz}}}\bigg)\frac{[\text{ADP}]_{\text{BL}}}{[\text{ADP}]^{*} + [\text{ADP}]_{\text{BL}}}.
$$
Here, $d_{\text{arz}} = d_{\text{agg}} + d_{\text{abl}}$ is the
thickness of the ADP reaction zone (ARZ) and $[ADP]^{*} = 1 \mu$M.
For some of our simulations, we instead considered ADP-independent
platelet activation in the plasma by setting
$A([\text{ADP}]_{\text{agg}}, [\text{ADP}]_{\text{BL}}) \equiv 1$, in
which case only the value of the rate constant $k^{act}_{0}$ matters.
Below we explore the effects of ADP-independent versus ADP-dependent
activation on aggregate growth and occlusion of the injury channel.

\subsection{Numerical Methods}
\label{sec:numerics}
Simulations of the ODE model consist of solving the HC system and the Brinkman-Stokes system to determine $G_{h}$ and $R_{M}$, respectively.  The differential equations describing the evolution of platelet species, ADP, and aggregate thickness and density are solved using MATLAB function \verb|ode15s|, which employs a variable-step, variable-order (VSVO) solver based on the numerical differentiation formulas (NDFs) of orders 1 to 5 for stiff systems \cite{shampine1997matlab}.  During each time step of the computation, we perform the following series of updates for the unknowns.
\begin{enumerate}
\item Assume $G_{h} = 1$ and solve the linear system in \Cref{sec:appB} to determine $u(y)$ and $R_{M}$.
\item Use $R_{M}$ as an input parameter for the HC system in \Cref{sec:appA} and solve for $G_{h}$.
\item Scale the flow velocity using the pressure gradient update; $\bar{u}(y) := G_{h}u(y)$. 
\item Use $u(y)$ and diffusivity constants ($D_{p}$, $D_{a}$) to determine boundary layers ($h^{plt}(x)$, $h^{adp}_{\text{BL}}(x)$, $h^{adp}_{\text{agg}}(x)$) for both aggregates.
\item Determine thicknesses of the PRZ ($d_{\text{prz}}$), compartments in the Stokes region tracking ADP ($d_{\text{abl}}$, $d_{\text{gap}}$), and the growing aggregate ($d_{\text{agg}}$) associated with the bottom and top walls.
\item Count the platelets activated within the previous time step and update the ADP release function $\sigma_{rel}$ for each aggregate.
\item Update mobile platelets and ADP concentrations associated with the top and bottom aggregates and reaction zones to account for advection using the velocity profile $u(y)$.
\item Update ADP concentrations to account for diffusion using ADP boundary layer thicknesses ($\bar{h}^{adp}_{\text{BL}}$, $\bar{h}^{adp}_{\text{agg}}$) of the top and bottom ARZs and aggregates.
\item Update the thicknesses of the aggregates $d_{\text{agg}}$, the volume fractions of bound platelets $\theta^{B}$, and hinderance function $W(\theta^{B} + \theta^{M})$ to account for platelet adhesion and cohesion events.
\item Repeated until desired time.
\end{enumerate}
The simulations described below were carried out for 20 minutes and all parameters values are found in \Cref{hca:dimensions}-\Cref{params:adh_coh} with specific details in \Cref{sec:appA} - \Cref{sec:appC}. Each simulation requires $\approx 50$ seconds of computational time. This computational cost can be substantially reduced by using a compiled language.

\section{Results}
\label{sec:results}
The presented mathematical model was calibrated and validated through
comparisons of model output with that of analogous MFA (\Cref{sec:appA_methods}) and PDE (\Cref{sec:appA_PDE}) models. Preliminary studies with the PDE model revealed that the pressure drop $\Delta P = P_{1} - P_{2}$ across the injury channel was not constant with respect to the y-direction. Therefore to achieve the most accurate model comparisons, it was necessary to compute an approximation of the flow rate $Q_{PDE}$ through the injury channel using the PDE model for all the following simulations. This changes the second step of our numerical scheme.  Specifically, we do not solve the HC system for the pressure gradient $G_{h}$. Instead, we assume
$$
G_{h} = \frac{Q_{PDE}(d_{b},d_{t},\theta^{B}_{b},\theta^{B}_{t})}{R_{M}L_{h}^{\text{eff}}},
$$
where $d_{b}, d_{t}$ and $\theta^{B}_{b}, \theta^{B}_{t}$ are the thicknesses and densities of the bottom and top aggregates, respectively. The effective length of the channel, $L_{h}^{\text{eff}} = 165 \ \mu \text{m}$ as described in \Cref{sec:appA_circuit}. Details regarding $Q_{PDE}$ are found in \Cref{fig:flow_map_comp}A-B and \cref{sec:appA_flowmap}. Note that the ODE model is not dependent on the use of the PDE flow map and the use of the ODE model beyond the scope of this work is independent of PDE model results.

Using this new calculation for the pressure gradient, we validated the fluid calculation associated with the ODE model in \cref{fig:flow_map_comp}C-D. Assuming ADP-independent platelet activation ($A([ADP]_{\text{agg}},[ADP]_{\text{BL}}) = 1$), we used MFA occlusion times to determine physiologically relevant values for the kinetic rate constant $k^{adp}_{0}$ (\cref{fig:agg_comp}A-B). Fixing a value for $k^{adp}_{0}$, we compared occlusion metrics \cref{fig:agg_comp}C-D from the ODE and PDE models. Next, we assessed and characterized occlusion of the injury channel as a function of the flow shear rate and $k^{adp}_{0}$ in \cref{fig:occlusion}A-B and then conducted similar experiments to test the sensitivity of model output to ADP-dependent platelet activation as shown in \cref{fig:dilution}A-D. Evidence of the effects of dilution of ADP due to advective and diffusive transport on aggregate formation is found in \cref{fig:proof}A-H.

\subsection{Model Validation and Calibration}
\label{sec:results_calibration}

\begin{figure}[!h]
  \centering
  \hspace{-5.0cm} {\bf{(A)}}  \hspace{5.5cm} {\bf{(B)}} \\
  \vspace{-0.4cm}
  \begin{multicols}{2}
  \includegraphics[width = 0.485\textwidth]{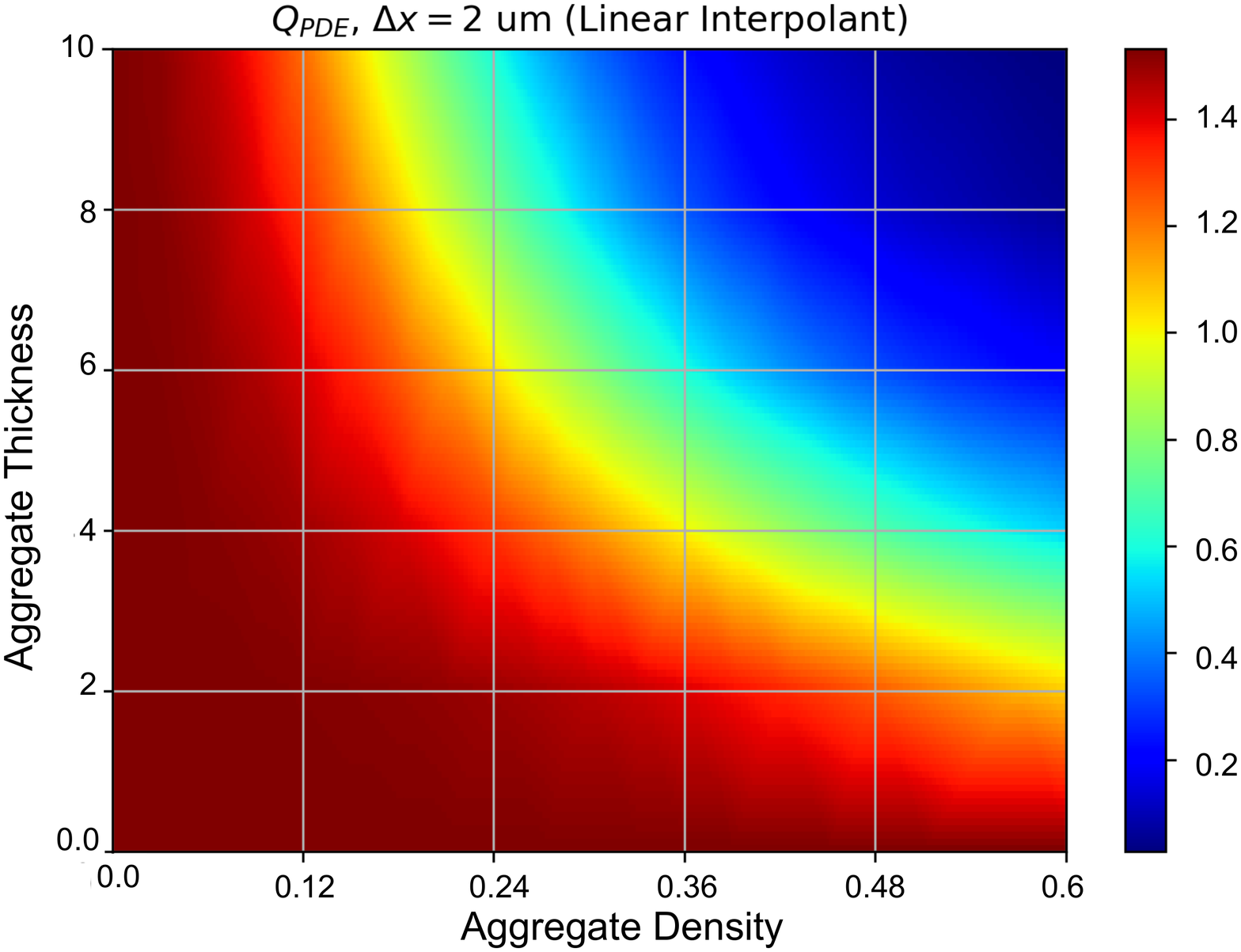}\\
   \includegraphics[width = 0.4625\textwidth]{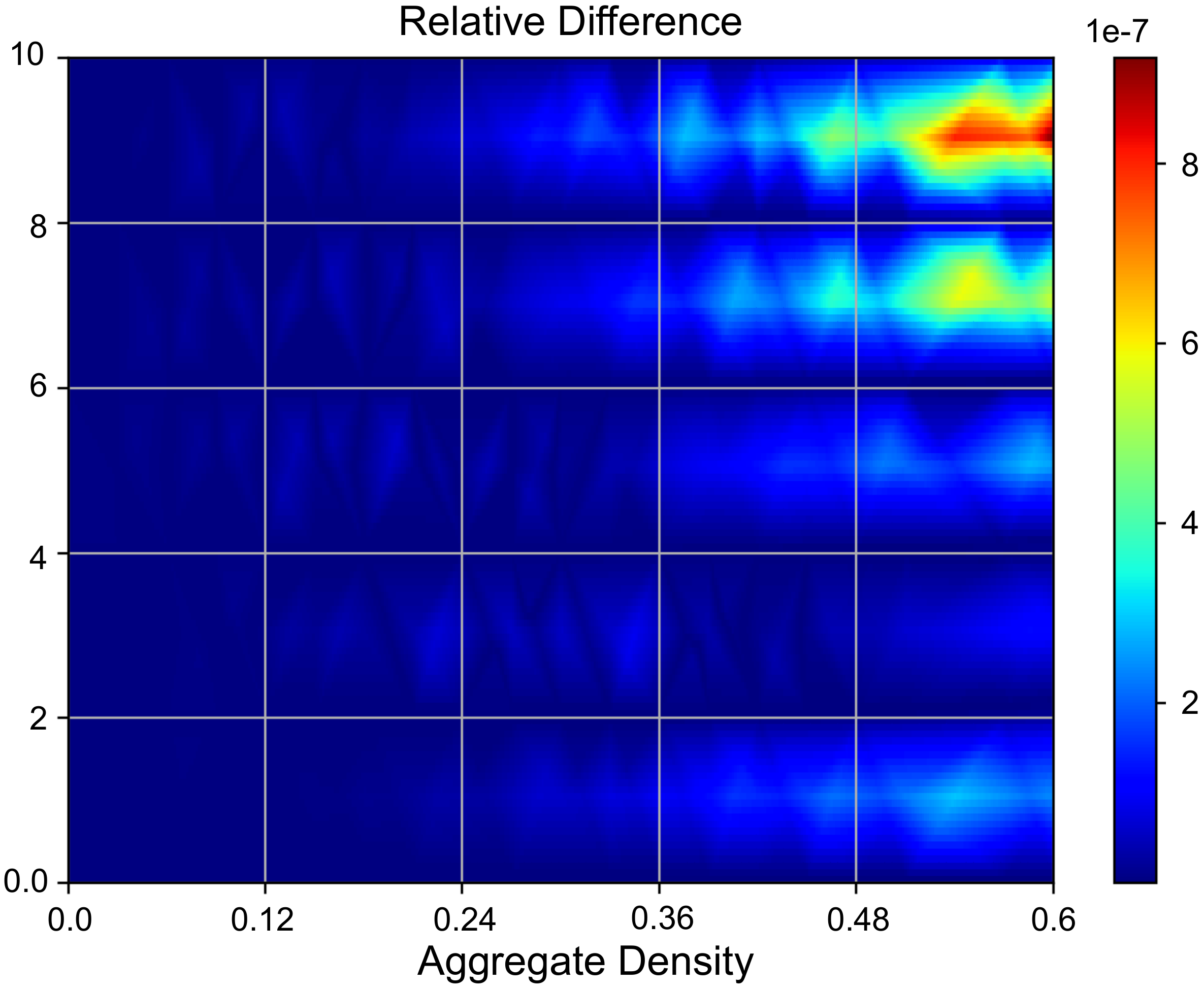}
 \end{multicols}
   \vspace{-0.5cm}
 \hspace{-5.0cm} {\bf{(C)}}  \hspace{5.5cm} {\bf{(D)}} \\
  \vspace{-0.4cm}
  \begin{multicols}{2}
  \includegraphics[width = 0.48\textwidth]{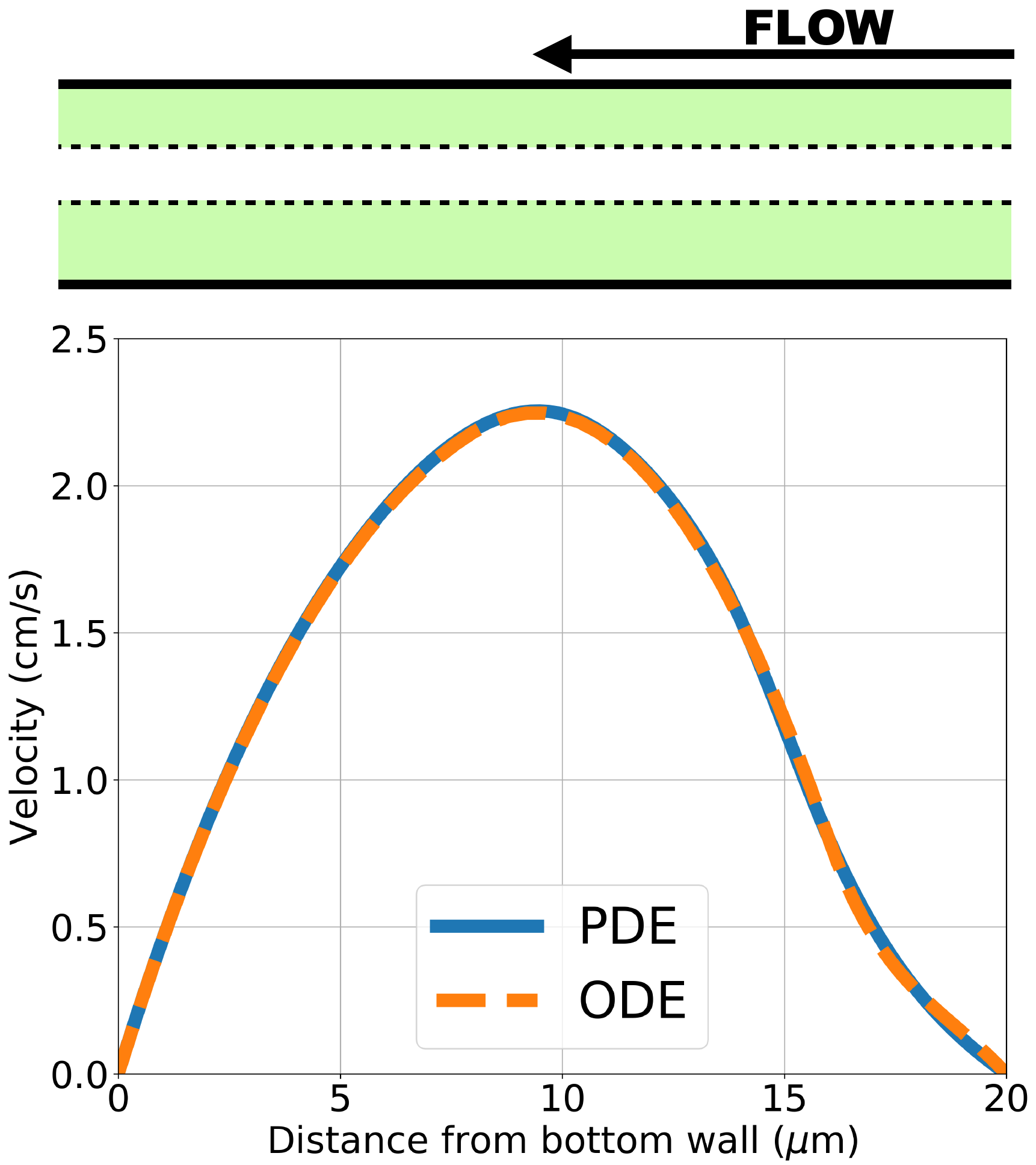}\\
   \includegraphics[width = 0.48\textwidth]{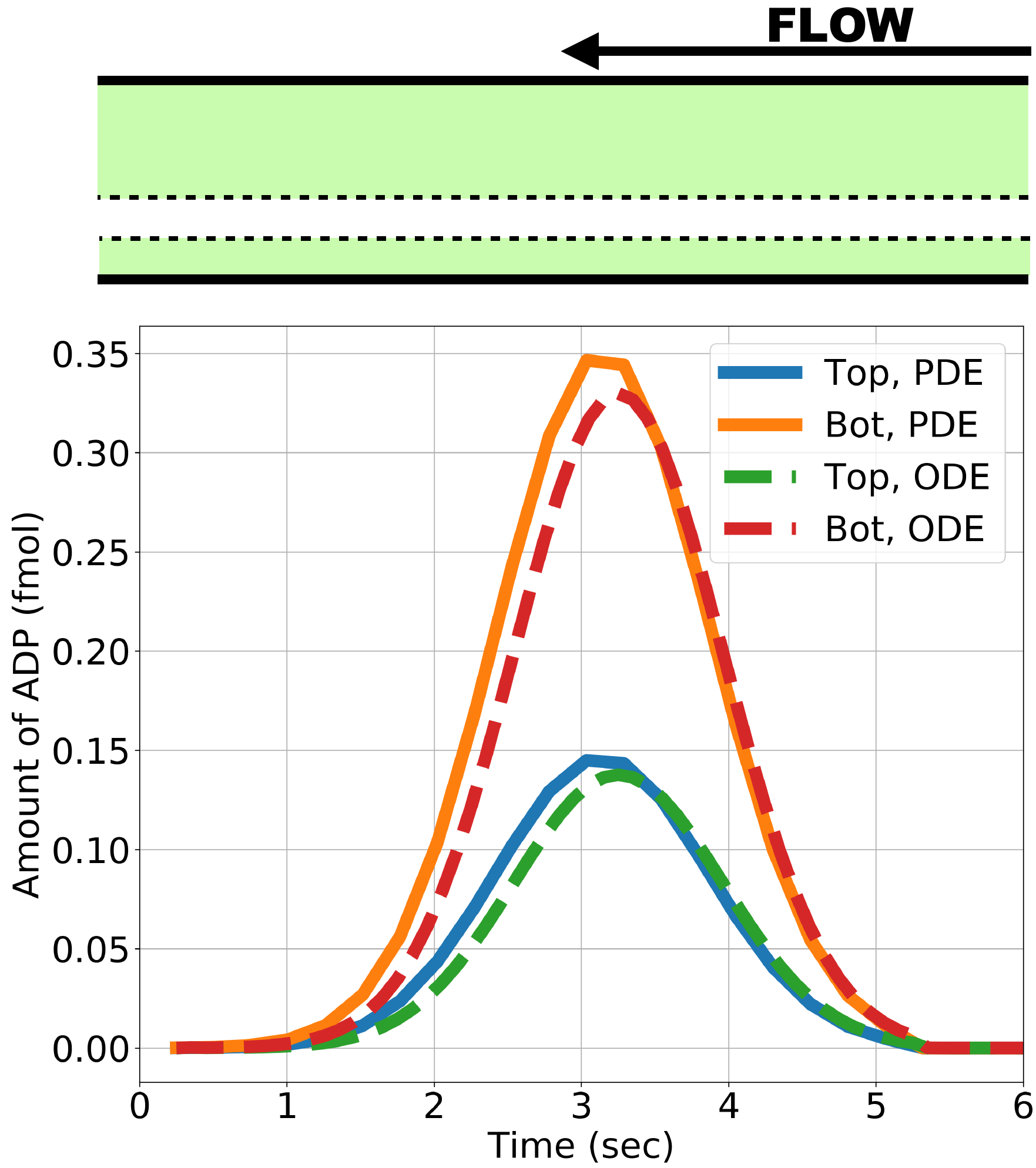}
 \end{multicols}
 \vspace{-0.7cm}
 \caption{{\bf{Model calibration and validation.}} (A) 3D flow rates
   $Q_{PDE}$ in prebound platelet simulations with the PDE model using
   spatial discretization $\Delta x = 2 \ \mu$m and (B) the relative
   difference between the flow rates generated with ($\Delta x = 2
   \ \mu$m) and ($\Delta x = 1 \ \mu$m). (C) Velocity profiles from ODE and PDE models for an asymmetric prebound
   platelet configuration ($d_{t} = 6 \mu$m, $\theta^{B}_{t} = 0.54$,
   $d_{b} = 8 \mu$m, $\theta^{B}_{b} = 0.06$).
   (D) Similar prebound platelet
   configuration ($d_{t} = 12 \mu$m, $\theta^{B}_{t} = 0.15$, $d_{b} =
   4 \mu$m, $\theta^{B}_{b} = 0.45$) experiment tracks the amount of
   ADP in the aggregate for the PDE and ODE models. Note
   $\theta^{\max} = 0.6$. }
  \label{fig:flow_map_comp}
\end{figure}

For this study, we simulated 2D flow through the injury channel with
fixed pre-formed platelet aggregates. For each simulation, a different
symmetric configuration of preformed aggregates was specified by
choosing the thicknesses $d_{\botagg} = d_{\topagg} \in \{2, 4, 6, 8, 10
\ \mu m\}$ and bound platelet densities $\theta^{B}_{\botagg} =
\theta^{B}_{\topagg} \in [0.0, 0.06, \cdots 0.6]$. The PDE model was
used to calculate the steady-state velocity field and two-dimensional
volumetric flow rate $Q_{PDE}$ for each configuration as detailed in \cref{sec:appA_flowmap}.  The `flow map'
$Q_{PDE}(d_{\botagg},\theta^{B}_{\botagg},d_{\topagg},\theta^{B}_{\topagg})$
obtained when the PDE calculation used a spatial discretization with
step size $\Delta x = 2 \ \mu$m is shown in \cref{fig:flow_map_comp}A,
and \cref{fig:flow_map_comp}B presents evidence that the flow rate
calculation is insensitive to further refinement in $\Delta x$.  The
maximum relative difference in flow rates for the two stepsizes, which
occured for an aggregate density of $\approx 0.5$ and with
almost-occlusive thicknesses of $\approx 9.5 \ \mu m$, was
approximately $8(10)^{-7}$.
\begin{figure}[!h]
  \centering
    \hspace{-5.0cm} {\bf{(A)}}  \hspace{5.5cm} {\bf{(B)}} \\
  \vspace{-0.3cm}
  \begin{multicols}{2}
\includegraphics[width = 0.455\textwidth]{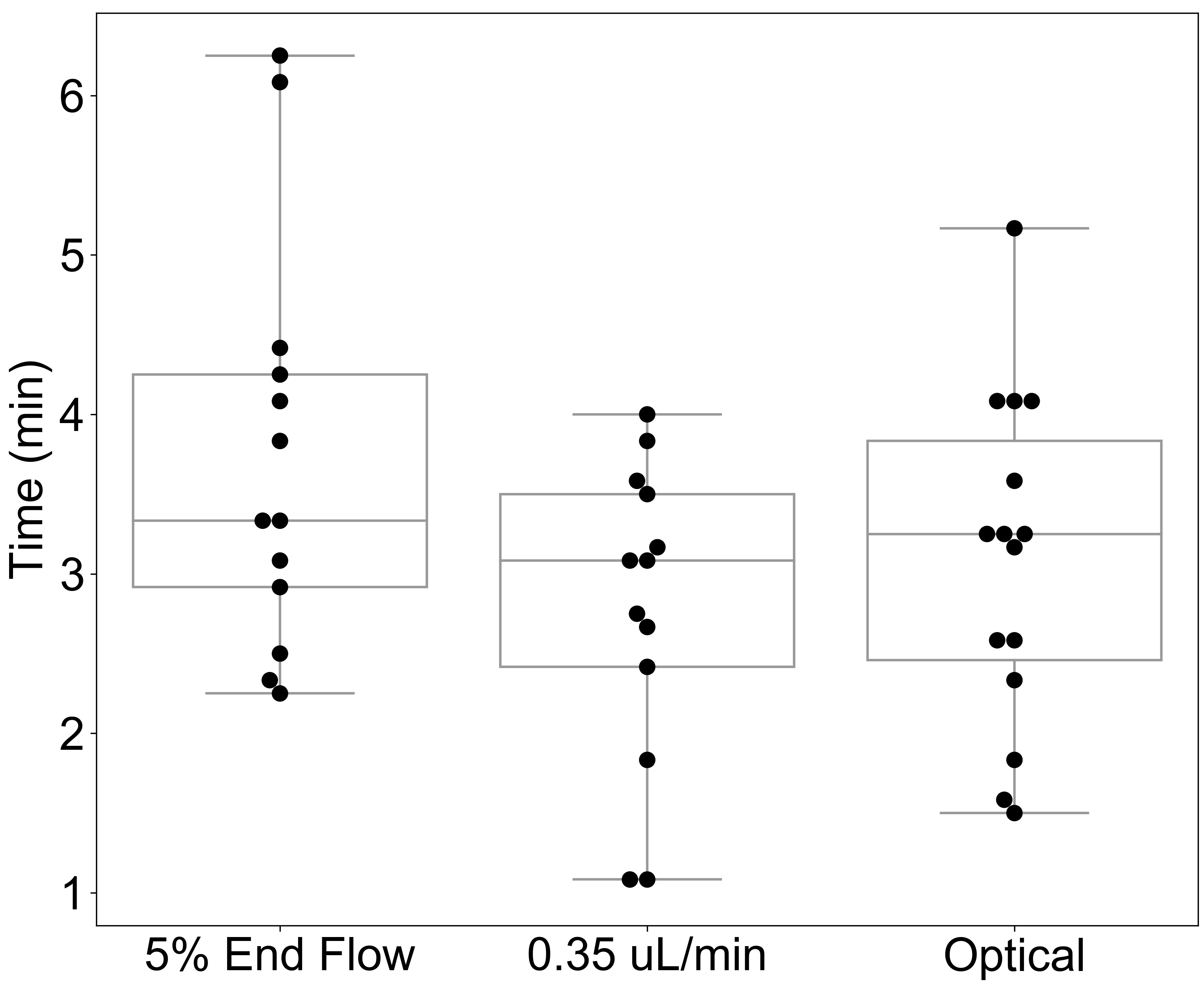}\\
\includegraphics[width = 0.485\textwidth]{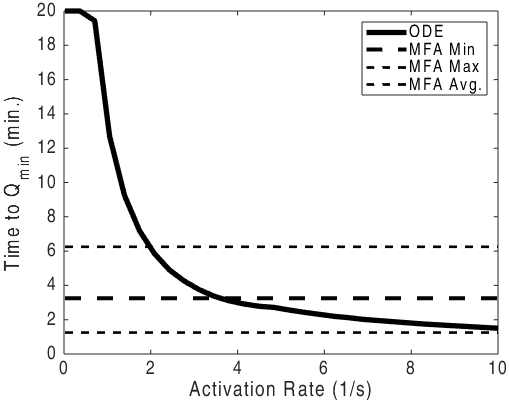}\\
\end{multicols}
  \vspace{-0.5cm}
  \hspace{-5.0cm} {\bf{(C)}}  \hspace{5.5cm} {\bf{(D)}} \\
\vspace{-0.3cm}
\begin{multicols}{2}
\includegraphics[width = 0.475\textwidth]{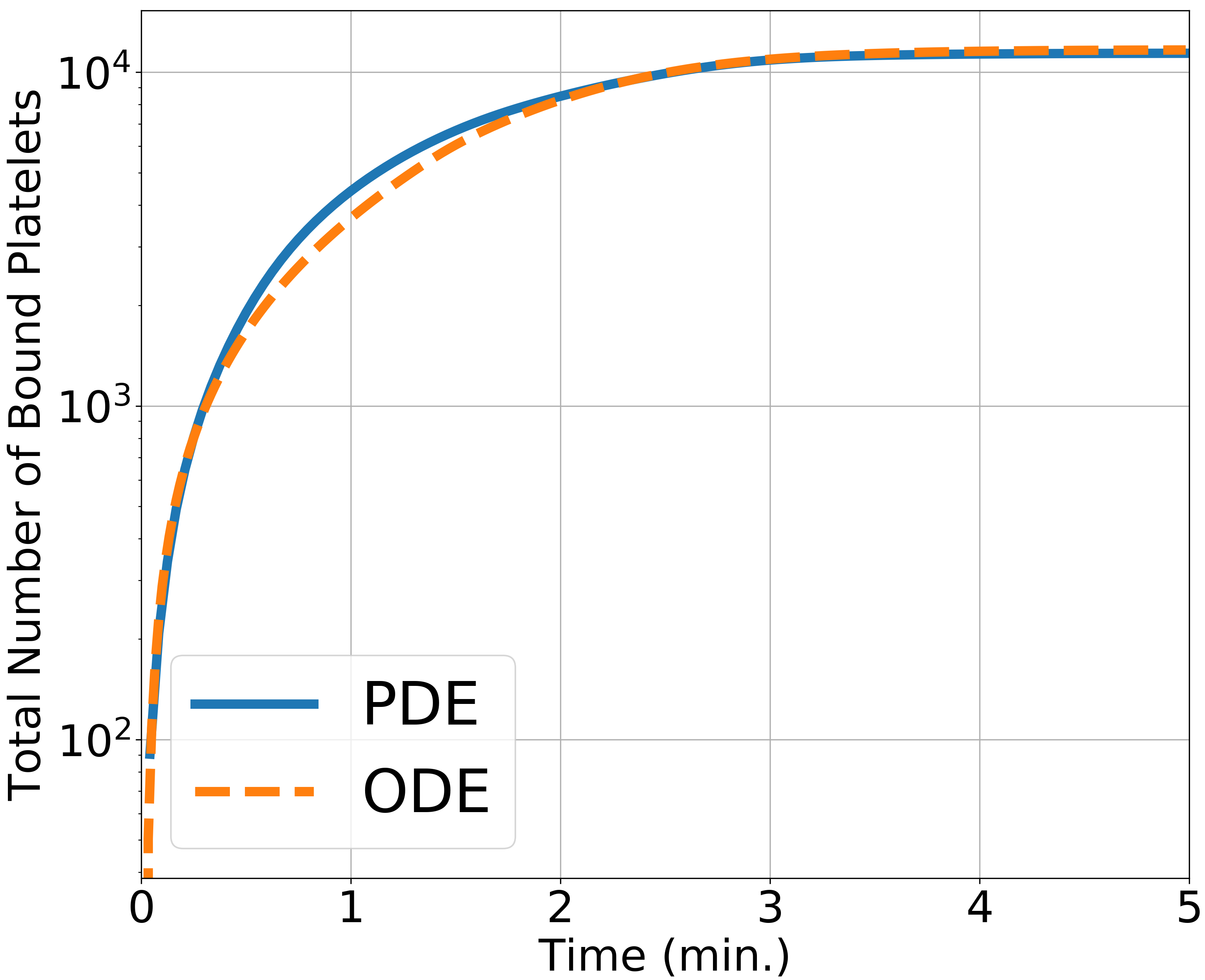}\\
\includegraphics[width = 0.475\textwidth]{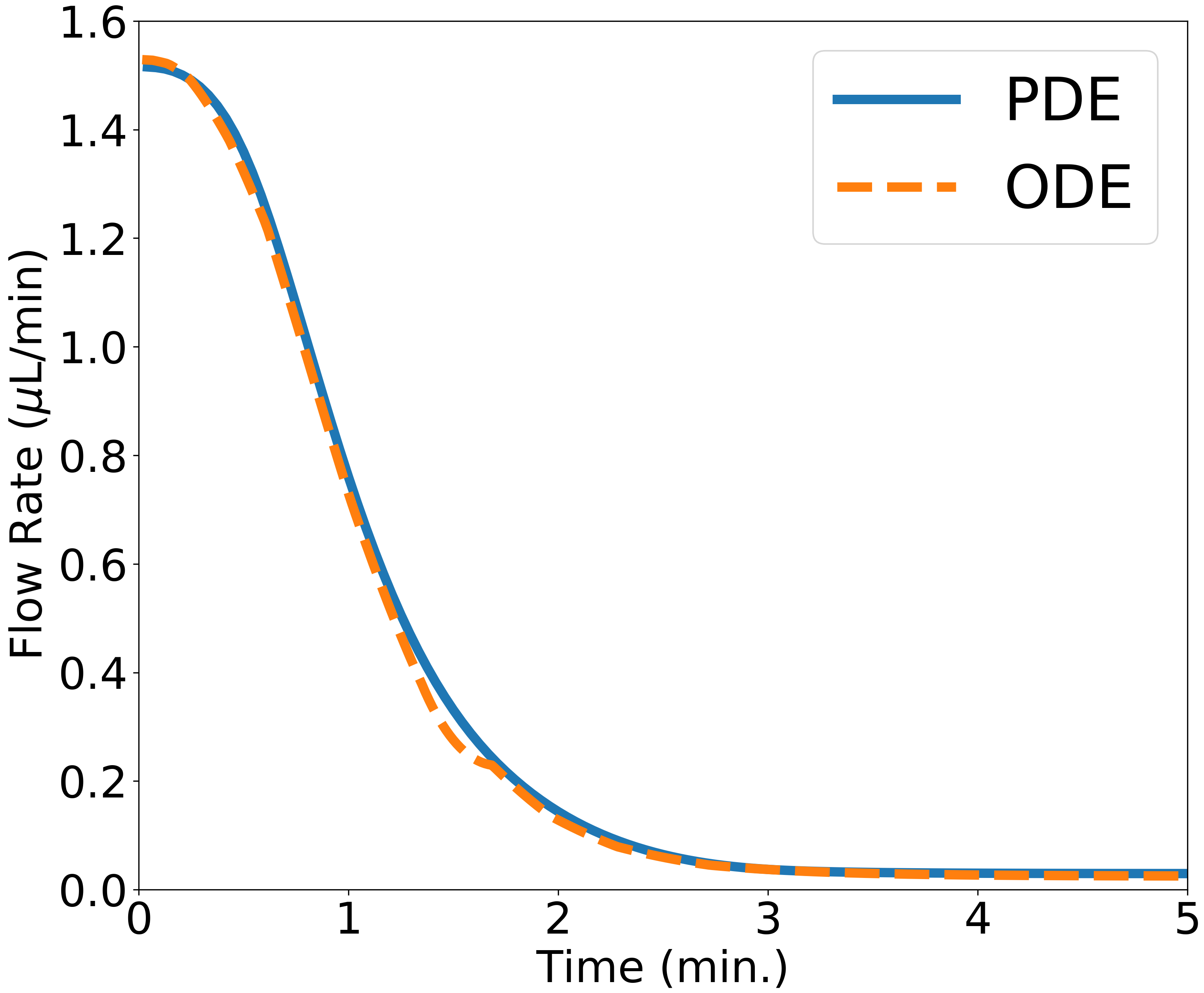}\\
\end{multicols}
\vspace{-0.75cm}
\caption{{\bf{Model calibration and validation.}}(A) Occlusion times
  generated by the MFA bleeding chip and (B) ODE model times to
  $Q_{\min} = 0.35 \ \mu$L/min as a function of the ADP-independent
  platelet activation rate $k^{act}_{0}$. Comparison of ODE (orange)
  and PDE (blue) model outputs with a fixed shear rate ($\approx 2400
  \ \text{s}^{-1}$) and ADP-independent platelet activation rate
  ($k^{act}_{0} = 3.4 \text{ s}^{-1}$). (C) Total number of bound
  platelets in the aggregate and (D) rate of blood flow through the
  injury channel.}
  \label{fig:agg_comp}
\end{figure}

\cref{fig:flow_map_comp}C shows velocity profiles from the ODE and PDE
models for asymmetric aggregates.  The ODE profile shows excellent
agreement in both Brinkman layers and the Stokes region with the PDE
profile obtained half-way through the injury channel.  The flow
through the top aggregate, which is thin ($d_{\topagg} = 4 \ \mu$m) but
dense ($\theta^{B}_{\topagg} = 0.54$) is slower than that in the thick
($d_{\botagg} = 8 \ \mu$m) yet loose ($\theta^{B}_{\botagg} = 0.06$) bottom
aggregate, because resistance to flow increases as aggregate density
increases.  

In \cref{fig:flow_map_comp}D, we show results from another asymmetric
pre-formed aggregate configuration, assuming that all of the platelets
in the aggregates became activated at time $t=0$.  Shown are the
amounts of ADP in each of the aggregates due to release of ADP by
these platelets in both the ODE and PDE models.  Less ADP is washed
downstream from the thin and dense bottom aggregate because the flow
in it is slower.  Consequently, the maximum concentration of ADP in
the bottom aggregate is higher than that in the thick and loose top
aggregate (not shown). It is important to note that the concentration
of ADP in each aggregate is subject to diffusive transport. The ODE
model accounts for lateral diffusion (perpendicular to the channel
walls) but not longitudinal diffusion.  The differences in the models'
outputs are largely due to longitudinal diffusion of ADP out of the
injury channel in the PDE model (results not shown).

One of our main objectives in this work was to develop a mathematical
model that can generate aggregates that grow to fill the injury
channel. This objective was motivated by experiments conducted in the
microfluidic bleeding chip model (MFA). \cref{fig:agg_comp}A shows
three metrics of occlusion times from experiments using the MFA.  The
median times for all three metrics of occlusion are approximately 3.25
minutes, with a minimum occlusion time of 1.15 minutes and a maximum
time of 6.25 minutes. The metric of occlusion best suited for
comparison with the ODE model output is the `time to $Q_{\min} = 0.35
\ \mu$L/min measurement'.  Using an ADP-independent platelet
activation regime
$(A([\text{ADP}]_{\text{agg}},[\text{ADP}]_{\text{BL}}) \equiv 1)$ in the
ODE model simulations yields the `time to Q$_{\text{min}}$' results obtained
for a range of platelet activation rates as shown in
\cref{fig:agg_comp}B.  For $k^{act}_{0} > 2.0 \ \text{s}^{-1}$, the
time to $Q_{\min}$ is within the bounds defined by the MFA results.
In \cref{fig:agg_comp}C-D, we compare the growth of aggregates using
the ODE model (orange) with output from the PDE model (blue).  Metrics
are the timecourse of the total number of bound platelets and of the
flow rate through the injury channel. For a fixed shear rate $\approx
2400 \text{ s}^{-1}$ and a fixed platelet activation rate of $3.4 \
\text{s}^{-1}$, model results fall within the range of the experimental values.  Note that in \cref{fig:agg_comp}B, the time to $Q_{\min}$ with the chosen activation rate is within the range (black, dashed lines)
seen with the MFA.  Additionally, both
computational models yielded aggregates that reduced the flow rate of
the injury channel in an amount of time similar to that seen in the
MFA results.

\subsection{Occlusion of Injury Channel and Flow Rate Reduction}
\label{sec:results_occlusion}
To better understand how shear rate and ADP-independent platelet activation affect aggregate formation, we performed simulations in which we varied both the shear rate and activation rate constant $k^{act}_{0}$.
 \begin{figure}[!h] \centering
  \hspace{-5.5cm} {\bf{(A)}}  \hspace{6.25cm} {\bf{(B)}} \\
  \vspace{-0.3cm}
   \begin{multicols}{2}
\includegraphics[width = 0.485\textwidth]{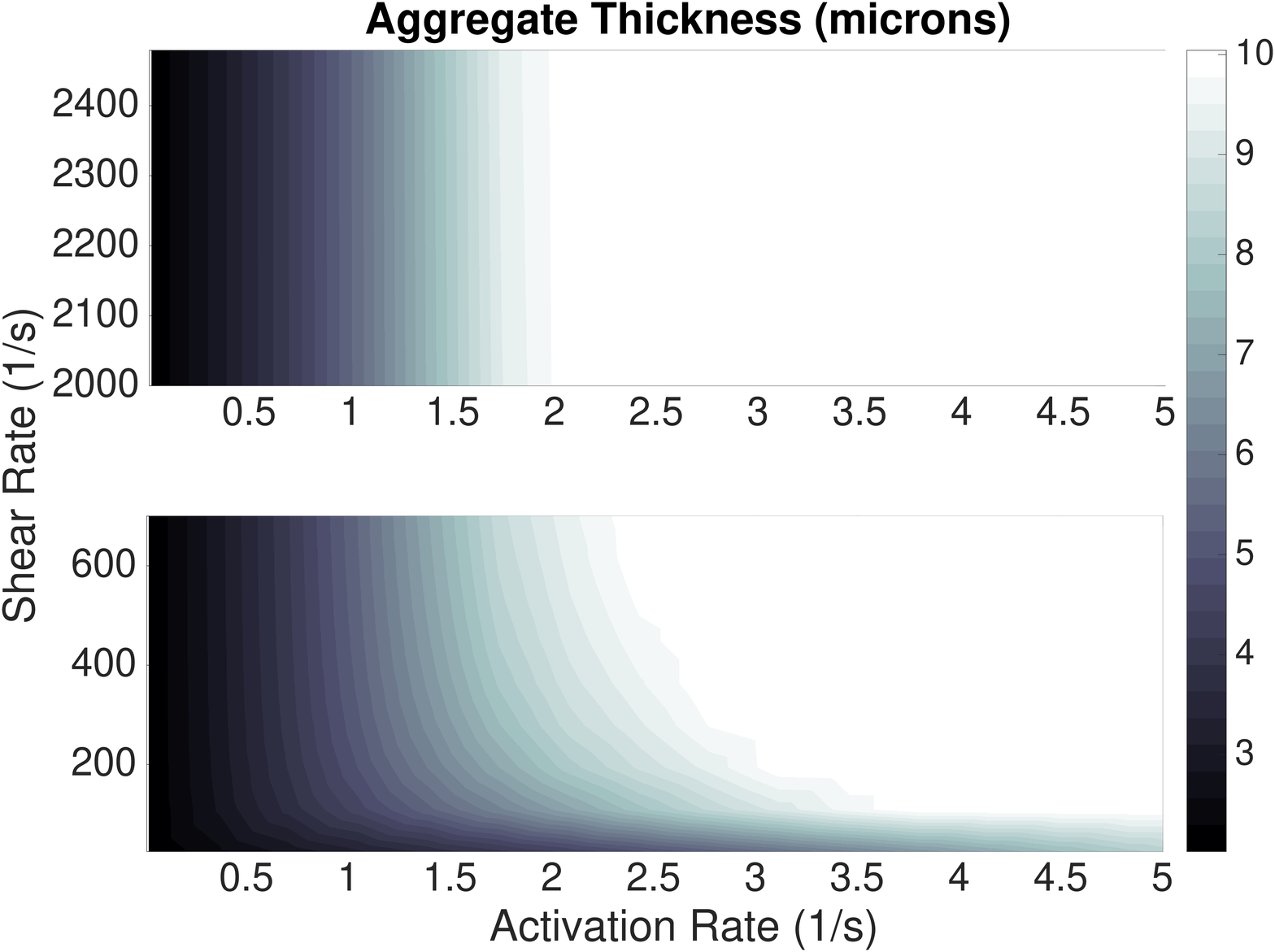}\\
\includegraphics[width = 0.485\textwidth]{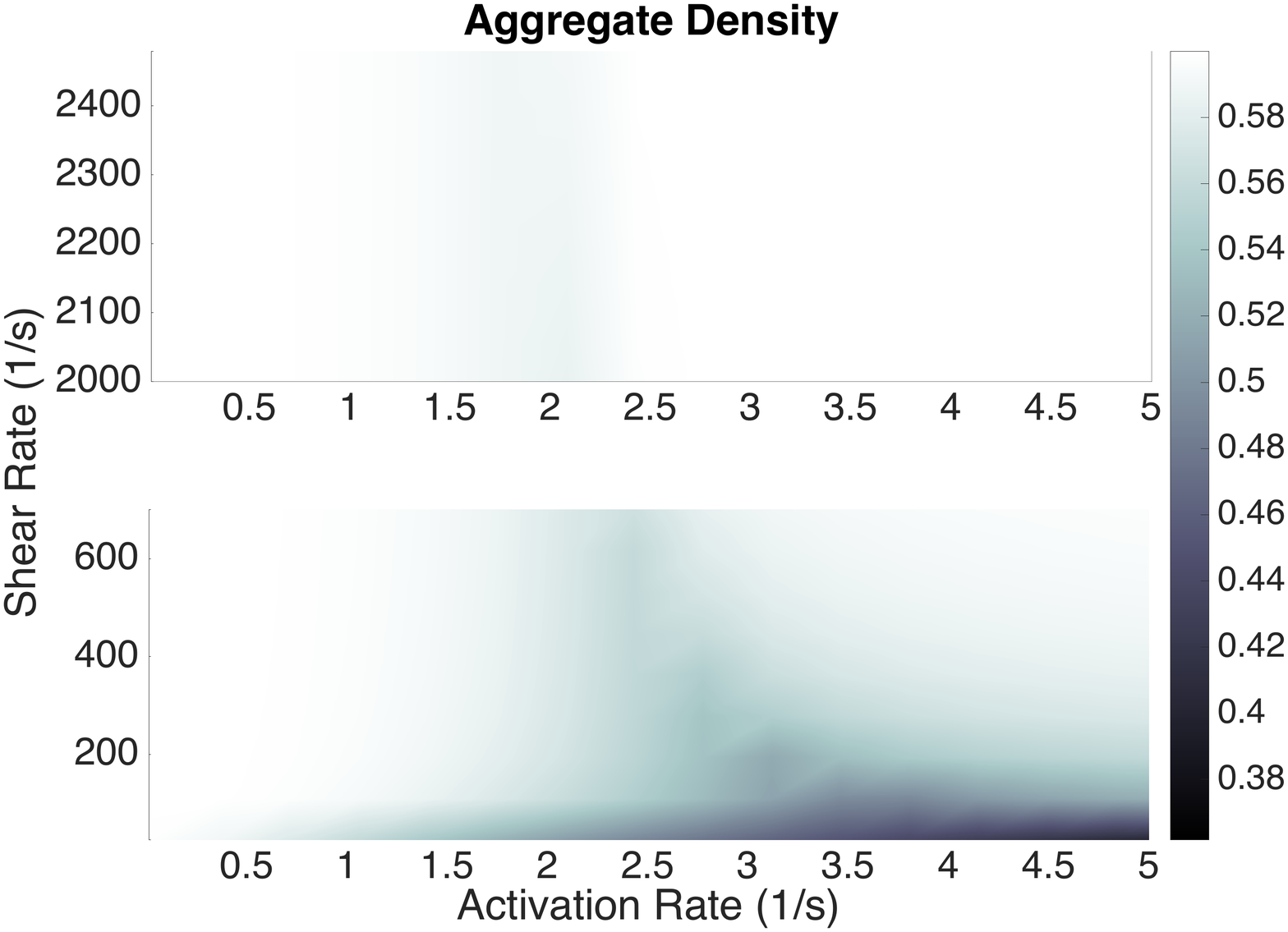}
\end{multicols}
\vspace{-0.75cm}
\caption{{\bf{Occlusion in the Injury Channel with ADP-independent
      activation.}} ODE model aggregate (A) thicknesses ($\mu$m) and
  (B) densities due to variations in ADP-independent platelet
  activation rates $k^{adp}_{act}$ and shear rate $\gamma$, after 10
  minutes of simulation time.}
   \label{fig:occlusion}
\end{figure}
Aggregate thickness $d_{\text{agg}}$ and aggregate density
$\theta^{B}$ after 10 minutes of simulation time are shown in
\cref{fig:occlusion}A-B. For shear rates $200 \text{ s}^{-1} < \gamma
< 2400 \text{ s}^{-1}$, occlusive aggregates formed if $k^{act}_{0} >
2.0\text{ s}^{-1}$ while significantly thinner but high density
aggregates form for $k^{act}_{0}$ below approximately 1.5 s$^{-1}$.
For shear rates less than $100 \text{ s}^{-1}$, aggregate thickness
increases as $k^{act}_{0}$ is increased, but in contrast to what is
seen for high shear rates, the aggregate density decreased as the
activation rate constant was increased from $1.0$ to $5.0
\text{ s}^{-1}$.  Hence the model predicts that occlusive aggregates
form under a wide range of shear rates provided platelet activation
rate is large enough, but the nature of the aggregates is sensitive to
activation rate especially at low shear rates.

\subsection{Dilution of Chemical Agonists}
\label{sec:results_dilution}
In this section we turn to the model's behavior when the platelet
activation rate is dependent on the ADP concentration $k^{plt}_{act} =
k^{act}_{0}$ $A([\text{ADP}]_{\text{agg}}, [\text{ADP}]_{\text{BL}})$.
If the ADP concentration is substantially below the transition
concentration $[\text{ADP}]^{*} = 1 \mu$M in this function, results
might be very different from those just reported with a constant rate
of platelet activation.

\begin{figure}[!h]
  \centering 
  \vspace{-0.25cm}
  \hspace{-5.5cm} {\bf{(A)}}  \hspace{6.25cm} {\bf{(B)}} \\
   \vspace{-0.25cm}
 \begin{multicols}{2}
 \includegraphics[width = 0.485\textwidth]{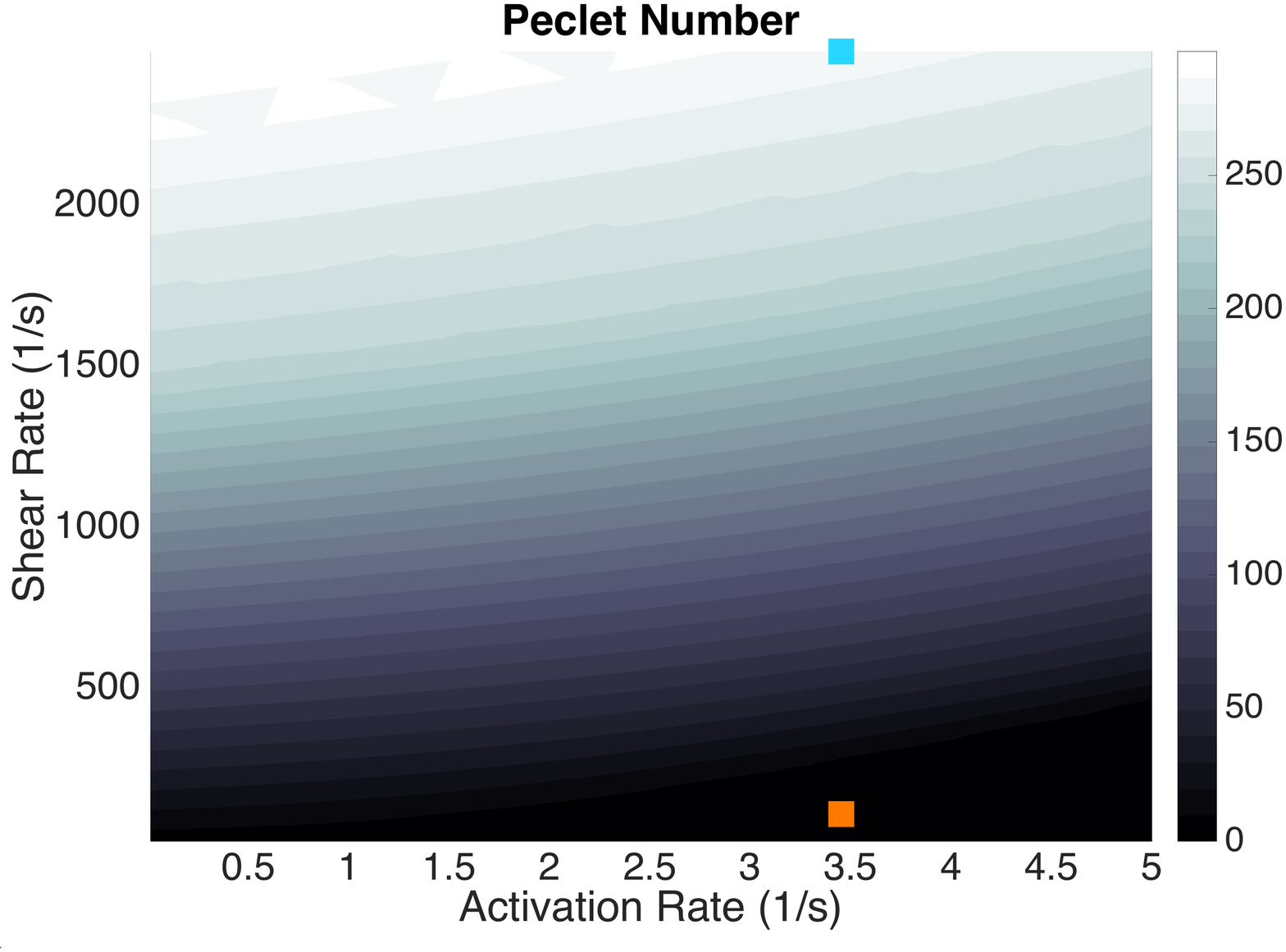}\\
\includegraphics[width = 0.485\textwidth]{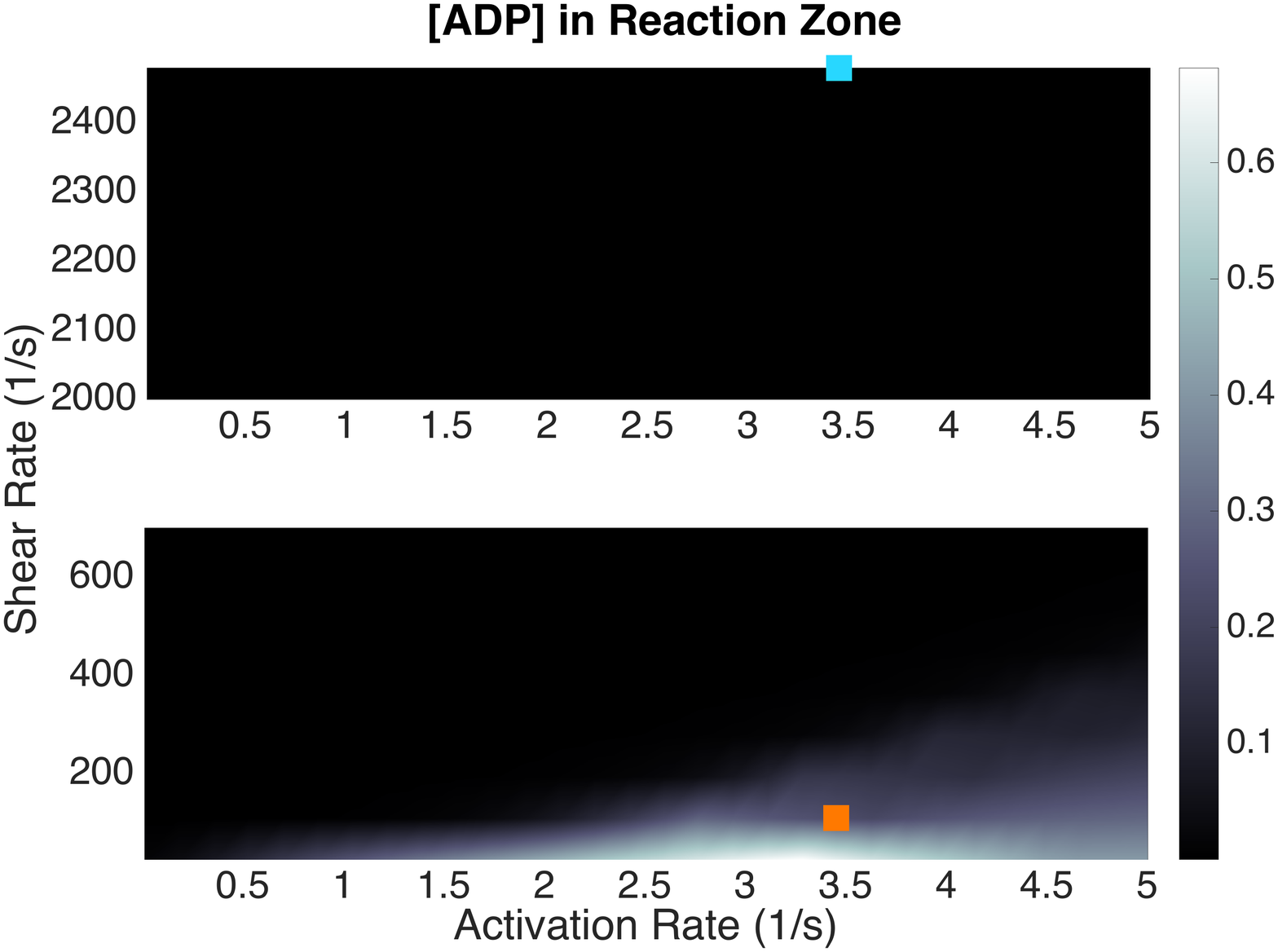}
  \end{multicols}
  \vspace{-0.25cm}
   \hspace{-5.5cm} {\bf{(C)}}  \hspace{6.25cm} {\bf{(D)}} \\
    \vspace{-0.25cm}
   \begin{multicols}{2}
  \includegraphics[width = 0.485\textwidth]{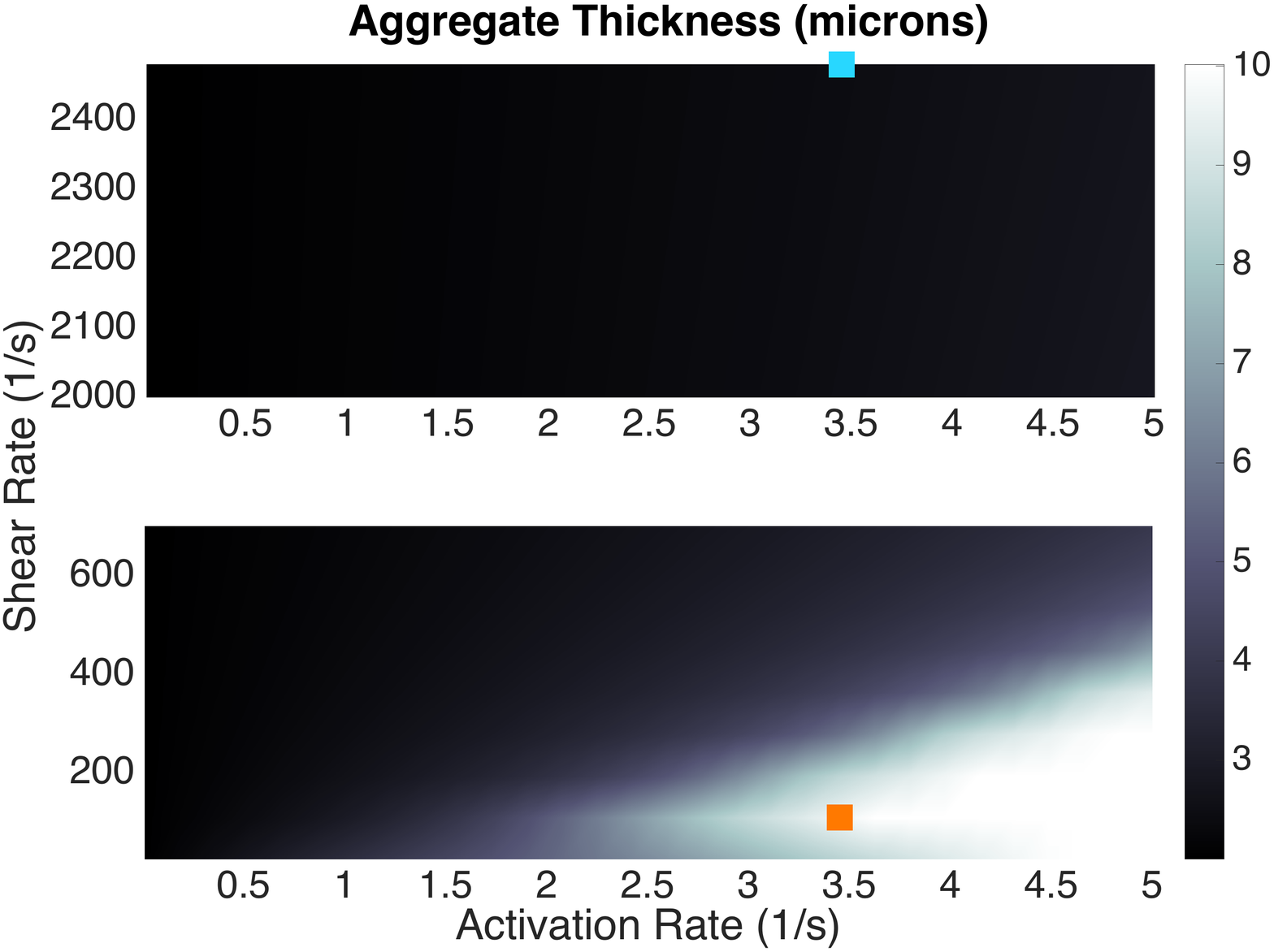}\\
   \includegraphics[width = 0.485\textwidth]{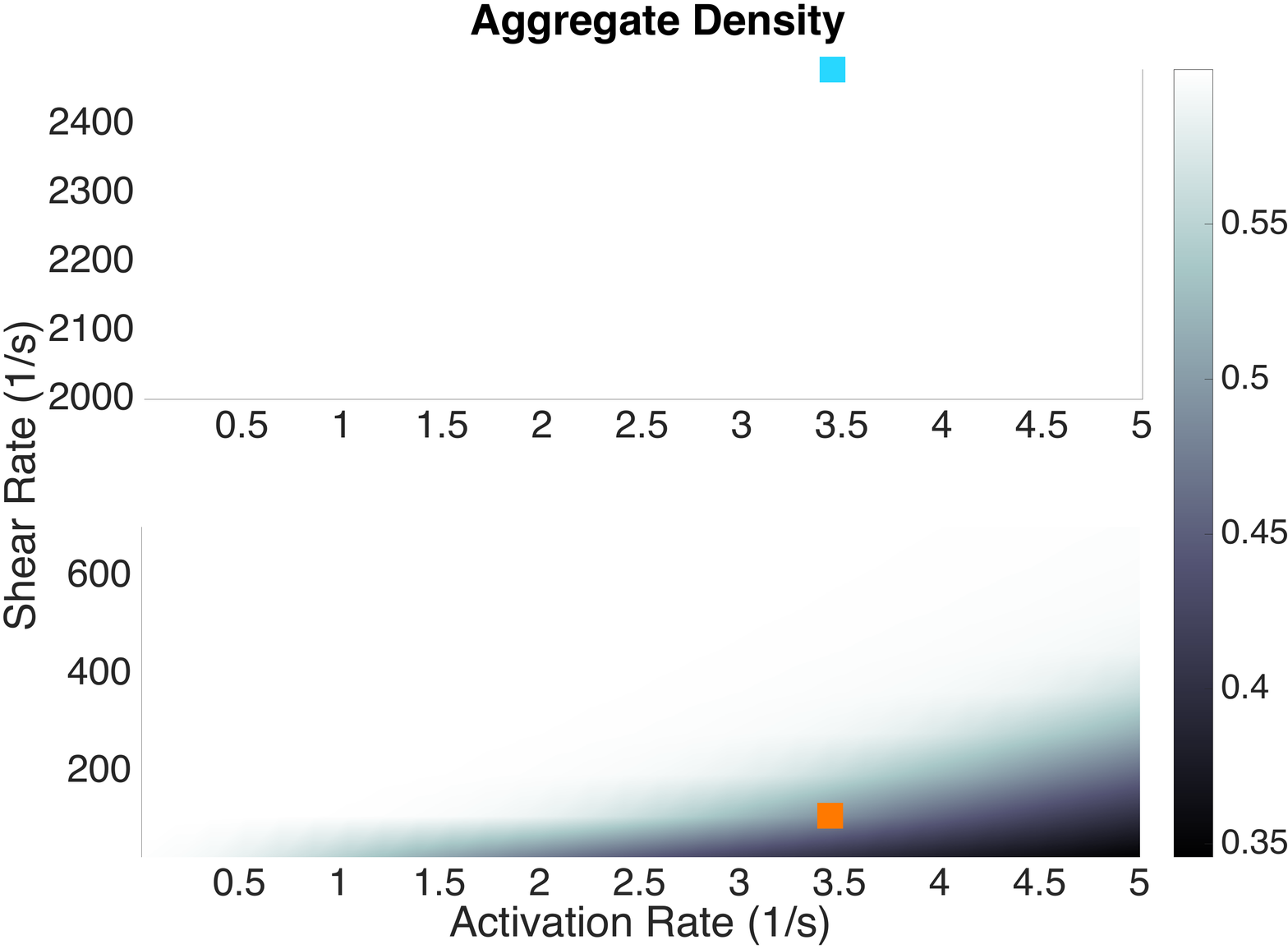}
  \end{multicols}
  \vspace{-0.5cm}
   \caption{{\bf{Aggregate Heterogeneity under variations in flow conditions and ADP-dependent platelet activation.}} (A) Peclet number describing the advective versus diffusive transport of soluble agonist ADP. (B) Concentration of ADP in the aggregate and associated boundary layer. (C) Aggregate thickness and (D) aggregate density as functions of shear rate and platelet activation rate constant. These results identified two flow regimes for a fixed platelet activation rate constant $k^{act}_{0} = 3.4 \ \text{s}^{-1}$; high shear (cyan square) and low shear (orange square). The metrics are taken after 10 minutes of simulation time.}
   \label{fig:dilution}
\end{figure}
 \begin{figure}
  \centering
   \begin{multicols}{2}
 \includegraphics[width = 0.45\textwidth]{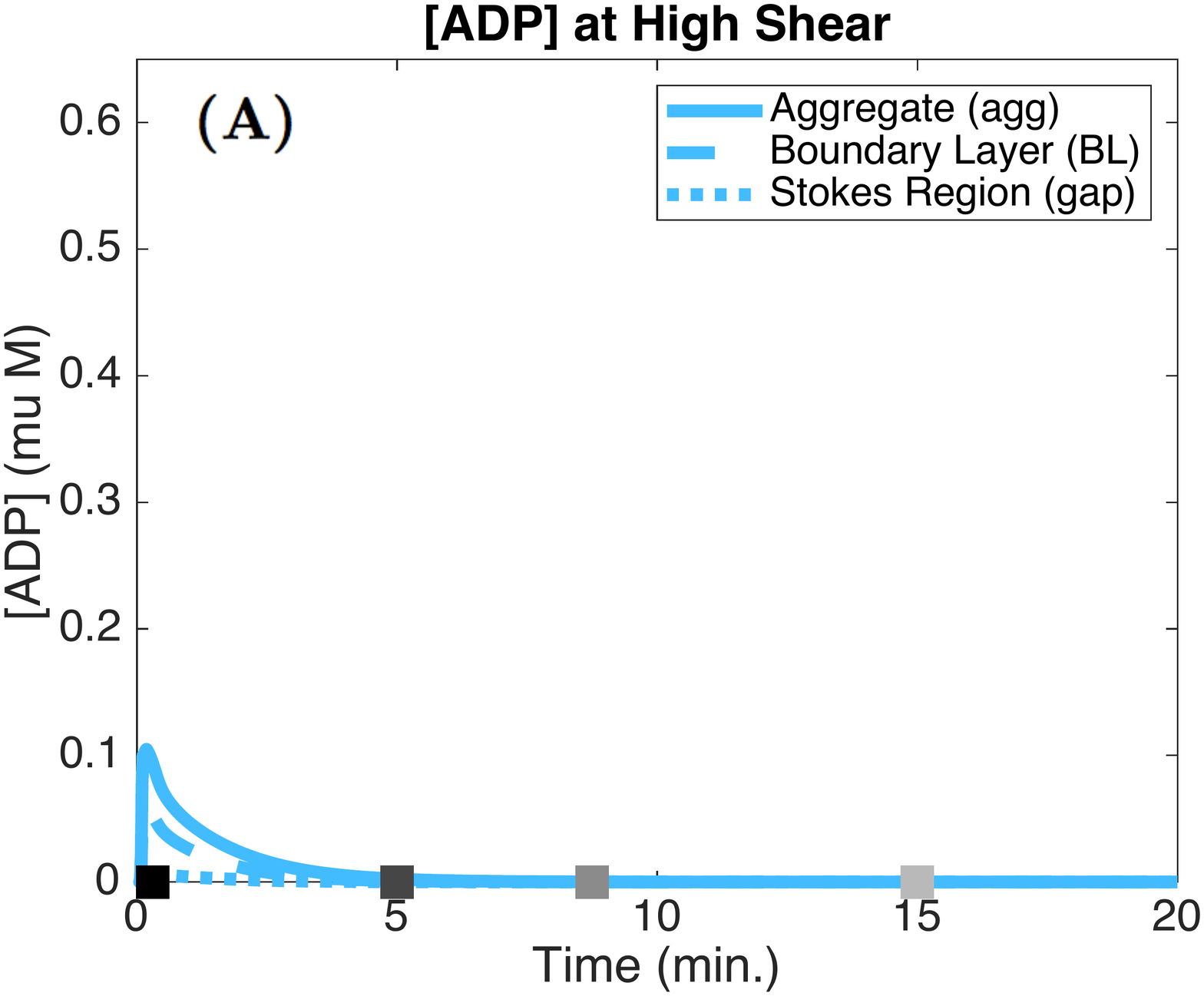}\\
  \includegraphics[width = 0.45\textwidth]{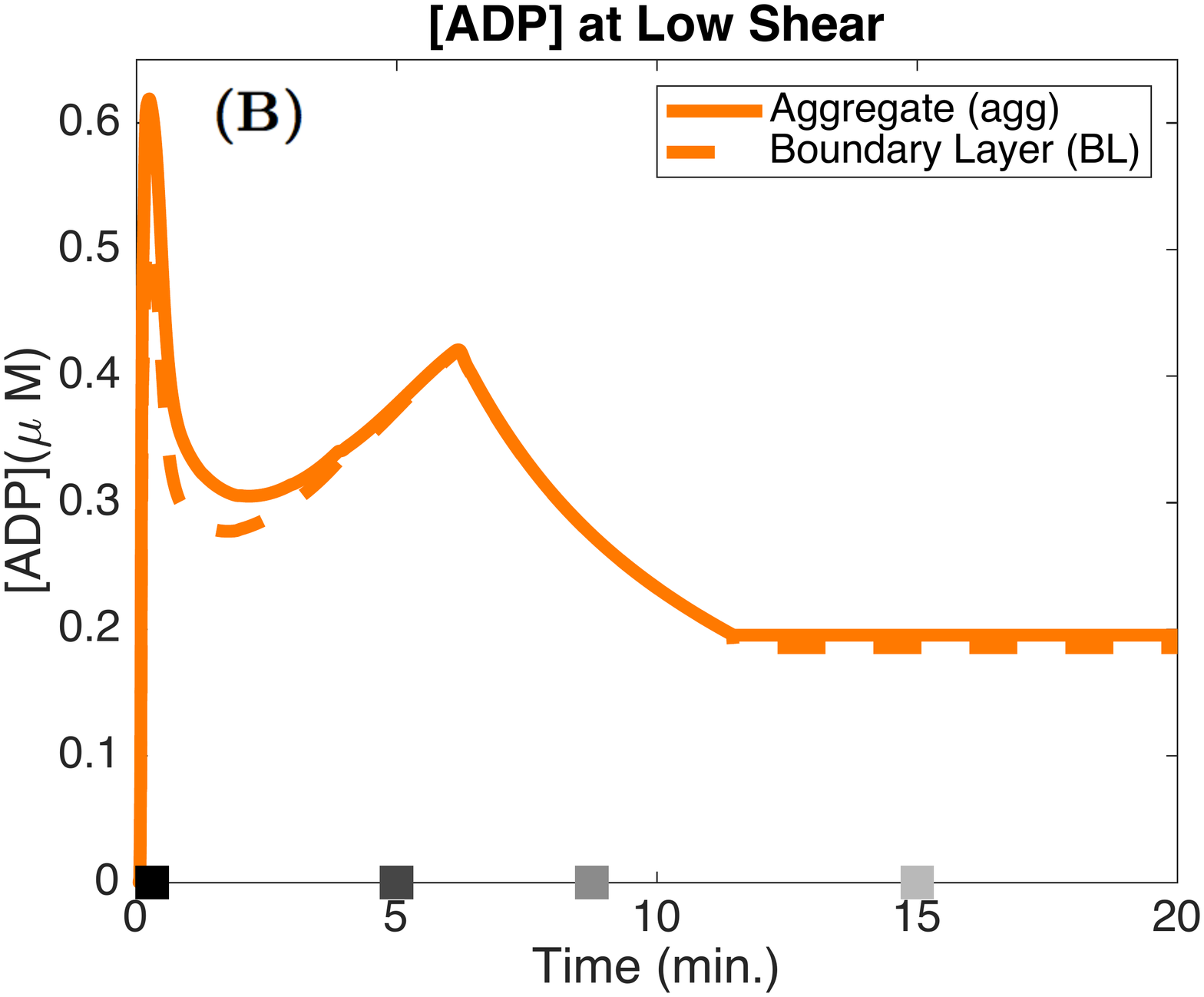}
  \end{multicols}
  \vspace{-0.75cm}
  \begin{multicols}{2}
  \includegraphics[width = 0.45\textwidth]{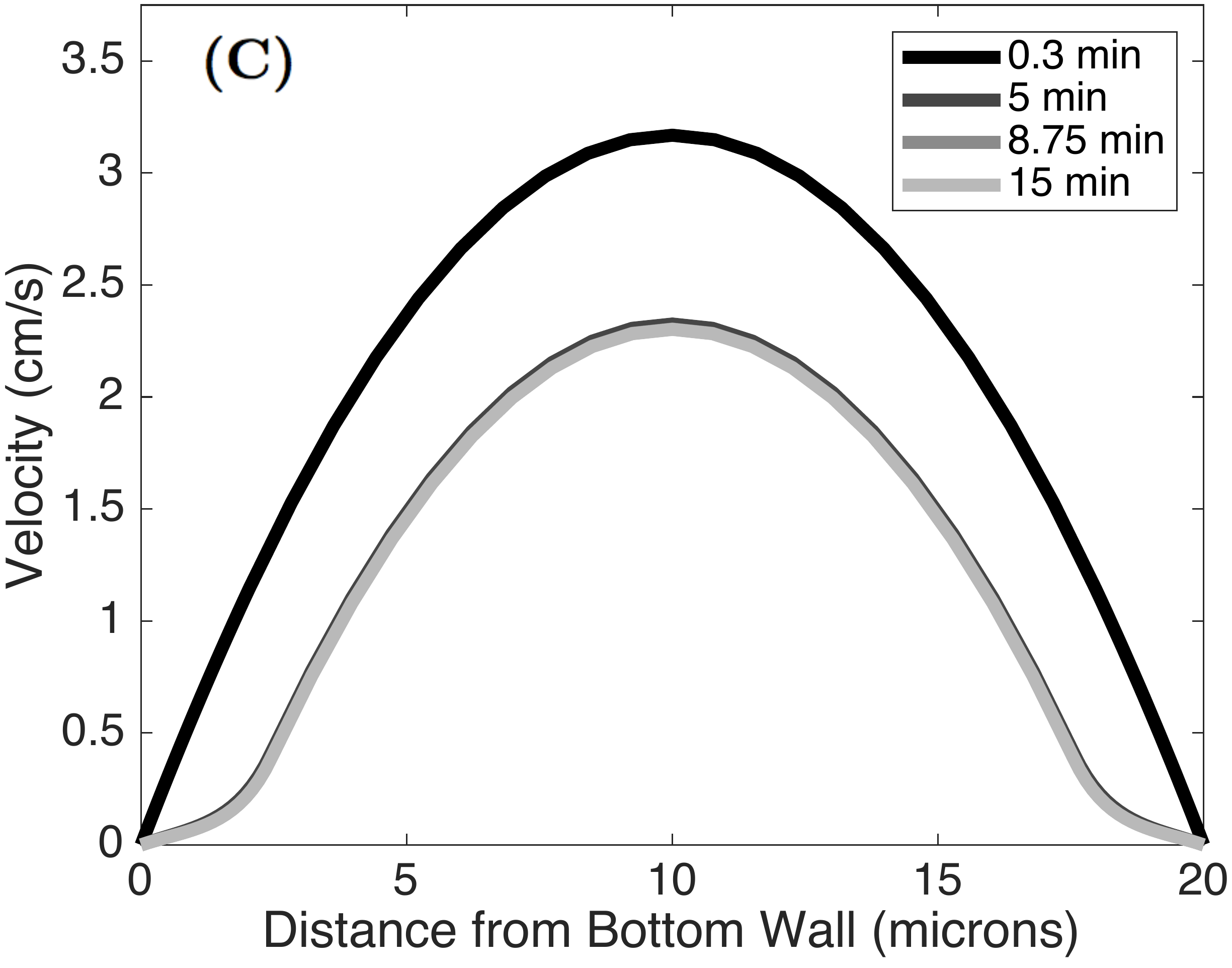}\\
 \includegraphics[width = 0.45\textwidth]{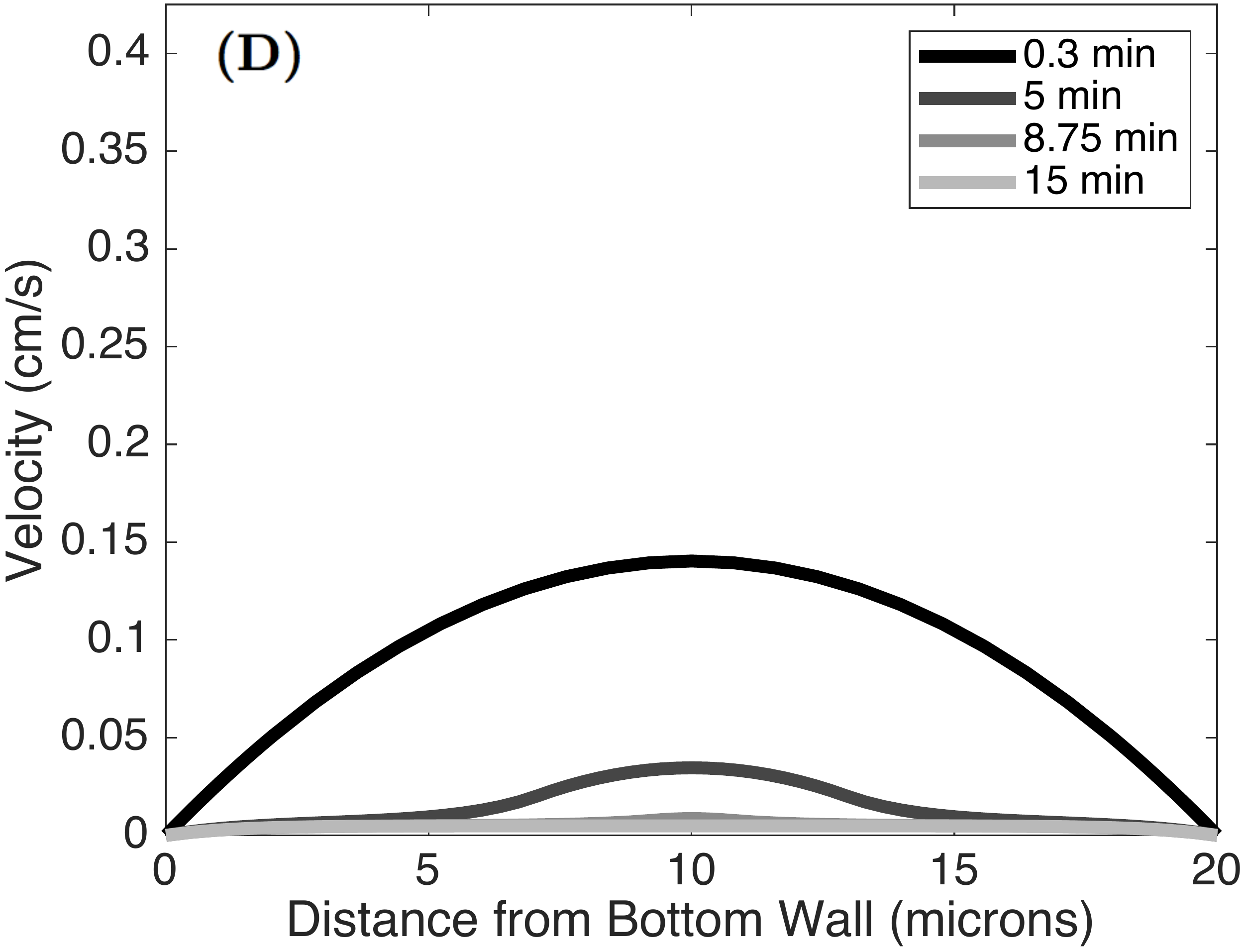}
 \end{multicols}  
 \vspace{-0.75cm}
\begin{multicols}{2}
 \includegraphics[width = 0.45\textwidth]{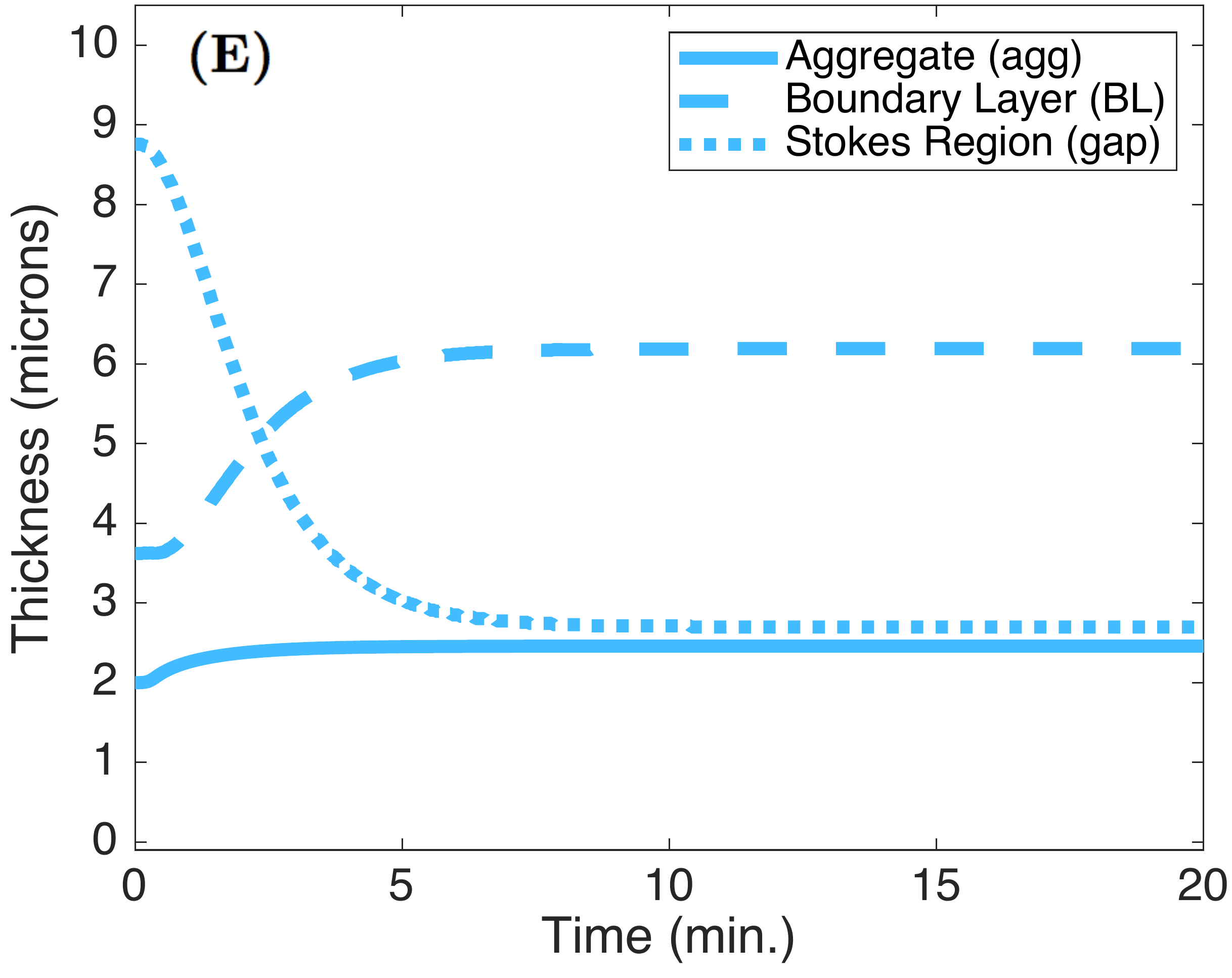}\\
  \includegraphics[width = 0.45\textwidth]{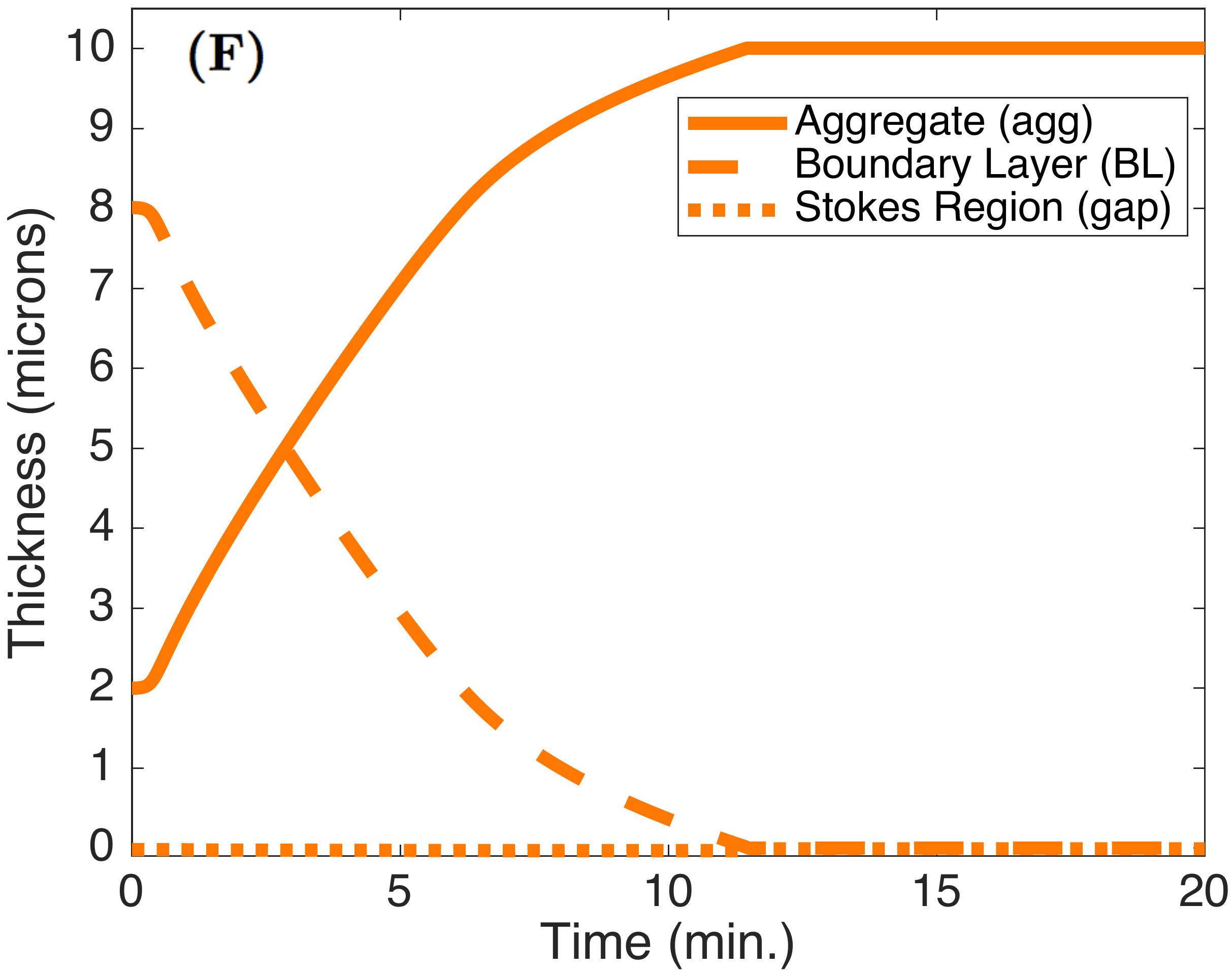}
  \end{multicols}
  \vspace{-0.75cm}
 \begin{multicols}{2}
 \includegraphics[width = 0.45\textwidth]{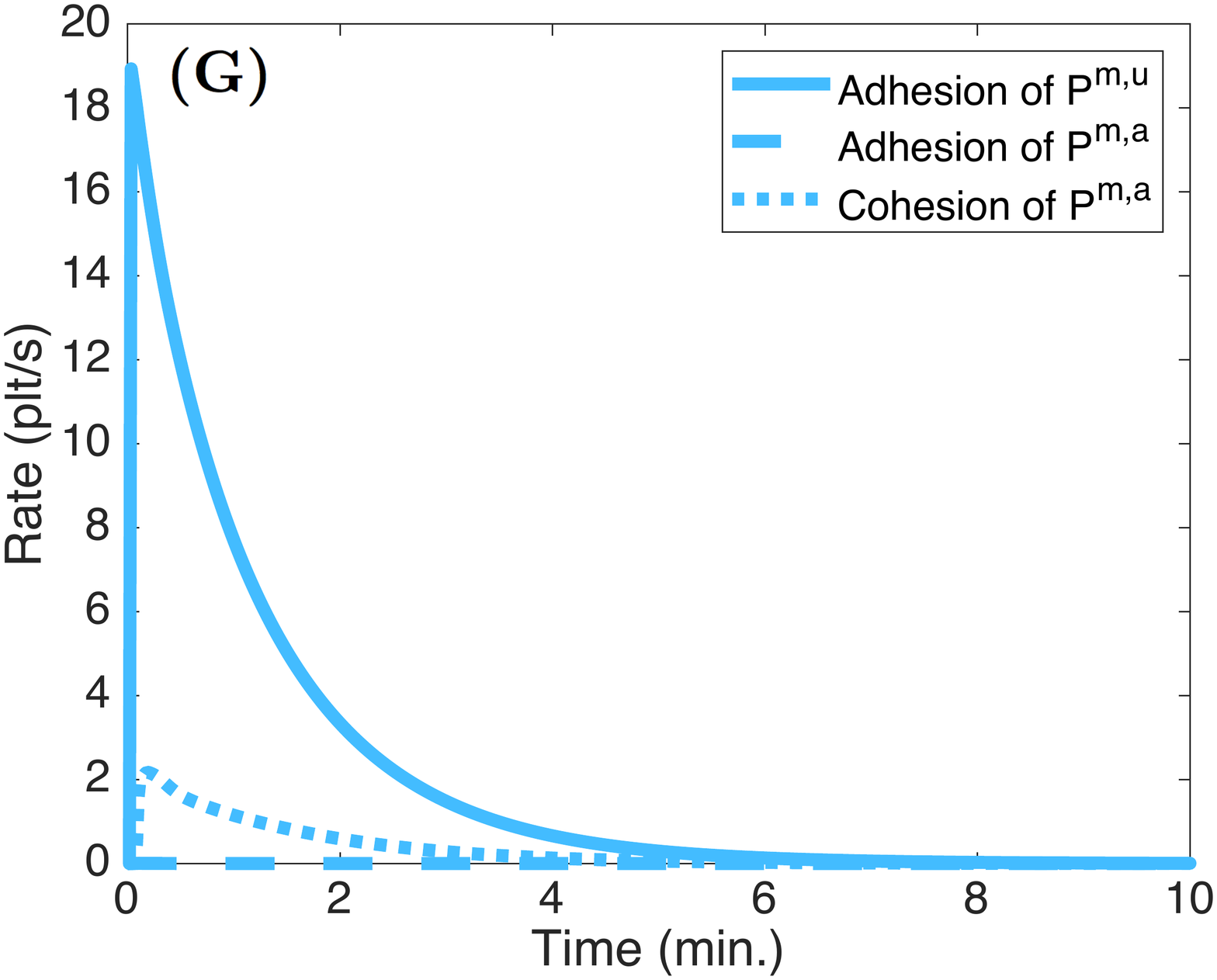}\\
  \includegraphics[width = 0.45\textwidth]{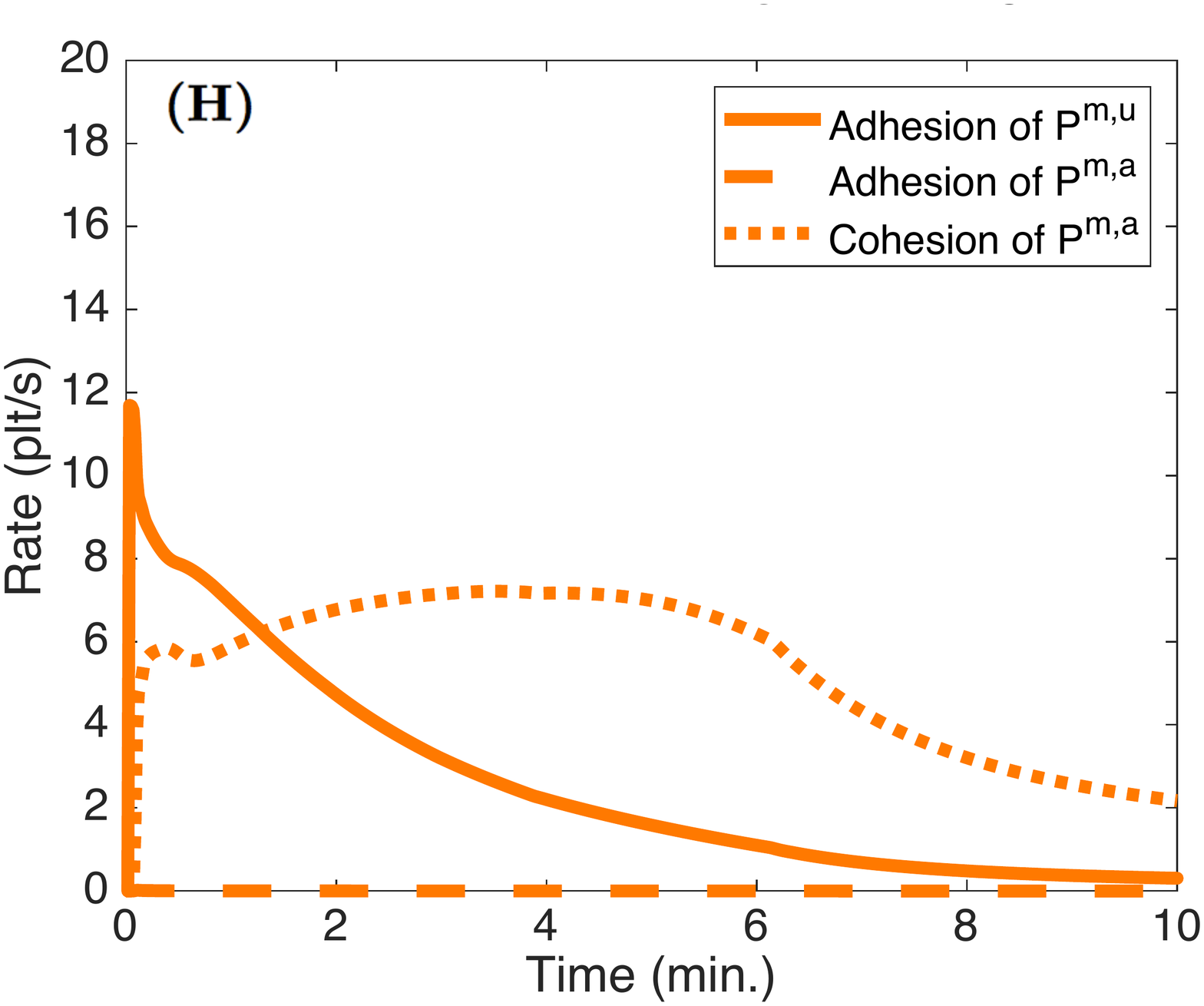}
  \end{multicols}  
  \vspace{-0.5cm}
    \caption{{\bf{Effect of ADP Dilution on Aggregate Formation.}} ADP concentrations in the aggregate, boundary layer, and gap under (A) high and (B) low shear conditions. Note there is no gap in the low shear conditions due to the thicknesses of the aggregate and boundary layer. The grey-scale squares denote the times at which the velocity profiles were determined. Velocity profiles at 0.3, 5, 8.75, and 15 minutes under (C) high and (D) low shear conditions.  Thicknesses of the aggregate, boundary layer, and gap under (E) high and (F) low shear conditions. The rates of adhesion and cohesion under high (G) and low (H) shear conditions.}
  \label{fig:proof}
\end{figure}
We explored how different forms of transport within the aggregates
affects their growth by calculating the ADP-related Peclet number
after 10 minutes of simulation time. This non-dimensional number is
defined as the ratio of the rate of advection by flow and the rate of
diffusion of ADP. For our purposes
$$
\text{Pe} = \frac{d_{\text{abl}} \ u(d_{\text{apz}})}{D_{a}},
$$
where $u(d_{\text{apz}})$ is the velocity at the edge of the ADP
boundary layer in the Stokes region ($d_{\text{apz}} = d_{\text{agg}} + d_{\text{abl}}$). \cref{fig:dilution}A shows that advective transport dominates for shear rates greater than $500 \text{ s}^{-1}$ for
activation rate constants $k_{0}^{act}$ ranging from $0.02 - 5.0
\text{ s}^{-1}$, but that, while still large, it is much lower for
shear rates less than $200 \text{ s}^{-1}$. We can see
that for most of the shear rate range, the concentration of ADP in
the PRZ after 10 minutes is much less that the activation transition
concentration $[\text{ADP}]^{*} = 1 \ \mu$M and that the shear rate
must be reduced to below $\approx 100 \text{ s}^{-1}$ for the ADP
concentration to reach $0.4 \ \mu$M.  As shown in
\cref{fig:dilution}B, increasing the value of $k_{0}^{act}$ results in
higher ADP concentrations. Plots of aggregate thicknesses and
densities in \cref{fig:dilution}C-D identify two flow regimes that for
a fixed activation rate constant yield occlusive aggregates under low
shear conditions and thin non-occlusive, but dense, aggregates under
high shear conditions.

To better understand the differences in aggregates that form under
high shear (cyan squares in \cref{fig:dilution}) and low shear (orange
squares in \cref{fig:dilution}) conditions, we consider the model output
in \cref{fig:proof}A-H.  The initial rate of adhesion is greater in
high shear case (\cref{fig:proof}G,H) because of the faster
replenishment of platelets by the faster flow.  Hence, the initial
rate of ADP release is also greater in this case.  Despite this, the
ADP concentrations are more than 6-fold lower (\cref{fig:proof}A,B)
because the faster flow carries ADP away more rapidly.  In the low
shear case, the two ADP boundary layers (ABL) are so thick
(\cref{fig:proof}E,F) that there is no gap region in that case.  For
both shear conditions, the velocity through the injury channel
decreases as the aggregates form. The reduction in peak velocity is
approximately 30\% after 15 minutes in the high shear case
(\cref{fig:proof}C), while the relative reduction is much greater, at
approximately 80\%, under low shear (\cref{fig:proof}D).  This larger
reduction in velocity results in lower rate of advective transport and
consequentially increases the boundary layer thickness (orange, dashed) for ADP (\cref{fig:proof}F). As the aggregate grows, this layer becomes smaller until the injury channel occludes. Because platelet
activation by ADP continues in the low shear case, ADP continues to be
released and its concentration remains significant for the entire
period leading up to occlusion. 

\section{Discussion}
\label{sec:discussion}
We developed the first mathematical model of flow-mediated primary hemostasis in an extravascular injury which is able to track the process from initial platelet deposition to occlusion.  Model calibration with PDE and MFA models yielded excellent agreement in model metrics. ODE model comparisons with the PDE model velocity profiles, amounts of ADP, and total number of bound platelets validate ODE model components describing the fluid, platelets, ADP, and aggregate formation dynamics. The model met the difficult and nuanced challenge of capturing injury occlusion in a non-spatial description. We compared MFA occlusion times to ODE model output as a function of ADP-independent platelet activation rate ($A([ADP]_{\text{agg}},[ADP]_{\text{BL}})= 1$). \ck{Next,} we studied \ck{both} ADP-independent and ADP-dependent platelet activation and how both activation and flow affect aggregate structure. Metrics of transport, ADP concentration, and aggregate height \ck{and} density confirmed \ck{that ADP dilution significantly limited} aggregate formation. \ck{However, we have shown experimentally that platelet aggregates can fully occlude the injury channel in our bleeding chip (data not shown) in the absence of coagulation, and thus there is a need to extend and further refine our platelet model. In particular,} force-dependent bond formation and breaking in the rates of platelet adhesion and cohesion \ck{will likely be necessary to include. This will} require the modeling of more mechanisms of platelet activation but it will also allow for platelet detachment from a growing aggregate.

\subsection{Limitations \& Extensions}
Most of the model parameters have been either previously estimated from the literature or fit to the model output generated by the PDE and MFA models. The previously estimated parameters are found in \Cref{sec:appC} (\Cref{params:diff} and \Cref{params:transitions}). The parameter values we are less certain of are numbers corresponding to the functions describing adhesion and cohesion ($d_{se}$, $k_{b}$, and $s_{b}$) and the permeability constant $C_{K}$, which prescribes the growing resistance to flow within the aggregate. In reference to $d_{se}$, $k_{b}$ and $s_{b}$, the values were determined through tuning ODE model output 
with flow rates and the number of bound platelets in the aggregate associated with the PDE model. Details are found in \Cref{sec:calibration} (\Cref{params:adh_coh}). We determine a value for $C_{K}$ using the protocol described in \Cref{sec:permeable}. In addition to these constant parameters, the number density of mobile unactivated platelets at the inlet to the injury channel, $P^{up}(y)$, is prescribed. We assume that the shape of the distribution does not evolve as the top and bottom aggregates grow. \ck{However, t}he model framework allows for the change in flow to affect the evolution of $P^{up}(y)$, the upstream platelet distribution.

In this paper, we \ck{consider only one pathway} of platelet activation and do not consider coagulation. \ck{In reality, ADP, thrombin (resulting from coagulation)}, and subendothelial collagen elicit different \ck{types of} platelet activation responses. For example, thrombin cleaves \ck{protease-activated receptors on the platelet surface}, PAR-1 and PAR-4, \ck{and is a stronger activator of platelets than ADP} \cite{Andersen1999,Li2010}.  
There is increasing indication that partially activated and even unactivated platelets may be able to bind, if only transiently, to a growing aggregate. Such binding would increase \ck{a} platelet's time of exposure to other chemical agonists and possibly lead to more activation and faster aggregate growth. This \ck{further} motivates the extension of the model to include alternative mechanisms of platelet activation and binding/unbinding.

The processes of platelet binding and unbinding are dictated by the types of bonds that form between platelets and the vessel wall as well as amongst platelets. Along with the transport of platelets to the site of injury, the flow conditions in the local environment determine the likelihood that bonds will form and break. At low shear rates ($< 200$ (1/s)), unactivated platelets near the wall adhere to the collagen embedded in the subendothelial matrix \cite{jackson2003signaling,savage1996initiation}. Specifically, bonds between GPVI receptors and collagen form and trigger activation of integrin receptors ($\alpha_{2}\beta_{1}$ \& $\alpha_{IIb}\beta_{3}$). The platelet-collagen bonds bring the platelet to rest and allow for firm adhesion to the subendothelial surface. Adhesion of platelets is a multistep process at high shear rates ($> 500$ (1/s)). More specifically, unactivated and mobile platelets transiently bind to immobilized vWF molecules associated with a thrombus via their GPIb receptors \cite{Jackson2007,Ruggeri2009}. This process slows the platelets and allows time for activation of integrins that mediate the formation of strong, long-lived $\alpha_{IIb}\beta_{3}$-fibrinogen bonds \cite{Jackson2007,Ruggeri1999,Savage1998}. \ck{Future work will include model extensions of this framework that} differentiate platelets by the types of bonds that  form on their surfaces and \ck{by} drag \ck{on the aggregate that effects} individual bonds amongst the aggregated platelets.

In addition to extending our occlusive model of platelet aggregation to include more mechanisms of activation and bond formation and breaking, we can include the biochemical reactions of coagulation, further maximizing the versatility and reduced computational cost of the model. The coagulation cascade occurs on the surfaces of platelets producing a key end product thrombin. Thrombin converts fibrinogen to fibrin, which polymerizes forming a stabilizing mesh that supports the platelet aggregate. Our previous mathematical model of thrombosis \cite{Kuharsky2001,link2018local,link2019mathematical} can be linked to the above model, contributing minimal additional computational cost. Both model extensions will utilize the above framework to develop the first flow-mediated mathematical model of hemostasis in an extravascular injury that incorporates both platelet function and coagulation, resulting in a high throughput screening tool that can identify modifiers of occlusion and embolization.

\section*{Author Contributions}
\vspace{0.5cm}
{\scriptsize{
\begin{itemize}
\item[]{\bf{Conceptualization:}} Kathryn G. Link, Aaron L. Fogelson, Karin Leiderman 
\item[]{\bf{Experimental Data:}} Matthew G. Sorrells, Keith B. Neeves
\item[]{\bf{Funding acquisition:}} Aaron L. Fogelson, Karin Leiderman, Keith B. Neeves
\item[]{\bf{Model Development:}} Kathryn G. Link, Aaron L. Fogelson
\item[]{\bf{Validation \& Calibration:}} Kathryn G. Link, Matthew G. Sorrells, Nicholas A. Danes, Karin Leiderman, Keith B. Neeves, Aaron L. Fogelson
\item[]{\bf{Writing-original draft:}} Kathryn G. Link, Aaron L. Fogelson
\item[]{\bf{Writing-review \& editing:}} Kathryn G. Link, Aaron L. Fogelson, Karin Leiderman, Keith B. Neeves
\end{itemize}}}

\bibliographystyle{siamplain}
\bibliography{references.bib}

\appendix
\section{Flow Through Bleeding Chip}
\label{sec:appA}
\subsection{Circuit Analog System}
\label{sec:appA_circuit}
We use hydrodynamic analogs of Kirchoff's and Ohm's laws to formulate a linear system of equations that describes the flow through the hydraulic circuit (HC) shown in \cref{fig:fluid}A.
\begin{align}
Q_{1}-Q_{M} &= Q_{2}, \label{circuit_eq1}\\
Q_{3}-Q_{M} &= Q_{4},\\
P_{B} - P_{1} &= Q_{1}R_{1},\\
P_{W} - P_{2} &= Q_{3}R_{3},\\
P_{1} - P_{CB} &= Q_{2}R_{2},\\
P_{2} - P_{CW} &= Q_{4}R_{4},\\
P_{1} - P_{2} & = Q_{M}R_{M}\label{circuit_eq6}.
\end{align}
\noindent The inlet and outlet pressures $P_{B}$, $P_{W}$, $P_{CB}$, and $P_{CW}$ are known and we assume the resistance through the injury channel $R_{M}$ is an input parameter determined by the Brinkman-Stokes calculation. Additional input parameters include $R_{1}$, $R_{2}$, $R_{3}$, and $R_{4}$, which correspond to the resistances associated with the blood and wash channels. Equations \eqref{circuit_eq1}-\eqref{circuit_eq6} yields a system of seven equations with seven unknowns. We can write them as the following matrix system.
\begin{equation}
\begin{pmatrix}
1 & \text{-}1 & 0 & 0 & \text{-}1 & 0 & 0  \\
0 & 0 & 1 & \text{-}1 & \text{-}1 & 0 & 0  \\
R_{1}& 0 & 0 & 0 & 0 & 1 & 0 \\
0 & 0 & R_{3} & 0 & 0 & 0 & \text{-}1  \\
0  &R_{2} & 0 & 0 & 0 & 1 & 0  \\
0 & 0 & 0 & R_{4} & 0 & 0 & 1 \\
0 & 0 & 0 & 0 & \text{-}R_{M} & 1 & \text{-}1  \\
\end{pmatrix}
\begin{pmatrix}
Q_{1}\\
Q_{2}\\
Q_{3}\\
Q_{4}\\
Q_{M}\\
P_{1}\\
P_{2}
\end{pmatrix}
=
\begin{pmatrix}
0\\
0\\
P_{B}\\
P_{W}\\
-P_{CB}\\
-P_{CW}\\
0
\end{pmatrix}
 \end{equation}
Values of pressures and resistances used in the hydraulic circuit (HC) system are summarized in \Cref{hca:dimensions}, \Cref{hca:resistances}, and \Cref{hca:pressure}.  There are variable viscosities corresponding to the blood, wash and injury channels. Furthermore, the downstream end of the wash channel will be a mix of the blood and the wash when blood can flow freely from the blood channel through the injury channel. We account for these viscosities through the resistance terms found in \Cref{hca:resistances}. Assume that the viscosity of the downstream end of the wash channel is an average of the blood and wash viscosities. In the fluid simulations, we use a volume-averaged viscosity through the entire domain $\mu = 0.0267$ dynes$\cdot$s/cm$^{2}$. We also assume there is an effective length of the injury channel that spans outside its 150 $\mu$m length, corresponding to the laminar flow that enters and exits the injury channel.  The effective length of the channel, $L_{h}^{\text{eff}} = 165 \ \mu$m, is used in the calculation of the pressure gradient through the injury channel when using the PDE `flow map' output. 
  
 \begin{table}[!h]
 \footnotesize
\centering
\caption{{\bf{Dimensions of Hydraulic Circuit (HC).}} See \cref{fig:fluid}A-B for reference.}
\vspace{1.0pt}
\begin{tabular}{|l|l|}
\hline
Length from $P_{B}$ to $P_{1}$ (and $P_{W}$ to $P_{2}$) & $85 \ \mu$m\\
\hline
Length from $P_{1}$ to $P_{CB}$ (and $P_{2}$ to $P_{CW}$) & $85 \ \mu$m\\
\hline
Width of vertical channels & $100 \ \mu$m\\
\hline
Length of injury channel & $150 \ \mu$m\\
\hline
Width of injury channel & $20 \ \mu$m\\
\hline
Depth of injury channel & $60 \ \mu$m\\
\hline
\end{tabular}
\label{hca:dimensions}
\end{table}

\vspace{-0.5cm}
 
  \begin{table}[!h]
  \footnotesize
\centering
\caption{{\bf{HC Viscosities and Computed Resistances.}} See \cref{fig:fluid}A for reference.}
\vspace{1.0pt}
\begin{tabular}{|l|l|}
\hline
Blood Channel Viscosity & $3.6 \times 10^{-2}$ Poise\\
\hline
Upstream Wash Channel Viscosity & $1 \times 10^{-2}$ Poise\\
\hline
Injury Channel Viscosity & $2.67 \times 10^{-2}$ Poise\\
\hline
Downstream Wash Channel Viscosity & $2.3 \times 10^{-2}$ Poise\\
\hline
$R_{1}$, $R_{2}$ & $5.3672\times 10^{6}$ Pa s/m$^{2}$\\
\hline
$R_{M}$ (no aggregate) & $6.60825 \times 10^{8}$ Pa s/m$^{2}$\\
\hline
$R_{3}$ & $1.02 \times 10^{6}$ Pa s/m$^{2}$\\
\hline
$R_{4}$ & $2.346 \times 10^{6}$ Pa s/m$^{2}$\\
\hline
\end{tabular}
\label{hca:resistances}
\end{table}
   
   \vspace{-0.5cm}
   
 \begin{table}[!h]
 \footnotesize
\centering
\caption{{\bf{Pressures used in the HC.}} See \cref{fig:fluid}A for reference.}
\vspace{1.0pt}
\begin{tabular}{|l|l|}
\hline
Pressure Inlet (Blood, $P_{B}$) & $6.24364 \times 10^{2}$ Pa\\
\hline
Pressure Inlet (Wash, $P_{W}$) & $1.7418 \times 10^{2}$ Pa\\
\hline
Pressure Outlet (Blood, $P_{CB}$) & $2.8422 \times 10^{2}$ Pa\\
\hline
Pressure Outlet (Wash, $P_{CW}$) & $0$ Pa\\
\hline
Pressure Drop Across Injury $P_{1} - P_{2}$ (no aggregate) & $2.80851 \times 10^{2}$ Pa\\
\hline
\end{tabular}
\label{hca:pressure}
\end{table}

\vspace{-0.5cm}

 \begin{table}[!h]
 \footnotesize
\caption{{\bf{Parameters that characterize the geometry and flow, mass transfer, and reaction regimes in the injury channel.}} The average blood velocity $U$, $\nu$ kinematic viscosity, $D$ diffusivity of platelets, $\gamma$ wall shear rate. **$D_{h}$, hydraulic diameter $[2d_{h}L_{z}/(d_{h} + L_{z})]$.}
\centering
\vspace{1.0pt}
\begin{tabular}{|llll|}
\hline
{\bf{Parameters}} & {\bf{Expression}} & {\bf{Value}} & {\bf{Constants}} \\
\hline
Aspect ratio & $d_{h}/L_{z}$ & $0.333$ & $d_{h} = 20  \ \mu\text{m}$,  $L_{z} = 60  \ \mu\text{m}$\\
\hline
Relative injury size &  $L_{h}/d_{h}$ & $7.5$ &  $L_{h} = 150  \ \mu\text{m}$\\
\hline
Reynolds number ($R_{e}$) & $U d_{h} / \nu $ & $0.0283$ & \parbox{9em}{$\nu=1.5\times10^{-5} \text{m}^{2}/\text{s}$, $U=2.13\times10^{-2} \text{m/s}$} \\
\hline
Entrance Length ($L_{e}$) & $0.05 R_{e} D_{h}$ & $0.0425 \ \mu$m & $D_{h} = 30 \ \mu\text{m}$ \\
\hline
Peclet number ($P_{e}$) & $\gamma d_{h}^{2} / 6D_{p}$ & $6.4\times 10^{3}$ & \parbox{10em}{$\gamma = 2400 \ \text{s}^{-1}$, $D_{p} = 2.5 \times 10^{-5}$ m$^{2}$/s} \\
\hline
Mass transfer coeff. ($k_{m}$) & $(D^{2}\rho / 8 L_{h})^{1/3}$ & $8.2\times10^{-6}$ m/s & $\rho = 1060 \text{ kg/m}^{3}$\\
\hline
\end{tabular}
\label{hca:flow_params}
\end{table}

\subsection{Experimental Methods \& Materials}
\label{sec:appA_methods}
\subsubsection{Materials}
Bovine serum albumin (BSA) (A9418), 3,3'-dihexylox~acarbocyanine iodide (DiOC6) (D273) were from Sigma-Aldrich (St Louis, MO, USA). 3.2\% sodium citrate vacutainers (369714) and 21-gauge Vacutainer\textregistered  \ 21 Safety-Lok blood collection sets (367281) were from Becton Dickson (Frankwood Lakes, NJ, USA). Tridecafluoro-1,1,2,2-tetrahydrooctyltrichlorosilane (FOTS) (SIT8174.0) was from Gelest (Morrisville, PA, USA). Polydimethylsiloxane and crosslinker were obtained from Krayden (Denver, CO, USA). Collagen related peptides (CRP-XL) [GCO (GPO)10GCOG-amide], 5(6)-carboxyfluorescein, and VWF-III [GPCGPP)5GPRGQ ~ OGVMGFO(GPP)5GPC-amide] were obtained from Cambcol Laboratories (Cambridgeshire, UK). Glass slides (1'' x 3'') were obtained from Fisher Scientific (Lenexa, KS, USA). Phosphate-buffered saline (PBS) was prepared to 137 mM NaCl, 2.7 mM KCl, 10 mM Na2HPO4, 1.8 mMKH2PO4, and pH 7.4. Phe-Pro-Arg-chloromethylke ~ tone (PPACK) vacutainers were obtained from Haematologic Technologies, Inc  (Essex, VT). 

\subsubsection{Blood Collection and Preparation} 
Blood was collected from healthy donors by venipuncture into 4.5 mL PPACK vacutainers (final concentration of 75 $\mu$M) using a 21 gauge needle. Prior to collecting blood into the PPACK vacutainers, one vacutainer of blood was collected into a 3.2\% sodium citrate vacutainer and discarded. DiOC6 was added to blood to a concentration of 1 $\mu$M and was incubated at 37 $^{\circ}$C for 10 min prior to the microfluidic device assay. The study and consent process was approved from the Colorado Multiple Institutional Review Board in accordance with the Declaration of Helsinki. 

\subsubsection{Device Fabrication, Preparation, and Operation}
A PDMS microfluidic device was created using stand photolithography and soft lithography techniques. The device was made with an `H-shaped' geometry, with a `blood' channel (w=110 $\mu$m, h=50 $\mu$m) on the right, a `wash' channel on the left (w=110 $\mu$m, h=50$ \ \mu$m), and an `injury' channel (l=150 $\mu$m, w=20 $\mu$m, h=50 $\mu$m) connecting the two vertical channels (\Cref{fig:model}A). The device was plasma bonded to a PDMS coated glass slide, and a solution of the collagen related peptides CRP, GFOGER, and VWF-BP were patterned in the injury channel and wash channel at a concentration of 250 $\mu$g/mL for each peptide for 4 hours. Following patterning, the entire device was rinsed and blocked with 2\% BSA in PBS for 45 min. 

\subsubsection{Whole Blood Microfluidic Flow Assays}
Pressure controllers (Flu- ~ igent MFCS-EZ)  coupled to flow meters (Fluigent FRP) were used to perfuse blood and wash buffer at a flowrate of 5.5 $\mu$L/min and 10 $\mu$L/min, respectively, through the device. Images of thrombus in the injury channel were obtained using brightfield and epifluorescent microscopy on an inverted microscope (Olympus IX81, 20X Objective, NA 0.4, Hammamatsu Orca R2 Camera). Assays were run until the injury channel was fully occluded. The time to occlusion in the device was determined through a variety of ways. Optically, we defined the time to occlusion as the time point in which no red cells could be seen passing through the injury channel. Additionally, we used the flow meter measurements on the inlet and outlet of the device to compute the flow rate through the injury channel. This flow measurement was used to calculate two other occlusion metrics: the time to reach 10\% of the endpoint flow rate and the time needed to reach 0.35 $\mu$L/min flow.

\subsection{Creation of flow map}
\label{sec:appA_flowmap}

For comparisons with the PDE model, the ODE model utilizes a flow rate through the injury channel to prescribe the appropriate corresponding pressure drop and this is needed for all possible combinations of aggregate heights and densities. To acquire these approximations, we used a two-dimensional PDE model of flow through an extravascular injury channel that was initialized to be filled with various heights and densities of porous material to represent possible aggregate formations. For each of the various aggregates, the model was simulated until the flow came to an equilibrium state and resulting flow rates were recorded; the results were assimilated into what we call the `flow map'. Details of these calculations are as follows.

The geometric configuration for the 2D model is the `H' domain as depicted in \Cref{fig:model}A and as described in detail in our previous work \cite{danes2019density}. Briefly, there are two vertical channels through which blood and wash (buffer) flow. These channels are bifurcated by a horizontal injury channel and, with specified flow rates through the vertical channels, blood flows through the injury channel and out into the wash channel. Within the injury channel, porous aggregates can be placed and fixed to study pressure drops and flows, or the aggregates can grow dynamically according to platelet and ADP equations. Fluid at the fixed flow rate enters the domain at the top of the vertical channels and conditions on pressure at the bottom of the vertical channels are prescribed. For studies with fixed, initial distributions of porous aggregate, the aggregate is considered only on the interior of the horizontal injury channel and is placed in such a way that the nonzero-porosity is centered along a length $L_h - 2\Delta x$, where $\Delta x$ is the size of one computational grid cell. For these simulations, $\Delta x \approx 2 \mu$m. 

Fluid throughout the computational domain is modeled with the Navier-Stokes-Brinkman equations to account for unimpeded flow in the vertical channels and impeded flow through the porous aggregate within the injury channel. The equations take the form:
\begin{eqnarray}
\rho \left( \frac{\partial \vec{u}}{\partial t} + \vec{u}\cdot \nabla \vec{u}\right) &=& -\nabla p -\mu \nabla^2  \vec{u} -\mu \alpha(\theta^B)\vec{u}, \label{fluid_eq1} \\
 \nabla \cdot \vec{u} &=&0 \label{fluid_eq2},
\end{eqnarray}
where $\vec{u}$ and $p$ represent fluid velocity and pressure, respectively, $\mu$ is the dynamic viscosity and $\rho$ is the fluid density. The term $-\mu \alpha(\theta^B)\vec{u}$ represents a frictional resistance to the fluid motion due to the presence of the platelet aggregate with volume fraction of bound platelets, $\theta^B $. Equations \eqref{fluid_eq1}-\eqref{fluid_eq2} are solved numerically using the rotational projection method, as described previously \cite{danes2019density,timmermans1996approximate}. We assume a functional form of the friction term, $\alpha(\theta^B)$, that follows the Carman-Kozeny relation and the particular value of $C_K$ is specified in \Cref{sec:permeable}.
 
Flow rates that correspond to experimental setups at the top of the blood channel $Q_{b}$ and wash channel $Q_{w}$ are used to set the velocity boundary conditions at their respective inlets by using a parabolic velocity profile of the form:
$$ u_{y} =  \frac{3}{2} \frac{Q_{i}}{W} \frac{\left( x - w_{i} \right)}{W} \left( 2 - \frac{x - w_{i}}{W} \right) $$
where $u_{y}$ is the $y-$component (parallel to the vertical channels) of the velocity field $\vec{u}$, $W$ is the width of the blood channel, $w_{i}$ the center point of inlet channel corresponding to index $i$, and $Q_i$ is the given flow rate corresponding to index $i$, with $i=b,w$. Outlet pressure conditions are set using an open boundary condition of the form:
 
$$ \frac{\partial u_n}{\partial n} - p = s, \frac{\partial u_t}{\partial n} = 0 $$
where $u_n$ and $u_t$ are the normal and tangential components of the fluid velocity, respectively, $p$ is the fluid pressure, and $s$ is the prescribed (constant) pressure given from the circuit calculation, as described above.

To construct the flow map in terms of height and density of the initial aggregate distribution, the flow simulations above were carried out for $d_{\botagg},d_{\topagg} \in \{2,4,6,8,10~ \mu m\}$, where $d_{\botagg},d_{\topagg}$ are the height of the bottom and top aggregate, respectively, and for $\theta_{\botagg}^{B},\theta_{\topagg}^{B} \in [0.0,0.06, \cdots 0.6]$, where $\theta_{\botagg}^{B},\theta_{\topagg}^{B}$ are the density of the bottom and top aggregate, respectively. The resulting flow map acts as a ``look-up" table; given ODE model values of $d_{\botagg}$, $d_{\topagg}$, $\theta^{B}_{\botagg}$, $\theta^{B}_{\topagg}$ we do quadrilinear interpolation from table values to determine $Q_{PDE}(d_{\botagg},d_{\topagg},\theta^{B}_{\botagg},\theta^{B}_{\topagg})$.

\subsection{PDE Model of Platelet Aggregation}
\label{sec:appA_PDE}
In this section, we detail all the components of the PDE model of platelet aggregation developed in tandem with the ODE model described in this paper. The evolution equations for the three species of platelets are given by:

\begin{align} 
\frac{\partial P^{m,u}}{\partial t} &= 
\underbrace{- \nabla \cdot \left[ W(\theta^T) \left( \vec{u} P^{m,u} - D_{p} \nabla P^{m,u} \right) \right]}_{\text{advection and diffusion}} \nonumber \\ 
      & \underbrace{- k_{\text{adh}({\bf x})} \left( P_{\max} - P^{b,a} \right) P^{m,u}}_{\text{adhesion}} \nonumber \\
      &\underbrace{- k_{\text{act}}([\text{ADP}]) P^{m,u}}_{\text{activation by ADP}},
\end{align}
\begin{align} 
\frac{\partial P^{m,a}}{\partial t} &= - \nabla \cdot \left[ W(\theta^T) \left( \vec{u} P^{m,a} - D_{p} \nabla P^{m,a} \right) \right] \nonumber\\ 
      & - k_{\text{adh}}({\bf x}) \left( P_{\max} - P^{b,a} \right) P^{m,a} \nonumber \\
      &\underbrace{- k_{\text{coh}} ~ \eta (\bf{x})~ P_{\max} P^{m,a}}_{\text{cohesion}} ~+~ k_{\text{act}}([\text{ADP}]) P^{m,u}, 
\end{align}
\begin{align} 
\frac{\partial P^{b,a}}{\partial t} &=  k_{\text{adh}}({\bf x}) \left( P_{\max} - P^{b,a} \right) P^{m,a} \nonumber \\
      & + k_{\text{coh}}~ \eta(\bf{x})~ P_{\max} P^{m,a}, 
\end{align}
where $\theta^T = \big(\frac{ P^{m,u} + P^{m,a} + P^{b,a} }{ P_{\max} }\big)\theta^{\max}$, and $D_p$ , $k_{\text{adh}}({\bf x})$, $k_{\text{coh}}$, and  $W(\theta^T)$ are the platelet diffusion coefficient, spatially-dependent adhesion rate, cohesion rate coefficient, and hindered platelet flux coefficient, all used as previously described \cite{danes2019density,Leiderman2011}. The parameter, $\eta (\bf{x})$ is a local approximation to the bound platelet fraction that indicates both the location and density of bound platelets to mobile platelets as part of the cohesion rate; the cohesion rate is enhanced where the local bound platelet fraction is increased. We approximate this parameter on a triangular but structured mesh, taking advantage of the interpolated values of $\theta^B$, which we will denote as $\tilde{\theta}^B$, defined using the finite element method. Since $\theta^B$ is defined on the finite element mesh with piecewise linear interpolants, we can approximate $\theta^B$ anywhere in the domain by simply evaluating these interpolants. We allow $\eta(\bf{x})$ at nodes to depend on values of $\theta^B$ that are up to $h_{\text{p}}=2 \mu$m microns away. Given a nodal point $(x_j, y_j)$ on the triangular mesh, we use a weighted average that depends the most strongly on $\theta^B(x_j,y_j)$: 

\begin{eqnarray} \label{eq:9ptstencil}
 \eta(x_j,y_j) &=& \frac{1}{4} \theta^B(x_j, y_j) + \frac{1}{8} \Big[  \tilde{\theta}^{B}(x_j + h_{\text{p}}, y_j) + \tilde{\theta}^B(x_j - h_{\text{p}}, y_j) \nonumber \\
 &+&~ \tilde{\theta}^B(x_j, y_j + h_{\text{p}}) + \tilde{\theta}^B(x_j, y_j - h_{\text{p}}) \Big], \nonumber \\
 &+&  \frac{1}{16} \Big[  \tilde{\theta}^B(x_j + h_{\text{p}}, y_j+ h_{\text{p}} ) + \tilde{\theta}^B(x_j - h_{\text{p}}, y_j + h_{\text{p}}) \nonumber \\
 &+&~ \tilde{\theta}^B(x_j + h_{\text{p}}, y_j - h_{\text{p}}) + \tilde{\theta}^B(x_j - h_{\text{p}}, y_j - h_{\text{p}}) \Big]. \nonumber 
 \end{eqnarray}  
Tracking ADP concentration and its effects on platelet activation in the PDE model is similar to that in the ODE model. Activation occurs through
\begin{equation}  k_{\text{act}} = k_{0}^{adp} \frac{[ADP]}{[ADP] +[ADP]^*} \end{equation}
where $k_{0}^{adp} = 3.4 \text{ s}^{-1}$ is the activation rate and $[ADP]^*$ is the critical concentration of ADP for which there is significant activation of platelets by ADP.  ADP molecules are released by bound, activated platelets and are transported in the fluid by advection and diffusion. Hence, we track the ADP concentration with an advection-diffusion-reaction equation:
\begin{equation}
	\frac{\partial [\text{ADP}]}{\partial t} = -\nabla \cdot \left( \vec{u} [\text{ADP}] - D_{\text{a}} \nabla [\text{ADP}  ] \right)  + \sigma_{\text{release}}
\end{equation}
where $D_{\text{a}}$ is the diffusion coefficient for ADP, and $\sigma_{\text{release}}$ is the release rate, and they are defined exactly the same way as is done in the ODE model, described above. The coupled flow equations are the Navier-Stokes-Brinkman equations described above with the volume fraction due to bound platelets defined as $\theta^B = \big(\frac{P^{b,a}}{P_{\max}}\big)\theta^{\max}.$

\section{Stokes-Brinkman Velocity System}
\label{sec:appB}
To determine the velocity profile through the injury channel described in \Cref{sec:fluid_injury}, we solve a linear system of six equations and six unknowns. Let $y_{\topagg} = d_{h} - d_{\topagg}$.
\begin{align}
& A_{\botagg} + B_{\botagg} = -G_{h} \bigg(\frac{1}{\mu\alpha_{\botagg}}\bigg), \label{lin_1}\\
&A_{\topagg}\exp\big(\sqrt{\alpha_{\topagg}}d_{h}\big) + B_{\topagg}\exp\big(-\sqrt{\alpha_{\topagg}}d_{h}\big) = -G_{h} \bigg(\frac{1}{\mu\alpha_{\topagg}}\bigg),\\
&Ad_{\botagg} + B - A_{\botagg}\exp\big(\sqrt{\alpha_{\botagg}}d_{\botagg}\big) - B_{\botagg}\exp\big(-\sqrt{\alpha_{\botagg}}d_{1}\big) = G_{h} \bigg(\frac{1}{\mu\alpha_{\botagg}} + \frac{d_{\botagg}^{2}}{2\mu}\bigg),\\
&Ay_{\topagg} + B - A_{\topagg}\exp\big(\sqrt{\alpha_{\topagg}}y_{\topagg}\big) - B_{\topagg}\exp\big(-\sqrt{\alpha_{\topagg}}y_{\topagg}\big) = G_{h} \bigg(\frac{1}{\mu\alpha_{\topagg}} + \frac{y_{\topagg}^{2}}{2\mu}\bigg),\\
&A - A_{\botagg}\sqrt{\alpha_{\botagg}}\exp\big(\sqrt{\alpha_{\botagg}}d_{\botagg}\big) + B_{\botagg}\sqrt{\alpha_{\botagg}}\exp\big(-\sqrt{\alpha_{\botagg}}d_{\botagg}\big) = G_{h}\bigg(\frac{d_{\botagg}}{\mu}\bigg),\\
&A - A_{\topagg}\sqrt{\alpha_{\topagg}}\exp\big(\sqrt{\alpha_{\topagg}}y_{\topagg}\big) + B_{\topagg}\sqrt{\alpha_{\topagg}}\exp\big(\sqrt{\alpha_{\topagg}}y_{\topagg}\big) = G_{h} \bigg(\frac{y_{\topagg}}{\mu}\bigg)\label{lin_6}.
\end{align}
\noindent The pressure gradient $G_{h}$, and $\alpha_{\botagg}$, $\alpha_{\topagg}$, $\mu$, $d_{\botagg}$, $d_{\topagg}$, and the width of the injury channel, $d_{h}$ are prescribed. The unknowns are $A$, $B$, $A_{\botagg}$, $B_{\botagg}$, $A_{\topagg}$, $B_{\topagg}$. This system \eqref{lin_1}-\eqref{lin_6} can be written in the following form.
\begin{equation*}
\mathbb{C}
\begin{pmatrix}
A\\
B\\
A_{\botagg}\\
B_{\botagg}\\
A_{\topagg}\\
B_{\topagg}
\end{pmatrix} = \frac{G_{h}}{\mu}
\begin{pmatrix}
\text{-}\frac{1}{\alpha_{\botagg}}\\
\text{-}\frac{1}{\alpha_{\topagg}}\\
\frac{1}{\alpha_{\botagg}} + \frac{d_{\botagg}^2}{2}\\
\frac{1}{\alpha_{\topagg}} + \frac{y_{\topagg}^2}{2}\\
d_{\botagg}\\
y_{\topagg}
\end{pmatrix}
\end{equation*}
\noindent where the matrix $\mathbb{C}$ is defined as
\begin{equation*}
\mathbb{C} = \begin{pmatrix}
0 & 0 & 1 & 1 & 0 & 0\\
0 & 0 & 0 & 0 & e^{\sqrt{\alpha_{\topagg}}d_{h}} & e^{-\sqrt{\alpha_{\topagg}}d_{h}} \\
d_{\botagg} & 1 & \text{-}e^{\sqrt{\alpha_{\botagg}}d_{\botagg}} & \text{-}e^{-\sqrt{\alpha_{\botagg}}d_{\botagg}} & 0 & 0\\
y_{\topagg} & 1 & 0 & 0 & \text{-}e^{\sqrt{\alpha_{\topagg}}y_{\topagg}} & \text{-}e^{-\sqrt{\alpha_{\topagg}}y_{\topagg}}\\
1 & 0 & \text{-}\sqrt{\alpha_{\botagg}} e^{\sqrt{\alpha_{\botagg}}d_{\botagg}} & \sqrt{\alpha_{\botagg}} e^{-\sqrt{\alpha_{\botagg}}d_{\botagg}} & 0 & 0 \\
1 & 0 & 0 & 0 & \text{-}\sqrt{\alpha_{\topagg}}e^{\sqrt{\alpha_{\topagg}}y_{\topagg}} & \sqrt{\alpha_{\topagg}}e^{-\sqrt{\alpha_{\topagg}}y_{\topagg}}
\end{pmatrix}.
\end{equation*}

Note that the right-hand side of the system is linearly dependent on the pressure gradient $G_{h}$. Therefore, we assume $G_{h} = 1$ and solve the linear system of six equations for the six constants that prescribe the flow velocity profile. The velocity profile is used to calculate the resistance $R_{M}$ through the injury channel as described below
$$
R_{M} = \frac{L_{h}}{L_{z}\int_{0}^{d_{h}}u(y)dy}.
$$
The effects of the growing aggregate on the flow velocity through the injury channel are captured in the quantity $R_{M}$, which is an input parameter in the hydraulic circuit (HC) system describing the flow of blood through the bleeding chip. The true $G_{h}$ is determined by solving the HC system and is used to scale the velocity profile that was obtained with $G_{h}=1$.
\section{Model Parameters}
\label{sec:appC}

 \begin{table}[!h]
 \footnotesize
\centering
\caption{{\bf{Diffusion coefficients, ADP granule release, and upstream platelet number density.}} }
\vspace{1.0pt}
\begin{tabular}{|l|l|l|}
\hline
Species & Value & Reference\\
\hline
Platelets & $2.5 \times 10^{-7}$ cm$^{2}$/s & \cite{turitto1972platelet}\\ 
ADP & $5 \times 10^{-6}$ cm$^{2}$/s & \cite{grabowski1978platelet}\\
$\hat{A}$ & $2(10)^{-17}$ moles of ADP/plt&  \cite{Leiderman2011}\\
$P^{up,*}$ & $2.5(10)^{5}$ plt/$\mu$L&  \cite{Leiderman2011}\\
\hline
\end{tabular}
\label{params:diff}
\end{table}

 \begin{table}[!h]
 \scriptsize
\centering
\caption{{\bf{Platelet transitions.}} }
\vspace{1.0pt}
\begin{tabular}{lllll}
\hline
{\bf{Transition}} & {\bf{Initial State}} & {\bf{Final State}} & ${\bf{M^{-1}s^{-1}}}$ & ${\bf{s^{-1}}}$\\
\hline
\parbox{9em}{Unactivated platelet adhering to SE} & $P^{m,u}$ & $P^{b,a}$ & $k_{0}^{adh} = 2.0 \times 10^{10}$ \cite{Kuharsky2001,Leiderman2011} & \\
\hline
\parbox{8em}{Activated platelet adhering to SE} & $P^{m,a}$ & $P^{b,a}$ & $k_{0}^{adh} = 2.0 \times 10^{10}$ & \\
\hline
\parbox{8em}{Activated platelet cohering to bound platelet} & $P^{m,a}$ & $P^{b,a}$ & & \parbox{7em}{$k_{\text{coh}} \times P^{\max} = 1 \times 10^{4}$ \cite{Leiderman2011} }\\
\hline
\parbox{8em}{Platelet activation by ADP} & $P^{m,u}$ & $P^{m,a}$ & & $k_{0}^{adp} = 3.4$ \cite{gear1994platelet} \\
\hline
\end{tabular}
\label{params:transitions}
\end{table}

\subsection{Calibrated Parameters}
\label{sec:calibration}
Both adhesion and cohesion parameters detailed in \Cref{sec:coh_adh_dil} were tuned with the PDE and MFA models. Additionally, constants $c_{1}$ and $c_{2}$ used in the definitions of the diffusive fluxes of ADP described in \Cref{sec:adp_diff} were determined through calibration with the PDE model. Values are summarized in \Cref{params:adh_coh}.

 \begin{table}[!h]
 \footnotesize
\centering
\caption{{\bf{Calibrated parameters.}} }
\vspace{1.0pt}
\begin{tabular}{|l|l|l|l|}
\hline
{\bf{Description}} & {\bf{Parameter}} & {\bf{Value}} \\
\hline
Partition constant & $k_{b}$ & 20  \\
\hline
Partition constant & $s_{b}$ &  0.155  \\
\hline
Adhesion thickness & $d_{se}$ & 2 \ $\mu$m  \\
\hline
Gradient constants & $c_{1}, c_{2}$ & $1/2$, $1/4$  \\
\hline
\end{tabular}
\label{params:adh_coh}
\end{table}

\subsection{Aggregate Permeability}
\label{sec:permeable}
When considering aggregate permeability, the significant differences between `white' and `red' clots must be considered. A white clot is usually formed under high shear and is predominantly made up of platelets whereas a fibrin-rich red clot forms under low shear or static conditions. \Cref{params:comp_perm} lists the computed permeability comparison results from the literature. It is shown that white clots are significantly more permeable than red clots. In addition to the type of clot, the flow conditions and experimental setup generate significant variability. For example results in \cite{wufsus2013hydraulic} were generated by measuring the permeability of platelet-rich aggregates at various volume fractions. Specifically, the authors induced clotting with thrombin in suspension with known platelet densities and then measured the permeation of a buffer solution through the clots at defined pressure gradients (constant pressure).  Permeability values of  $1.5 \times 10 ^{-5} \pm 3.3 \times 10^{-6} - 6.1 \times 10^{-3} \pm 1.1 \times 10^{-3} \ \mu \text{m}^{2}$ were calculated from platelet-rich clots (PRC) with platelet volume fractions of $0.31 - 0.61$. In conclusion, both flow conditions and experimental set up yield a wide range of permeabilities.
 \begin{table}[!h]
 \footnotesize
\centering
\caption{{\bf{Permeability comparison.}} }
\vspace{1.0pt}
\begin{tabular}{|l|l|l|l|}
\hline
{\bf{Clot Type}} & {\bf{Permeability [$\mu$m$^{2}$]}} & {\bf{Reference}} \\
\hline
Red & $ 5.45 \times 10^{-6}$ & Muthard and Diamond (2012) \cite{colace2012thrombus}  \\
\hline
Red & $ 1.5 \times 10^{-5}$ & Wufsus et al. (2013) Platelet 61\% \cite{wufsus2013hydraulic}  \\
\hline
Red & $ 6.1 \times 10^{-3}$ & Wufsus et al. (2013) Platelet 31\% \cite{wufsus2013hydraulic}   \\
\hline
White & $ 0.2$ & Kobayashi et al. (2016) \cite{KOBAYASHI2016} \\
\hline
White & $2.8571 $ & calibrated with PDE/MFA models   \\
\hline
\end{tabular}
\label{params:comp_perm}
\end{table}
Given the variability in permeability data found in the literature, we tuned the constant $C_{K}$ using input from both PDE and MFA models. More specifically, we assume a functional form of the Brinkman coefficient that follows the Carman-Kozeny relation \cite{mccabe2005unit}
$$
\alpha(\theta^{B}) = C_{K}\frac{(\theta^{B})^{2}}{(1-\theta^{B})^{3}}, \ \ \ \ C_{K} = 3.5 \times 10^{7} \text{ cm}^{-2}.
$$
This particular value of $C_{K}$ was determined by performing flow simulations with the PDE model in which the injury channel was completely filled with aggregate at a minimum porosity until it matched the experimentally observed flow rate of $0.12 \ \mu$L/min.

\section{Non-uniform Aggregate Formation}
\label{sec:appD}
\begin{figure}[!h]
\centering
 \vspace{-0.25cm}
 \hspace{-5.5cm} {\bf{(A)}}  \hspace{6.25cm} {\bf{(B)}} \\
\begin{tabular}{cc}
\includegraphics[width=0.45\textwidth]{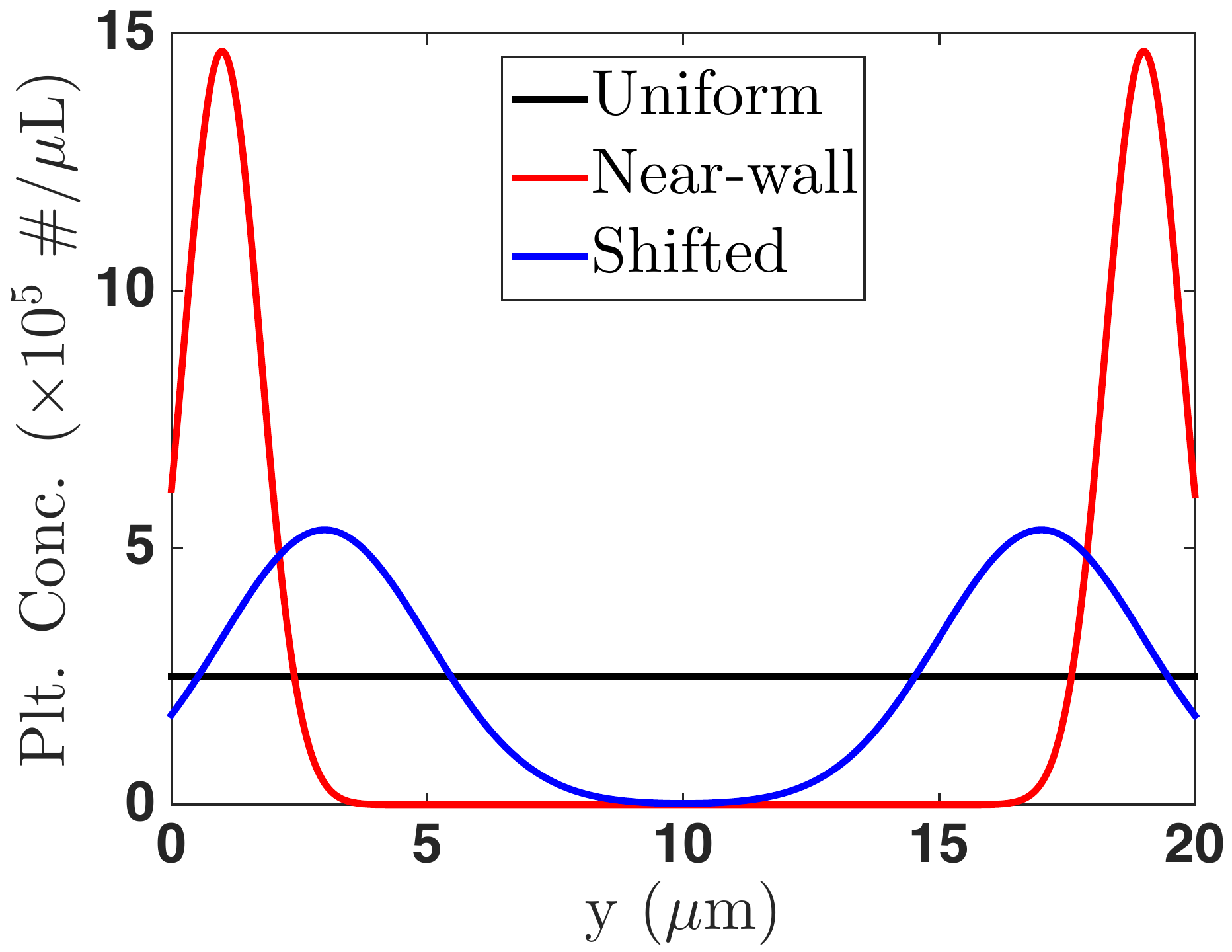}&
\includegraphics[width=0.45\textwidth]{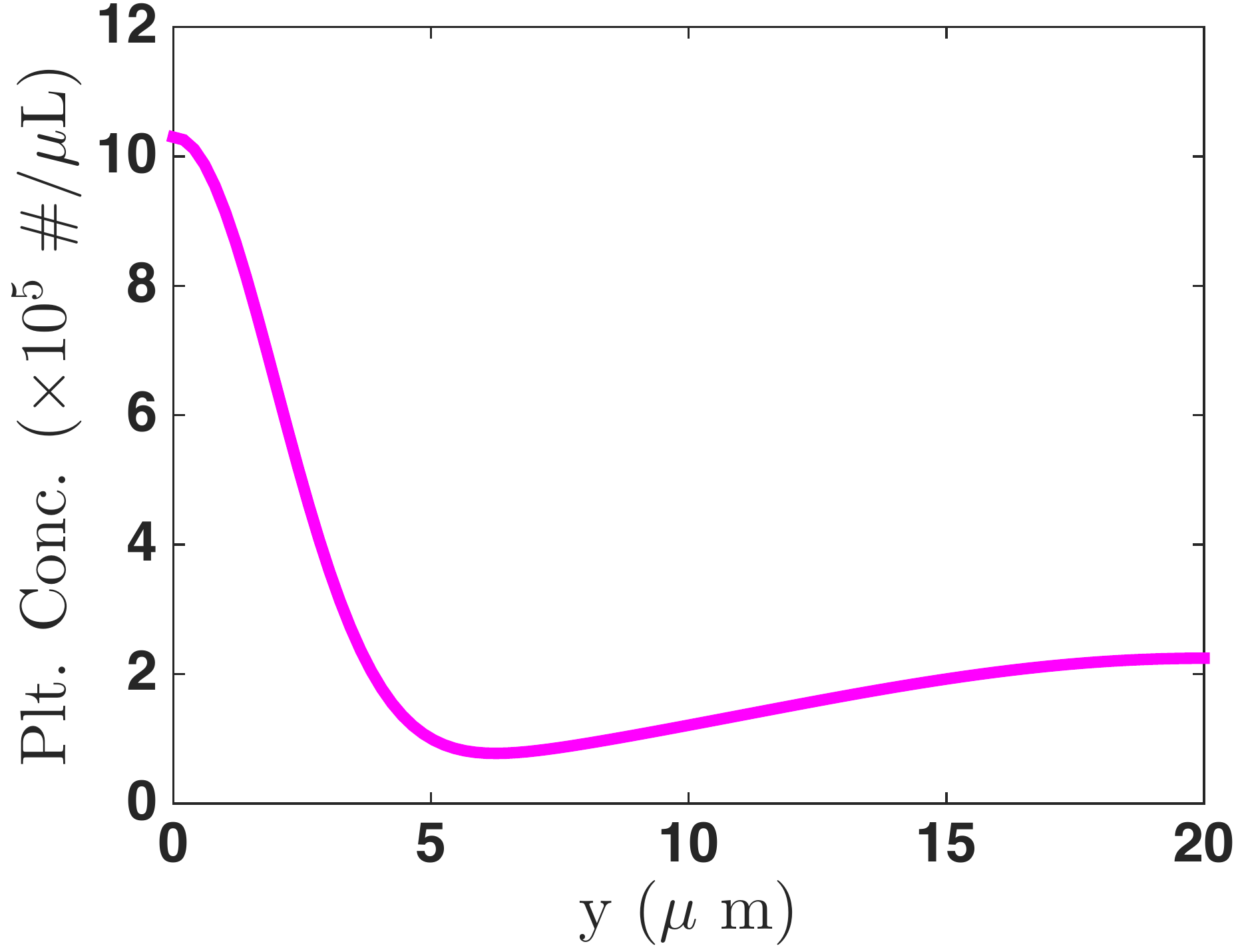}
\end{tabular}
\caption{{\bf{Upstream Platelet Distributions:}} (A)Uniform ($P^{up}(y) = P^{up,*}$), near-wall, and shifted upstream platelet distributions. These are motivated by {\it{in vivo}} \cite{tangelder1985distribution} and {\it{in vitro}} \cite{eckstein1991model,eckstein1988conditions,tilles1987near,yeh1994estimated} observations. Additional details are found in \cite{Leiderman2011}.}
\label{Pup_dist}
\end{figure}

We considered uniform, near-wall, and shifted upstream platelet distributions $P^{up}(y)$ found in \Cref{Pup_dist}A and compared the formation of the resulting aggregates. Additionally, we examined the effects of an asymmetric upstream platelet distribution shown in \Cref{Pup_dist}B. The distributions are defined by the distance $y$ from the bottom wall of the injury channel. Therefore, $y = 20 \ \mu$m is the location associated with the top wall. Both sets of {\it{in silico}} experiments utilized ADP-independent activation where $A([ADP]_{\text{agg}},[ADP]_{\text{BL}}) = 1$. 

\begin{figure}[!h]
\centering
\begin{tabular}{cc}
\hspace{-4.0cm} {\bf{(A)}} & \hspace{-4.0cm} {\bf{(B)}} \\
\includegraphics[width=0.45\textwidth]{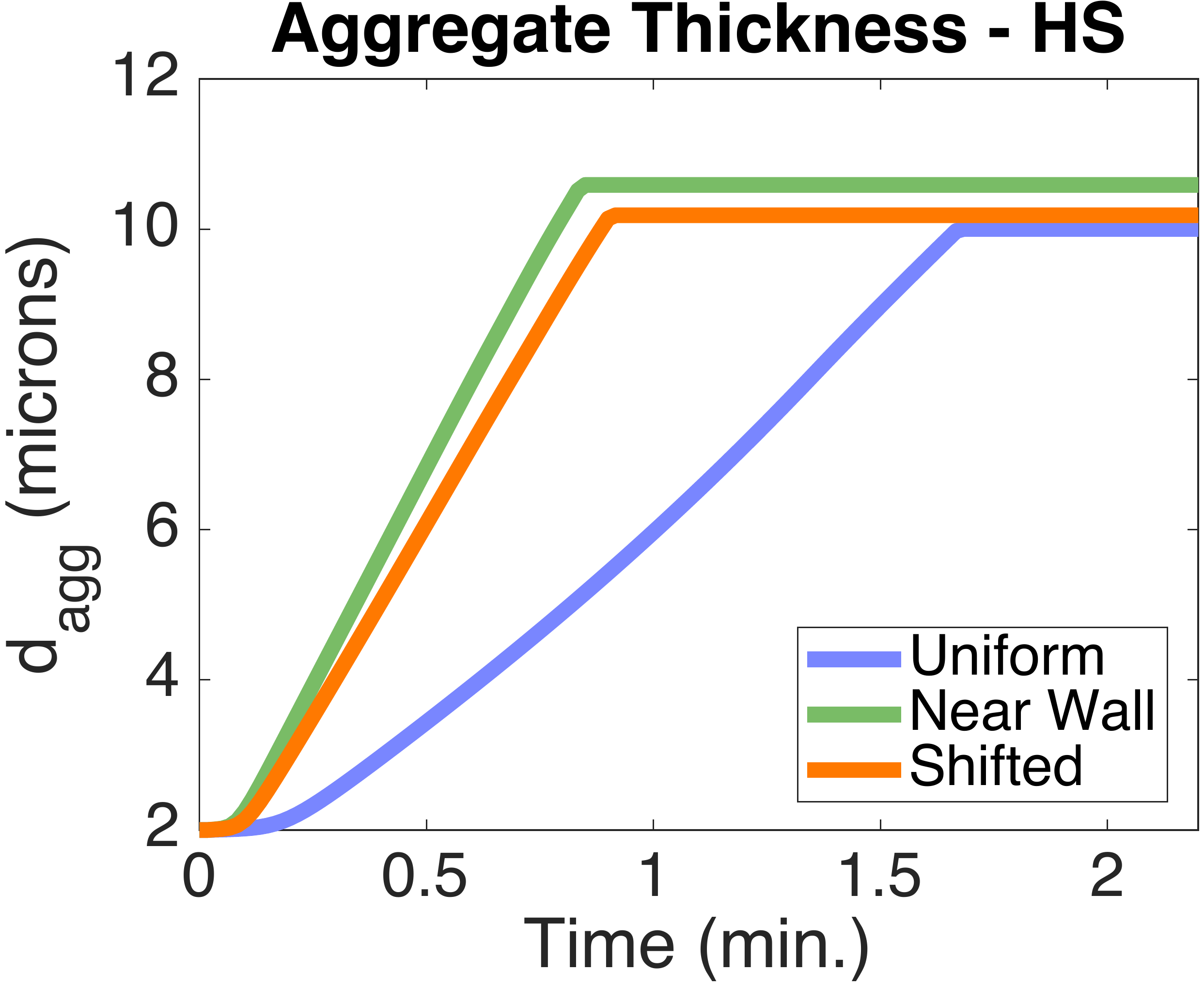}&
\includegraphics[width=0.45\textwidth]{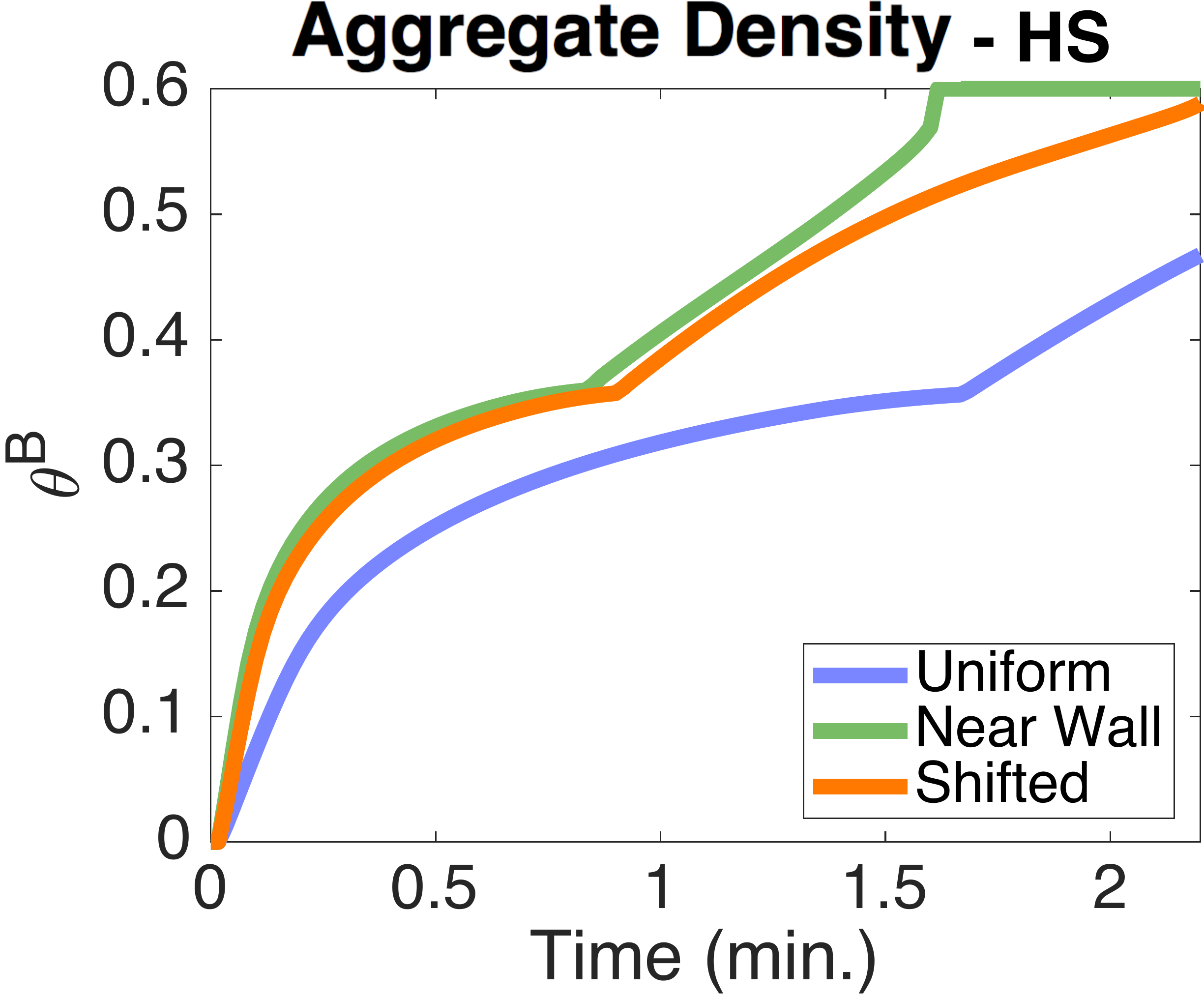}
\end{tabular}
\begin{tabular}{cc}
\hspace{-4.0cm} {\bf{(C)}} & \hspace{-4.0cm} {\bf{(D)}} \\
\includegraphics[width=0.45\textwidth]{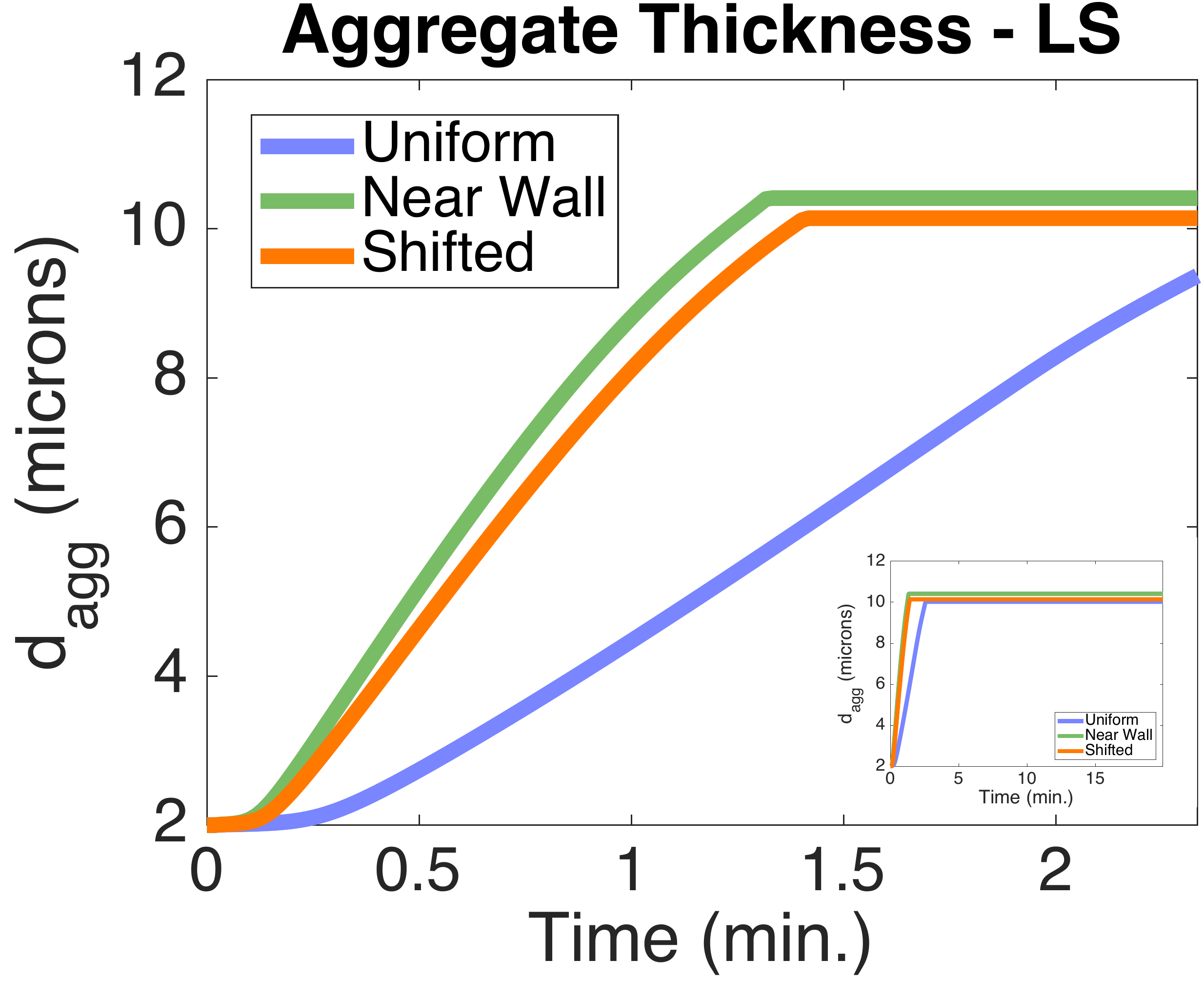}&
\includegraphics[width=0.45\textwidth]{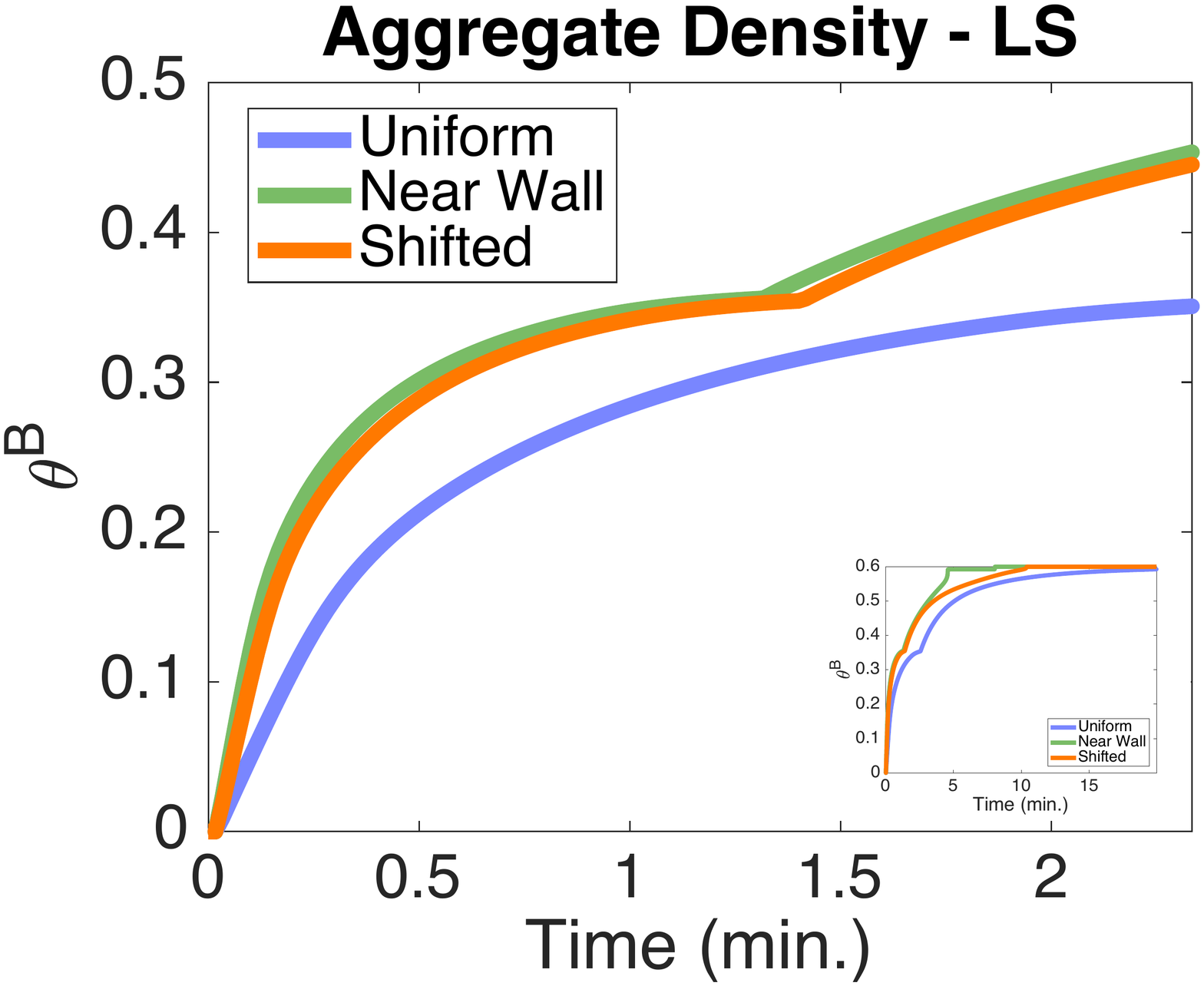}
\end{tabular}
\caption{{\bf{Symmetric Aggregate Formation:}} Given different symmetric upstream platelet distributions $P^{up}(y)$: uniform, near-wall, and shifted (see \Cref{Pup_dist}A) with ADP-independent platelet activation $A([ADP]_{\text{agg}}, [ADP]_{\text{BL}}) = 1$ under high shear (HS) conditions $(\gamma = 2400 \text{ s}^{-1}$) {\bf{A)}} the aggregate thickness {\bf{B)}} and density vary as functions of time. Similar comparisons under low shear (LS) conditions $(\gamma = 100 \text{ s}^{-1}$) yield aggregates of varying thickness {\bf{(C)}} and density {\bf{(D)}}. Insets show results after 20 minutes of simulation time confirming occlusion.}
\label{symmResults}
\end{figure}

Under both high and low shear conditions, aggregates simulated with the uniform platelet distribution are the slowest growing in both thickness and density (\Cref{symmResults} A-B). The use of a near-wall platelet distribution, which has the highest concentration of platelets closest to the walls, results in the fastest occlusion time of approximately 1.75 minutes.  The shifted platelet distribution produced aggregates that occlude approximately 30-45 seconds slower. A similar trend is seen in the low shear conditions \Cref{symmResults} C-D, however occlusion times are longer. The results from these symmetric cases yield aggregates with varying thicknesses and densities, identifying the upstream platelet distribution $P^{up}(y)$ as a modifier of aggregate formation. Additionally, we were motivated to perform experiments that address the effects of asymmetric upstream platelet distribution on aggregate formation. 

\begin{figure}[!h]
\centering
\begin{tabular}{cc}
\hspace{-4.0cm} {\bf{(A)}} & \hspace{-4.0cm} {\bf{(B)}} \\
\includegraphics[width=0.45\textwidth]{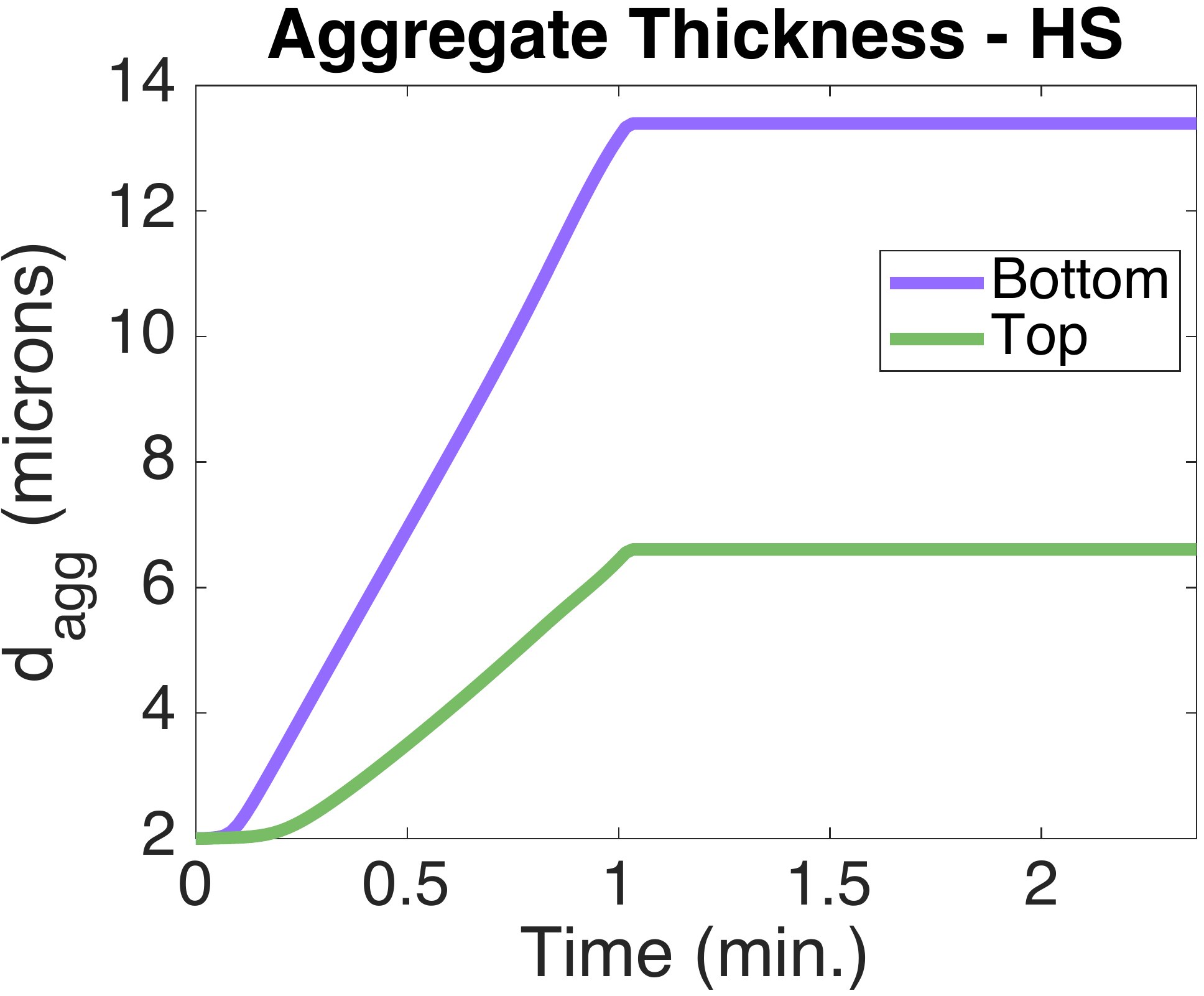}&
\includegraphics[width=0.45\textwidth]{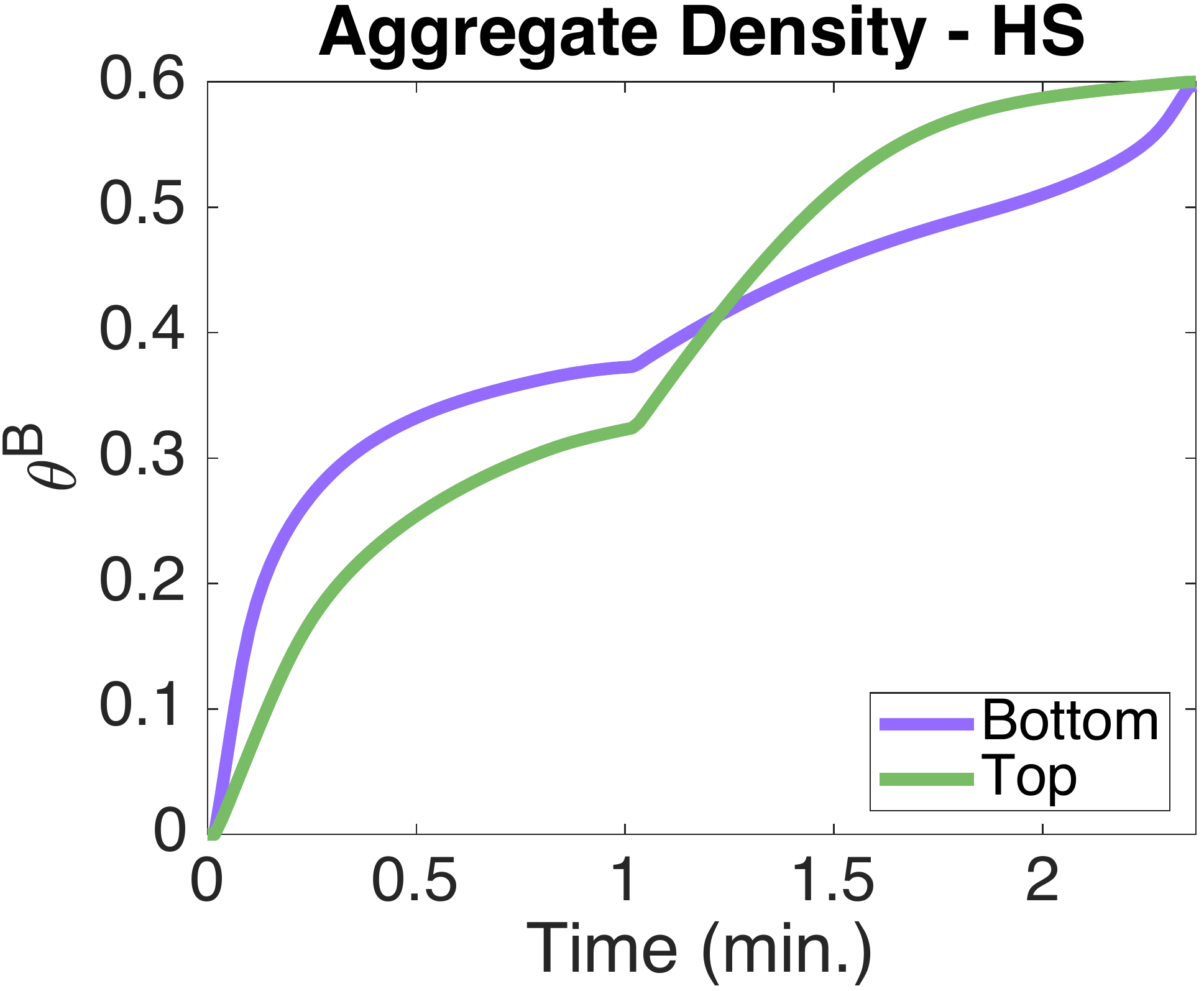}
\end{tabular}
\begin{tabular}{cc}
\hspace{-4.0cm} {\bf{(C)}} & \hspace{-4.0cm} {\bf{(D)}} \\
\includegraphics[width=0.45\textwidth]{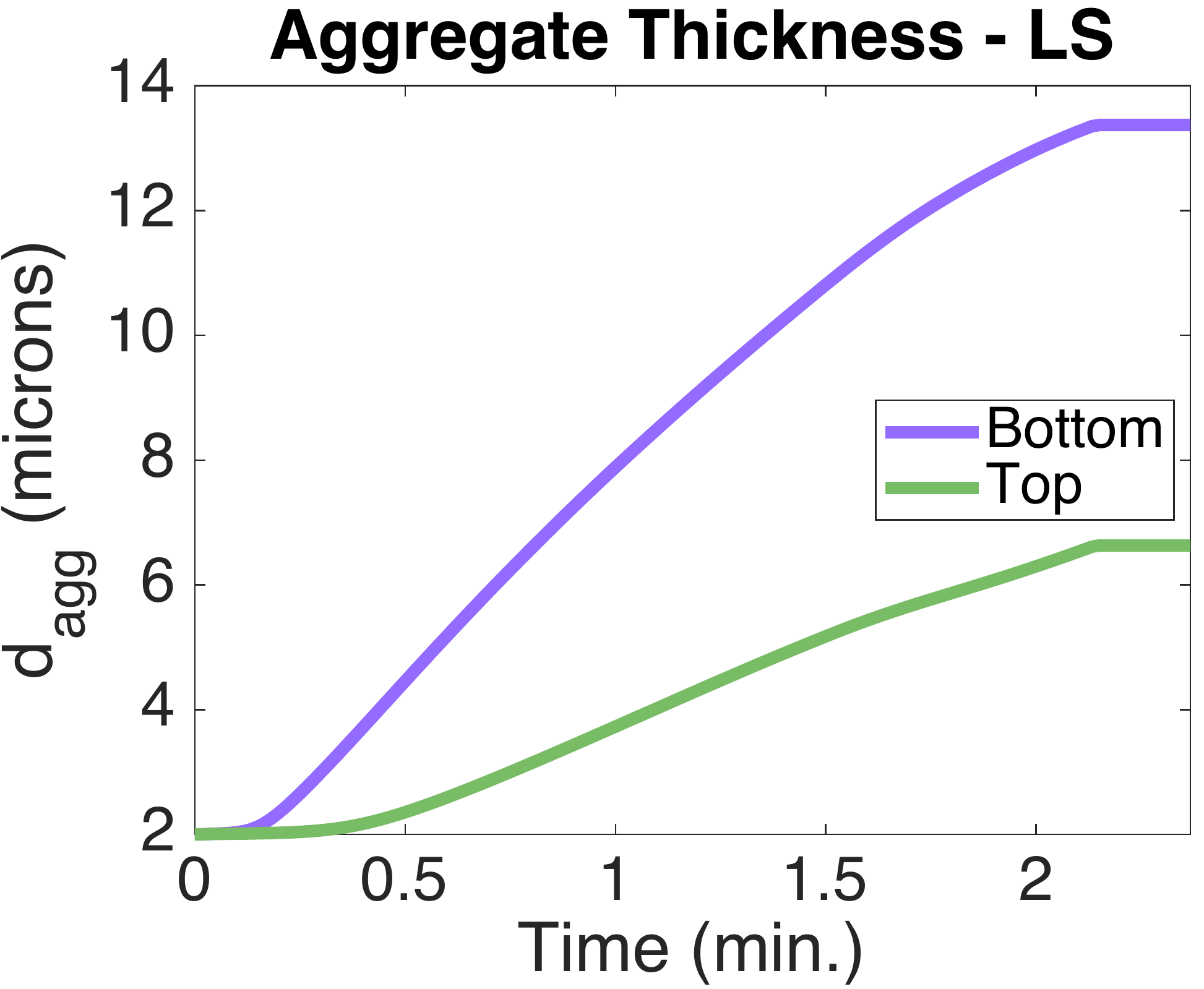}&
\includegraphics[width=0.45\textwidth]{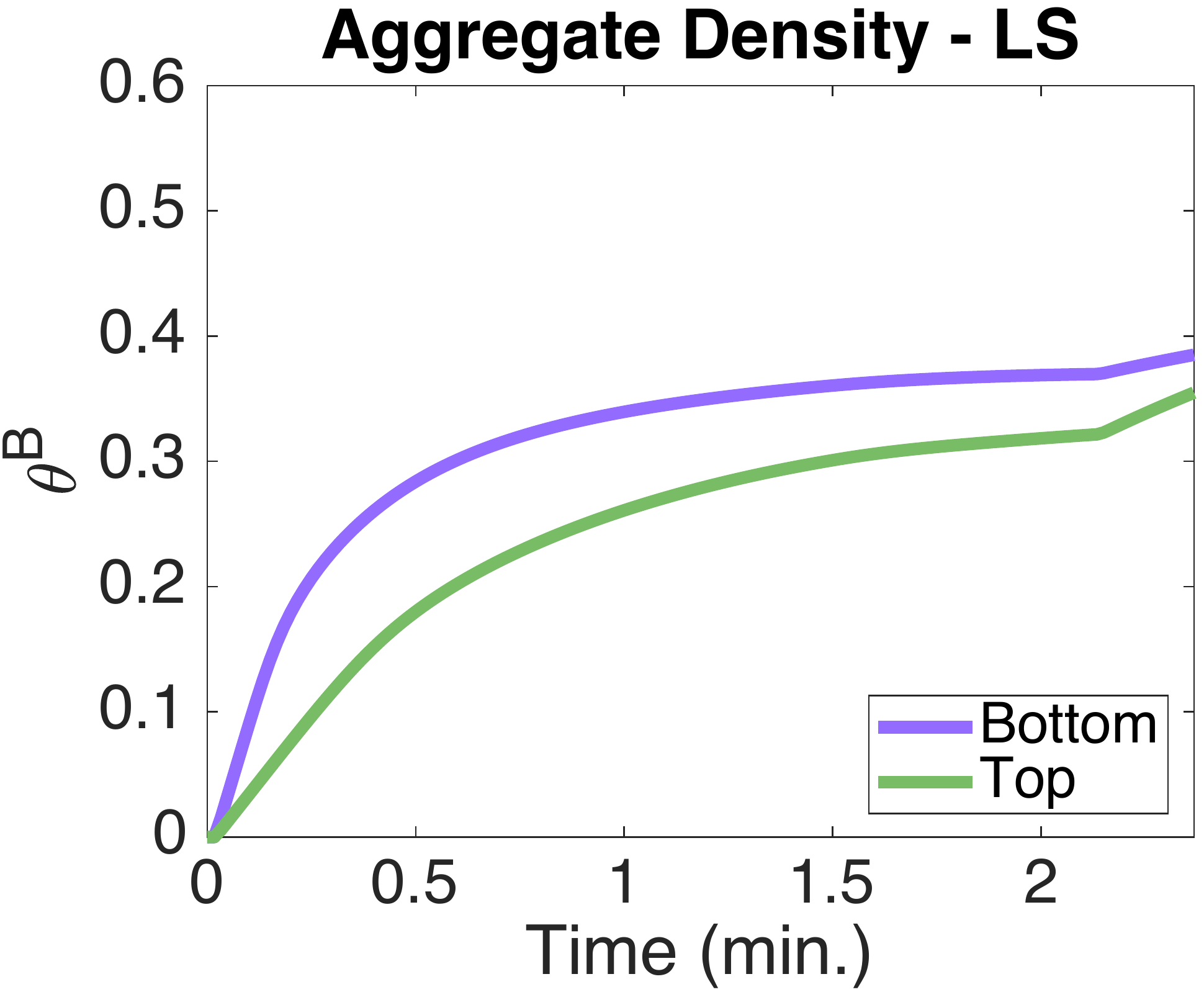}
\end{tabular}
\vspace{-0.2cm}
\caption{{\bf{Asymmetric Aggregate Formation:}} Given an asymmetric upstream platelet distribution (see \Cref{Pup_dist}B) with ADP-independent platelet activation ($A([ADP]_{\text{agg}}, [ADP]_{\text{BL}} ) = 1$) and high shear conditions ($\gamma = 2400 ^{-1}$), the aggregate thickness {\bf{A)}} and aggregate density {\bf{B)}} as functions of time. Similar comparisons under low shear (LS) conditions $(\gamma = 100 \text{ s}^{-1}$) yield aggregates of varying thickness {\bf{(C)}} and density {\bf{(D)}}.}
	\label{asymmResults}
\end{figure}

Figure \ref{asymmResults} A-D show aggregate thicknesses and densities associated with the bottom (purple) and top (green) walls. We chose an asymmetric distribution found in \Cref{Pup_dist} B, where there is a greater concentration of platelets near the bottom wall of the injury channel. The aggregate grows asymmetrically in both thickness and density. In high shear (HS) conditions, the bottom aggregate region is thicker and denser after 1.25 minutes. There is, however, a significant increase in the density of the top aggregate after 1.25 minutes, resulting in a thin but dense top aggregate. Similar behavior is seen in results under low shear (LS) conditions, however, there is a delay. Therefore, the model is able to produce aggregates of different orientations, thicknesses and densities by varying model inputs such as upstream platelet distribution and flow conditions.

\end{document}